\documentclass[pdflatex,sn-mathphys-num]{sn-jnl}

\usepackage{graphicx}%
\usepackage{multirow}%
\usepackage{amsmath,amssymb,amsfonts}%
\usepackage{amsthm}%
\usepackage{mathrsfs}%
\usepackage[title]{appendix}%
\usepackage{xcolor}%
\usepackage{textcomp}%
\usepackage{manyfoot}%
\usepackage{booktabs}%
\usepackage{algorithm}%
\usepackage{enumitem}
\usepackage{subfigure}
\usepackage{algorithmicx}%
\usepackage{algpseudocode}%
\usepackage{listings}%
\usepackage{bm}

\theoremstyle{thmstyleone}%
\theoremstyle{thmstyletwo}%

\theoremstyle{thmstylethree}%

\begin{document}

\title[Article Title]{Seamlessly Natural: Image Stitching with Natural Appearance Preservation}


\author[1]{\fnm{Gaetane Lorna N.} \sur{TCHANA}}\email{gaetane.tchana@facsciences-uy1.cm}

\author[1]{\fnm{Damaris Belle M.} \sur{Fotso}}\email{makwate.fotso@facsciences-uy1.cm}
\author[2]{\fnm{Antonio} \sur{Hendricks}}\email{a.hendricks1@ufl.edu}

\author*[2]{\fnm{Christophe} \sur{Bobda}}\email{cbobda@ece.ufl.edu}

\affil[1]{\orgname{University of Yaounde I}, \orgaddress{\city{Yaoundé}, \postcode{812}, \country{Cameroon}}}

\affil[2]{\orgname{University of Florida, ECE}, \orgaddress{\city{Gainesville}, \postcode{32611}, \state{FL}, \country{US}}}


\abstract{This paper introduces \textbf{SENA (SEamlessly NAtural)}, a geometry-driven image stitching approach that prioritizes \textbf{structural fidelity} in challenging real-world scenes characterized by parallax and depth variation. Conventional image stitching relies on homographic alignment, but this rigid planar assumption often fails in dual-camera setups with significant scene depth, leading to distortions such as visible warps and spherical bulging.

SENA addresses these fundamental limitations through three key contributions. First, we propose a hierarchical affine-based warping strategy, combining global affine initialization with local affine refinement and smooth free-form deformation. This design preserves local shape, parallelism, and aspect ratios, thereby avoiding the hallucinated structural distortions commonly introduced by homography-based models. Second, we introduce a geometry-driven adequate zone detection mechanism that identifies parallax-minimized regions directly from the disparity consistency of RANSAC-filtered feature correspondences, without relying on semantic segmentation. Third, building upon this adequate zone, we perform anchor-based seamline cutting and segmentation, enforcing a one-to-one geometric correspondence across image pairs by construction, which effectively eliminates ghosting, duplication, and smearing artifacts in the final panorama.

Extensive experiments conducted on challenging datasets demonstrate that SENA achieves alignment accuracy comparable to leading homography-based methods, while significantly outperforming them in critical visual metrics such as shape preservation, texture integrity, and overall visual realism.}




\maketitle
\section{Introduction}
\label{sec:intro}

The rise of digital imaging technology has fundamentally transformed how individuals engage with digital content and their surrounding environments \cite{nghonda2023enable}, directly enabling the development of advanced immersive experiences such as virtual and augmented reality. A critical requirement for these platforms is the ability to capture complex scenes from multiple viewpoints to provide users with continuous, seamless panoramic visuals. This is typically achieved by synthesizing and merging image data acquired simultaneously from two or more cameras in a process known as image stitching. Most image stitching methods establish correspondences between images using salient visual features \cite{lowe2004distinctive,bay2006surf,ji2022research,liu2023utilization,potje2024xfeat}. These correspondences define a geometric transformation that is used to warp images into a common coordinate frame. The final panorama is then obtained by merging overlapping regions, either directly or with additional mechanisms such as seam selection or blending to reduce visual artifacts. Contributions in the state-of-the-art are most often focused on the refinement or acceleration of one or more of these individual stages.

  \begin{figure}[H]
  \centering
  \setlength{\tabcolsep}{2pt} 
  \renewcommand{\arraystretch}{0.3} 
  \begin{tabular}{ccc}
    \begin{minipage}[b]{0.28\linewidth}
      \centering
      \includegraphics[width=\linewidth]{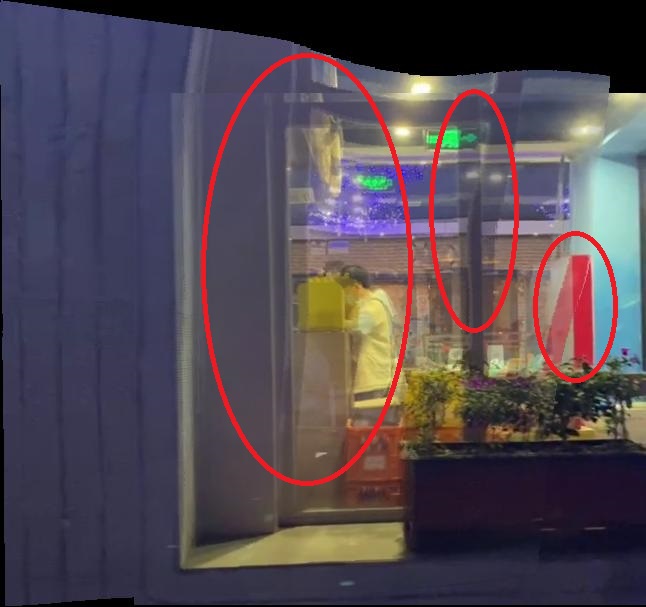}
      \\[1pt] APAP
    \end{minipage} &
    \begin{minipage}[b]{0.29\linewidth}
      \centering
      \includegraphics[width=\linewidth]{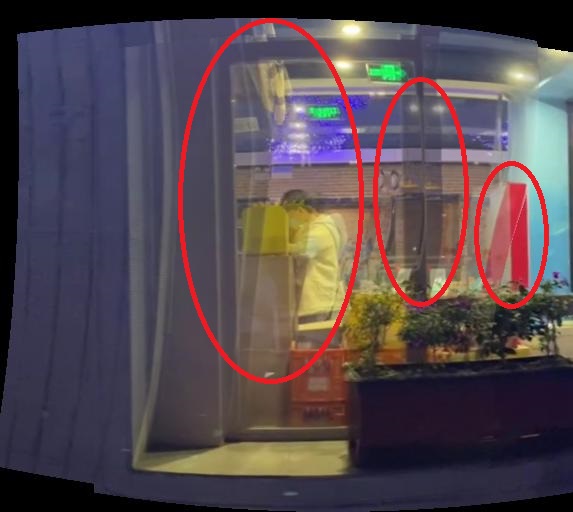}
      \\[1pt] ELA
    \end{minipage} &
    \begin{minipage}[b]{0.30\linewidth}
      \centering
      \includegraphics[width=\linewidth]{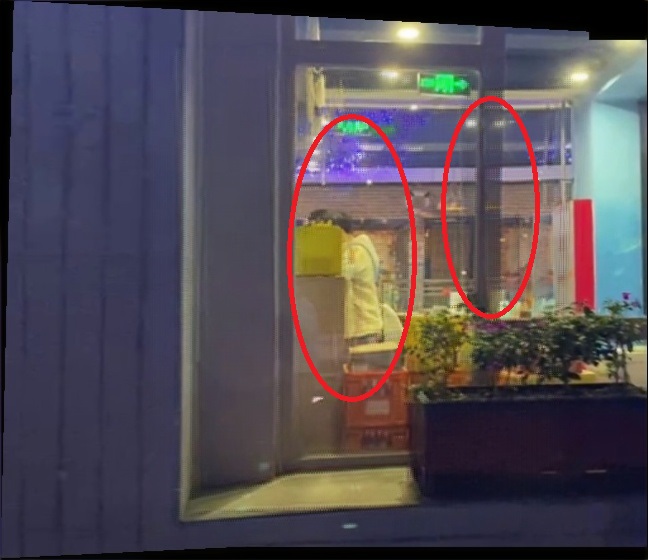}
      \\[1pt] UDIS
    \end{minipage} \\[2pt]
    \begin{minipage}[b]{0.30\linewidth}
      \centering
      \includegraphics[width=\linewidth]{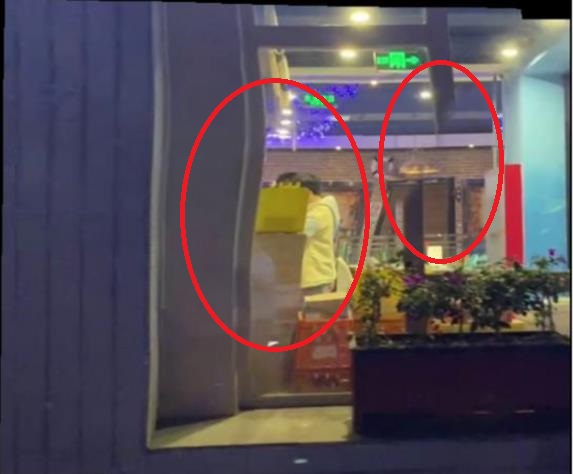}
      \\[1pt] UDIS++
    \end{minipage} &
    \begin{minipage}[b]{0.30\linewidth}
      \centering
      \includegraphics[width=\linewidth]{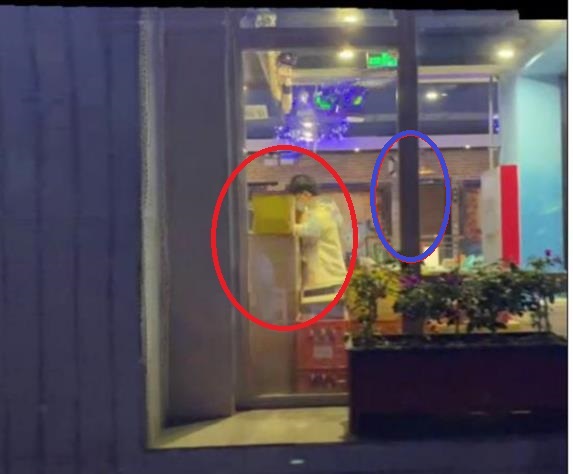}
      \\[1pt] SEAMLESS
    \end{minipage} &
    \begin{minipage}[b]{0.30\linewidth}
      \centering
      \includegraphics[width=\linewidth]{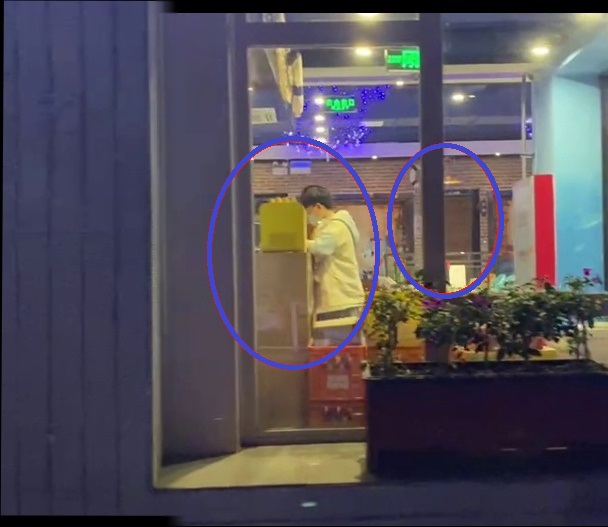}
      \\[1pt] OURS
    \end{minipage}
  \end{tabular}
 \caption{Limitations of Existing Stitching Methods}
  \label{limitations_SOA}
\end{figure}

Geometric warping plays a central role in stitching quality, with homography remaining the dominant transformation model. Owing to its eight degrees of freedom, homography can accurately model full perspective effects and provides exact alignment for planar scenes or pure camera rotations. However, real-world scenes are rarely planar and often contain significant depth variation. In such cases, homography attempts to reconcile non-planar motion using a planar model, leading to over-flexible geometric distortions such as stretching, bending of straight lines, and shape deformation.

A large body of work in image stitching focuses on improving geometric alignment through increasingly flexible warping models. Hybrid deformation models~\cite{wen2022structure} estimate multiple homographies for different scene regions or depth layers to better preserve structural consistency. Other methods replace global homography with local spline-based registration, such as thin-plate spline (TPS) warps, to increase alignment flexibility~\cite{nie2023parallax,chen2024seamless}. While these approaches often improve overlap alignment, their additional degrees of freedom frequently introduce excessive geometric deformation, leading to stretched textures, bent lines, and loss of structural realism. Even combinations of homographic and affine transformations~\cite{li2023combined} reduce but do not eliminate these artifacts, particularly in scenes with strong parallax.
Figure~\ref{limitations_SOA} illustrates these limitations in representative state-of-the-art methods (e.g., APAP \cite{zaragoza2013projective}, ELA \cite{li2017parallax}, UDIS \cite{nie2021unsupervised}, UDIS++ \cite{nie2023parallax}, and SEAMLESS \cite{chen2024seamless}), where visual artifacts such as ripples, ghosting, and information loss are evident (highlighted in red). These artifacts primarily arise from the inherent tendency of homography-based models to compensate for depth induced parallax through non-physical geometric distortions.

In parallel, many stitching methods address misalignment artifacts not by correcting geometry, but by carefully selecting where images are merged. Seam-based approaches define a stitching path through the overlapping region that minimizes visual discrepancy according to a predefined cost function. Representative works~\cite{huang2021semantic,qin2021image,chai2022upscaling,garg2020stitching} guide seam placement using cues such as intensity differences, gradients, semantic labels, or feature consistency. Although effective in visually hiding some misalignments, these methods often rely on heavy preprocessing pipelines—including semantic segmentation, feature classification, or depth inference which increase computational cost and sensitivity to errors. Furthermore, seam optimization alone does not guarantee geometric consistency across the seam, leaving duplication or ghosting unresolved in many cases.
\\
Overall, stitching methods still face three major challenges. First, global or overly flexible warps (e.g., homography-only or spline-based models) often introduce geometric distortions and compromise structural fidelity. Second, seam selection strategies based on cost-function minimization frequently depend on complex preprocessing steps such as semantic segmentation or feature classification, which increases sensitivity to inaccuracies. Third, even when alignment is locally accurate, blending across regions often lacks consistency, leading to duplication, stretching, or ghosting that reduce the visual realism of the stitched panorama.
\\
This work addresses the above limitations through three contributions. (1) We introduce a hierarchical deformation strategy that combines global affine initialization (using RANSAC-filtered correspondences), local affine refinement within the overlap region, and a smooth free-form deformation (FFD) field regulated by seamguard adaptive smoothing. This multi-scale design preserves structural fidelity while avoiding the excessive distortions of global models and the instability of spline-based warps. (2) within the same seam-selection paradigm, we propose an adequate zone detection strategy that departs from prior methods based on semantic segmentation. Instead of such complex preprocessing, we analyze the disparity consistency of feature correspondences, providing a lightweight, model-free criterion that robustly identifies parallax-minimized regions. (3) we perform anchor-based segmentation aligned with the detected adequate zone, ensuring structural consistency across image pairs and enabling seamless stitching. Unlike prior approaches that stop at defining a seamline and then rely on blending to hide artifacts, our method partitions both images into corresponding vertical slices anchored by refined keypoints. This guarantees one-to-one geometric correspondence across segments, reducing duplication and ghosting in the final panorama.

\section{Related work}
\label{sec:Related}
Research in image stitching has explored a wide range of strategies, from feature-based methods to deep learning approaches, yet three major limitations persist across the state-of-the-art.

\begin{itemize}

\item \textbf{Geometric Distortion from Global Warps:} The dominant homography model Zheng et al.\cite{zheng2019novel}, Yadav et al. \cite{yadav2018selfie}, Jia et al. \cite{jia2021leveraging} is frequently used because its 8 degrees of freedom allow it to model full perspective effects. However, when applied to real-world scenes with parallax and depth variation, the homography is forced to reconcile conflicting motions, which often results in non-uniform distortions like spindle-shaped warps, unnatural stretching, or spherical bulging. Advanced hybrid warps, such as the "as-projective-as-possible" (APAP) \cite{zaragoza2013projective}, and elastic warping improve flexibility but risk overfitting or over-flexibility, which can produce local stretching artifacts.

\item \textbf{Reliance on Complex Preprocessing}: Many seamline optimization methods(\cite{huang2021semantic}, \cite{qin2021image}, \cite{chai2022upscaling}) formulate a cost function over the overlapping region and search for a seam that minimizes this cost, ideally passing through visually consistent areas. However, these approaches often depend on complex preprocessing steps, such as semantic segmentation or depth estimation, to guide the cost map, increasing computational cost and sensitivity to errors and inaccuracies.

\item \textbf{Structural Assumptions and Computational Overhead}:
\begin{itemize}

\item \textbf{Plane or Multi-Homography} Models Jia et al. \cite{jia2021leveraging}, \cite{wen2022structure} attempt to reduce single homography distortion by fitting local projective models per plane or blending dual homographies. These methods, however, rely on strong assumptions about scene structure (e.g., two planes or projective-consistent regions) and can become brittle in complex or irregular depth geometries.

\item \textbf{Structure-Preserving Warps} successfully reduce distortions and preserve salient structures by incorporating complex optimization frameworks and constraints (e.g., collinearity constraints). These methods, however, are often computationally demanding and remain sensitive to poor feature distribution.

\item \textbf{Learning-Based Approaches} use deep learning for tasks like transformer-based warping and optical flow with inpainting Nie et al \cite{nie2021unsupervised,nie2023parallax,chen2024seamless}. While they generalize well and show strong performance, they demand heavy training requirements, reliance on large datasets, and risk hallucinating content or propagating errors, reducing interpretability compared to geometric models.
\end{itemize}
\end{itemize}

Overall, prior methods still face three major challenges: (1) global or overly flexible warps compromise structural fidelity and introduce geometric distortions; (2) seam selection strategies based on cost-function minimization often rely on complex preprocessing steps (like semantic segmentation or feature classification), increasing sensitivity to inaccuracies that reduce the visual realism of the stitched panorama; and (3) dependence on strong scene assumptions (planarity, dual planes, or consistent surface normals).

\section{Seamless and Structurally Consistent Image Stitching}

\label{sec:formatting}
\begin{figure*}[h]
   \centering
   \includegraphics[width=\textwidth]{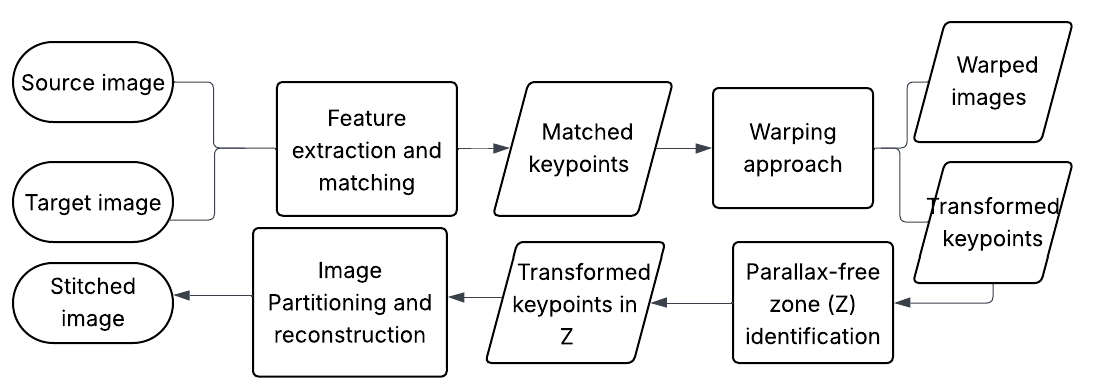}
    \caption{Flowchart of the proposed approach}
    \label{overall_flowchart}
\end{figure*}

To overcome the excessive distortion and misalignment common in image stitching, we propose a three-stage framework that emphasizes local adaptability, geometry-driven parallax handling, and structurally consistent reconstruction. The method transforms the source image into the domain of the target image while rigorously preserving geometric structure.

\subsection{The Three Stage Framework}
\begin{itemize}
    \item Local Image Warping: We deliberately move beyond global homographies and spline-based deformations. The source image is coarsely aligned using a global affine transformation (estimated via RANSAC). This is followed by a refinement stage where the overlap is subdivided into local grids, and distinct affine models are fitted to local feature correspondences. This process generates a Smooth Free-Form Deformation (FFD) field, which is blended and adaptively smoothed using a seamguard strategy that utilizes a ramp mask and match density weighting to prevent discontinuities at boundaries. This preserves overall structural fidelity while reducing geometric distortion.
    
    \item Adequate Parallax Minimized Zone Identification: In contrast to seam selection methods relying on semantic segmentation, we introduce a model free strategy. This method analyzes disparity variations and geometric consistency among matched features to isolate a stable stitching zone that inherently minimizes parallax artifacts.
    
    \item Image Partitioning and Reconstruction: Within the identified stable zone, an ordered chain of refined keypoints defines the optimal stitching line. Both images are then partitioned into corresponding vertical slices anchored by these keypoints. This anchor-based segmentation enforces structural consistency between the two images by construction: every vertical slice in one image has a direct, aligned counterpart in the other. By tying the seamline to these geometric anchors, our method eliminates the duplication, ghosting, and misalignment issues often associated with blending.
\end{itemize}
The overall flow of the proposed method is provided in Figure \ref{overall_flowchart}.

\subsection{Locally adaptive image warping}
\label{subsec:proposed_warping}
Our approach generates a content-aware, seam-guarded deformation field to precisely align a source image ($I_
s$) to a target image ($I_t$) using only sparse feature matches. The pipeline is structured in six sequential steps, strategically combining global alignment with local adaptivity. This methodology ensures both structural fidelity and robustness through confidence-weighted blending and specific gating mechanisms.

\subsubsection*{Step 1: Global Affine Estimation via RANSAC}
We begin by detecting, extracting, and matching features using the XFeat algorithm \cite{potje2024xfeat}, which is selected for its effective balance between computational efficiency and matching accuracy. From the resulting set of matched keypoints $\{(p_{s_i}, p_{t_i})\}$, we compute a robust global affine transformation $A_{\text{glob}} \in \mathbb{R}^{3\times3}$ using the RANSAC algorithm. This initial step serves two critical functions: establishing a coarse alignment and filtering outliers. Only the inlier matches, as determined by RANSAC, are retained for all subsequent local refinement steps.

\subsubsection*{Step 2: Overlap Region and Local Grid Cell Generation}

This step defines the area of alignment and prepares it for localized warping.

Using the global affine transformation $A_{\text{glob}}$, the four corners of the source image ($I_s$) are projected into the coordinate space of the target image ($I_t$), creating a quadrilateral. This quadrilateral is then clipped to the bounds of $I_t$
using the Sutherland-Hodgman algorithm to precisely define the polygonal overlap region. A binary mask, $M_{\text{tgt}}^{\text{overlap}}$, is generated from this resulting polygon.

Next, a uniform $G_x \times G_y$
grid is overlaid onto the bounding box of this overlap region. For each resulting grid cell ($c_j$), the following parameters are computed:

\begin{itemize}
    \item A binary mask $M_{c_j}$, derived from the intersection of the grid cell with $M_{\text{tgt}}^{\text{overlap}}$.

    \item A centroid $\mathbf{c}_j = (c_{x_j}, c_{y_j})$, calculated using image moments.

    \item A bounding box ($(x_0^{(j)}, x_1^{(j)}, y_0^{(j)}, y_1^{(j)})$), utilized for subsequent spatial weighting and diagnostics.
\end{itemize}

\subsubsection*{Step 3: Local Affine Refinement with Adaptive Transformation Selection \& Confidence Scoring
}

For each grid cell $c_j$, a local affine transformation $T_j$ is fitted using the feature matches that fall within the cell and its immediate neighbors. This fitting employs ridge regression with a regularization $\lambda_1$, which biases the local solution toward the initial global transformation $A_{\text{glob}}$ to ensure stability.

\paragraph{Spatial Confidence Metric}
Instead of relying on statistical residuals or covariance, a novel confidence score is defined for each local affine transformation based on the density and spatial distribution of its supporting feature points.
The Gaussian weight $w_i^{(j)}$ of a feature point $i$ relative to the cell center $(c_j)$, is computed as:

 $w_i^{(j)} = \exp\left( -\frac{ \| \mathbf{p}_{t_i} - \mathbf{c}_j \|^2 }{ 2 \sigma_j^2 } \right)$ 

Here, the Gaussian standard deviation is defined by the scaling factor $\alpha$ applied to the diagonal length of the cell's bounding box $\text{diag}(c_j)$, where $\sigma_j = \alpha \cdot \text{diag}(c_j)$. The overall confidence score is then computed based on this weighted density.

The confidence score is computed as:

\begin{equation}
\text{conf}_j = \max\left( \kappa_{\min},\ \min\left( \kappa_{\max},\ \frac{ \sum_i w_i^{(j)} }{ \beta \cdot \max_i w_i^{(j)} } \right) \right)
\label{eq:confidence_metric}
\end{equation}
where $\kappa_{\min}, \kappa_{\max}, \beta$ are clamping and normalization constants. This score reflects the “weight mass” of matches in the cell — a heuristic robust to sparsity and uneven distributions.

\paragraph{Adaptive Transformation Selection via Composite Diagnostic Score:}

The system employs an adaptive strategy to maintain geometric stability. If the initial local affine transformation $T_j$ produces unstable geometry—diagnosed using metrics such as Root Mean Square Error (RMSE), the determinant, the condition number, or the displacement from the global affine $A_{glob}$—the transformation is refitted using a stronger regularization parameter $\lambda_2 > \lambda_1$. The superior version of the transformation is then selected based on a \textbf{composite instability score}.

\begin{equation}
\text{score}(T) = \text{RMSE} + \omega_{\text{cond}} \cdot \text{cond} + \omega_{\det} \cdot \max(0, \tau_{\det} - |\det|) + \omega_{\delta} \cdot \delta_{\text{mean}}
\label{eq:composite_score}
\end{equation}
where $\omega_{\text{cond}}, \omega_{\det}, \omega_{\delta}, \tau_{\det}$ are weighting and threshold parameters, and $\delta_{\text{mean}}$ is the mean displacement between the outputs of the local affine transformation and the global affine transformation, computed over an $N_g \times N_g$ evaluation grid uniformly sampled within the cell’s bounding box. This embedded selection mechanism ensures geometric stability without manual intervention.

\subsubsection*{Step 4: Free-Form Deformation Field via Confidence-Weighted Local Transformation Blending}

We construct a deformation field on an $N_y \times N_x$ lattice over the output canvas. For each lattice point $\mathbf{u} = (u, v)$, we compute its corresponding coordinate in target space $\mathbf{p}_t = (u - o_x, v - o_y)$, and its base source coordinate via global affine: $\mathbf{p}_{\text{base}} = \Pi(A_{\text{glob}}^{-1} \cdot [\mathbf{p}_t, 1]^T)$, where $\Pi(\mathbf{x}) = (x/z, y/z)$ is perspective division. For each local transformation $T_j$, we compute its mapped source coordinate $\mathbf{p}_{\text{local}}^{(j)} = \Pi(T_j^{-1} \cdot [\mathbf{p}_t, 1]^T)$, and the displacement $\Delta \mathbf{p}^{(j)} = \mathbf{p}_{\text{local}}^{(j)} - \mathbf{p}_{\text{base}}$.

\paragraph{Confidence-Weighted Spatial Blending}
The final displacement $\Delta \mathbf{p}$ is a normalized blend of all local transformation displacements, weighted by both spatial proximity and transformation confidence:
\begin{equation}
\Delta \mathbf{p} = \sum_{j=1}^J w_j(\mathbf{p}_t) \cdot \Delta \mathbf{p}^{(j)}
\quad \text{where} \quad
w_j(\mathbf{p}_t) = \frac{
\text{conf}_j \cdot \exp\left( -\frac{ \| \mathbf{p}_t - \mathbf{c}_j \|^2 }{ 2 \sigma_f^2 } \right)
}{
\sum_{k=1}^J \text{conf}_k \cdot \exp\left( -\frac{ \| \mathbf{p}_t - \mathbf{c}_k \|^2 }{ 2 \sigma_f^2 } \right)
}
\label{eq:ffd_blending}
\end{equation}
where $\sigma_f = \alpha_f \cdot \text{mean}(\text{cell diagonals})$, and $\alpha_f$ is a spatial decay factor. Displacements are clipped to $[-d_{\max}, d_{\max}]$ and lightly smoothed with Gaussian blur ($\sigma_l$) before upsampling to full canvas resolution via bicubic interpolation.

\subsubsection*{Step 5: Seam Guarding via Dual-Channel Gating}

The final step in suppressing artifacts near boundaries or in sparse regions is the modulation of the full-resolution deformation field $\Delta \mathbf{p}$ using a \textbf{multiplicative gate ($\mathbf{G}$)} derived from two primary signals:

\begin{itemize}
 \item \textbf{Geometric Ramp ($\mathbf{R}_{\text{canvas}}$)}: A smootherstep function applied to the signed distance field of the overlap region, with bandwidth proportional to image diagonal ($b = \rho \cdot \text{diag}_{\text{img}}$).
 \item \textbf{Match Density Map ($\mathbf{D}_{\text{canvas}}$)}: A Gaussian-blurred heatmap of inlier keypoint locations, normalized to $[0,1]$, with kernel standard deviation ($\sigma_d$).
\end{itemize}

\paragraph{Dual-Gated Seam Suppression}
The final gating mask combines both signals multiplicatively:
\begin{equation}
G(\mathbf{u}) = \underbrace{ \left[ S\left( R_{\text{canvas}}(\mathbf{u}) \right) \right]^{\gamma_p} }_{\text{geometric falloff}} \cdot \underbrace{ \left( \gamma_{\min} + (1 - \gamma_{\min}) \cdot S\left( D_{\text{canvas}}(\mathbf{u}) \right) \right) }_{\text{density-aware modulation}}
\label{eq:seam_guard}
\end{equation}
where $S(t) = 6t^5 - 15t^4 + 10t^3$ is the smootherstep function, $\gamma_p$ controls ramp steepness, and $\gamma_{\min}$ is the minimum gate value. The guarded deformation field is then smoothed:
\[
\Delta \mathbf{p}_{\text{guarded}}(\mathbf{u}) = \text{GaussianBlur}_{\sigma_g} \left( \Delta \mathbf{p}(\mathbf{u}) \cdot G(\mathbf{u}) \right)
\]
This is not post-warp blending; rather, it is a mechanism for pre-warp suppression of unreliable deformations. The final displacement field $\Delta_p$ is modulated by the gate $G$: 
\[
\Delta \mathbf{p}_{\text{gated}} = \mathbf{G} \odot \Delta \mathbf{p}
\]

This multiplicative gating ensures that the full-resolution displacement is only applied in geometrically stable and feature-supported areas. The combination of geometric and photometric (match-density based) gating for Free-Form Deformation (FFD) fields is, to our knowledge, unprecedented in the image stitching literature.

\subsubsection*{Step 6: Final Warping and Output}

The final source-to-canvas mapping is computed by combining the inverse global affine transformation with the gated deformation field. The source image $I_s$ is then warped into the final canvas space using bilinear interpolation. Concurrently, the target image $I_t$ is pasted onto the output canvas using the pre-computed offset $(o_x, o_y)$.

For evaluation purposes, the final transformed coordinates of the original inlier matches are computed. This involves a simple translation to canvas coordinates for target points and applying the full warp (global affine + FFD + multiplicative gate) with bilinear interpolation of the displacement field for source points.

These collective innovations enable our method to produce seamless, artifact-free alignments even when operating with sparse, uneven, or noisy feature matches.

\subsection{Optimal stitching line}

\subsubsection{Determination of an Adequate Parallax-Free Zone}
Identifying an adequate stitching area relies on finding a region containing a high density of reliable feature correspondences governed by a dominant geometric transformation. Initially, regions with low information content, which are quantifiable by low local image variance, are systematically excluded as they produce unstable matches. However, scenes with significant depth often exhibit motion parallax, resulting in a multi-modal distribution of disparity vectors that complicates the search for a consistent seam.

To robustly handle this parallax, our method focuses on isolating the most extensive and stable surface. This is achieved by identifying the dominant motion group, which corresponds to the true inliers for a stable stitch. Specifically, the algorithm locates regions where keypoint disparities are statistically consistent (exhibiting low local variance) and where their local mean converges to the global mean $\mu$ of the primary motion mode. This approach allows the algorithm to robustly handle parallax by isolating the most extensive and stable surface.

Given a set of matched keypoint pairs 
(($x_S,y_S$),($x_T,y_T$)), where $x_S$ and $y_S$ denote the coordinates in the source image and $x_T$ and $y_T$ denote the coordinates in the target image, respectively, keypoints are initially sorted using two primary spatial criteria to account for the camera geometry.

\begin{itemize}
    \item First, $x_S > x_T$, reflecting the relative frontal position of the camera of the source image.
    \item $(y_S > y_T) \lor (y_T > y_S)$, for one camera physically positioned above the other in most cases.
\end{itemize}  

The matched keypoints are partitioned into classes $C_i$ based on their abscissa values (x-coordinates). Each class corresponds to a spatial range, R, which is defined as R=width/20. For every keypoint within a class, we compute its disparity as ($x_S-x_T$) and subsequently calculate the mean disparity for that class.

We then iterate through the resulting list of mean disparities (one value per class) and group the corresponding classes into clusters based on the statistical consistency of these mean disparity values. This clustering process isolates the most dominant motion groups, enabling the identification of the parallax-minimized "adequate zone."

\begin{algorithm}[h!]
    \caption{Threshold-Based Disparity Clustering}
    \label{algorithm1}
    \begin{algorithmic}[1]
        \Require $D$: A list of mean disparities $d_1,d_2,\dots,d_n$, where $d_i$ corresponds to the mean disparity of class $i$. 
        \Require $v$: A positive, user-defined threshold value for disparity difference.
        \Ensure $C$: A set of clusters, where each cluster is a list of consecutive classes from $D$.

        \State Initialize an empty list of clusters $C$
        \State Initialize the current cluster $C_{\text{current}}$ with the first class $d_1$
        \For{$i = 2$ to $n$}
            \If{$|d_i - d_{i-1}| \leq v$}
                \State Add class $i$ to $C_{\text{current}}$
            \Else
                \If{$C_{\text{current}}$ contains at least two classes}
                    \State Add $C_{\text{current}}$ to the list of clusters $C$
                \EndIf
                \State Start a new cluster with class $i$: $C_{\text{current}} \gets \{d_i\}$
            \EndIf
        \EndFor
        \If{$C_{\text{current}}$ contains at least two classes}
            \State Add $C_{\text{current}}$ to the list of clusters $C$
        \EndIf
        \State \Return $C$
    \end{algorithmic}
\end{algorithm}

The optimal stitching cluster is selected based on a scoring metric that evaluates the consistency and reliability of each cluster against three key parameters:

\begin{itemize}
    \item Standard Deviation ($\sigma_k$): This measures the internal coherence of the cluster. A smaller standard deviation indicates that the individual class mean disparities are tightly grouped around the cluster mean, suggesting a more consistent depth plane.
    
    \item Cardinality ($C_k$): This is the total number of keypoints contained within all classes of the cluster. High cardinality indicates that the cluster is supported by a large amount of data, which directly increases the reliability and robustness of its mean disparity calculation.
    \item Disparity Deviation ($\Delta\mu_k$): This is defined as the absolute difference between the mean disparity of the cluster ($\mu_{c,k}$) and the global mean disparity of all classes ($\mu_g$).
  \end{itemize}

To prevent the selection of clusters with low reliability (low cardinality), high incoherence (high standard deviation), or minimal deviation from the global mean (low significance), we define a weighted score, $w_k$ , for each cluster k given by Equation \ref{eq:ratio}.

\begin{equation}
w = \frac{C}{(\sigma + \lambda \cdot \Delta\mu) + \epsilon}
\label{eq:ratio}
\end{equation}

where: 
\begin{itemize}
    \item $\sigma$: standard deviation of the disparities within the cluster, 
    \item $C$: cardinality (number of keypoints in the cluster), 
    \item $\lambda$: weighting factor controlling the influence of the disparity deviation, 
    \item $\Delta\mu = |\mu_c - \mu_g|$: absolute difference between the mean disparity of the cluster  $\mu_c$ and the global mean disparity $\mu_g$, 
    \item $\epsilon$: a small constant added to avoid division by zero.
\end{itemize}

The cluster with the highest calculated score, $w_{max}$, is selected as the optimal cluster.
The "adequate zone" is then defined as the combination of all the classes within this selected cluster. 

\subsubsection{Keypoint Chain Refinement}

Once an adequate overlapping area is identified, the initial set of keypoint correspondences cannot be used directly, as this raw data is inherently unreliable and contains significant outliers or misaligned matches. Using these erroneous points to guide partitioning would create inconsistent divisions between the two images, leading directly to visible stitching artifacts such as ghosting and duplications. Therefore, refining this initial set of keypoints is a critical prerequisite.

A dedicated algorithm is employed to generate two clean, synchronized, and ordered lists of keypoints, forming a coherent "keypoint chain". For brightness consistency, a selection step is performed, determining each keypoint's intensity and retaining only those with similar levels. An initial anchor is established at the keypoint closest to the image edge. From there, the algorithm iteratively matches each point in the first image with its closest corresponding point in the second based on Euclidean distance. A crucial constraint is applied during this process: each match must be unique in its horizontal (x) position. This filtering step prunes the ambiguous or conflicting matches that cause structural inconsistencies, yielding a refined set of high-confidence pairs.

The optimal stitching line in each image is defined by the path connecting the first element to the last element of the refined set in the corresponding pair.

\subsection{Partitioning and Reconstruction}
\subsubsection{Image Partitioning}

\begin{figure}[hbt]
    \centering
       \begin{minipage}{0.45\textwidth}
  Source image  Target image
        \centering
        \includegraphics[width=\linewidth]{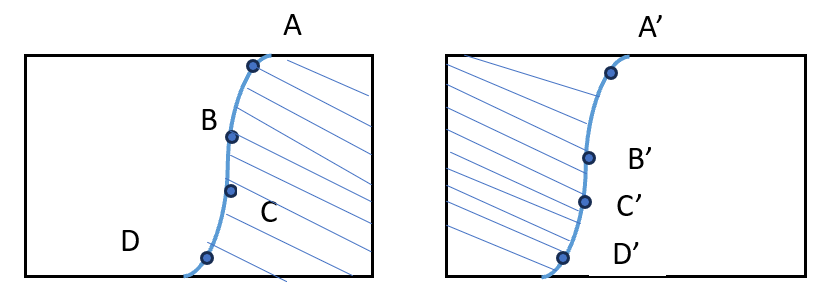}
        \caption{Optimal stitching line}
        \label{opt_line}
    \end{minipage}
    \hfill
    \begin{minipage}{0.45\textwidth}
     If $x_A > x_B$ and $x_A’ > x_B’$:
        \centering
        \includegraphics[width=\linewidth]{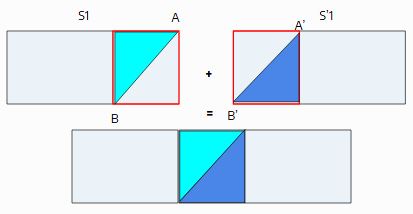}
        \caption{Complementary segments}
        \label{case2}
    \end{minipage}
   
\end{figure}

The image partitioning process begins with the optimal stitching line, which is defined by an ordered set of n reliable keypoint pairs (see Figure \ref{opt_line}. These points function as anchors for segmenting each image into $S=n+1$ vertical slices.

\begin{figure}[hbt]
    \centering
    \begin{minipage}{0.45\textwidth}
    If $x_A < x_B$ and $x_A’< x_B’$:
        \centering
        \includegraphics[width=\linewidth]{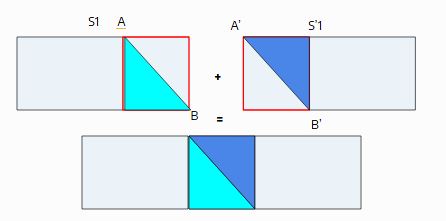}
        \caption{Complementary segments}
        \label{case1}
    \end{minipage}
      \begin{minipage}{0.45\textwidth}
        If $x_A > x_B$ and $x_A’< x_B’$
        \centering
        \includegraphics[width=\linewidth]{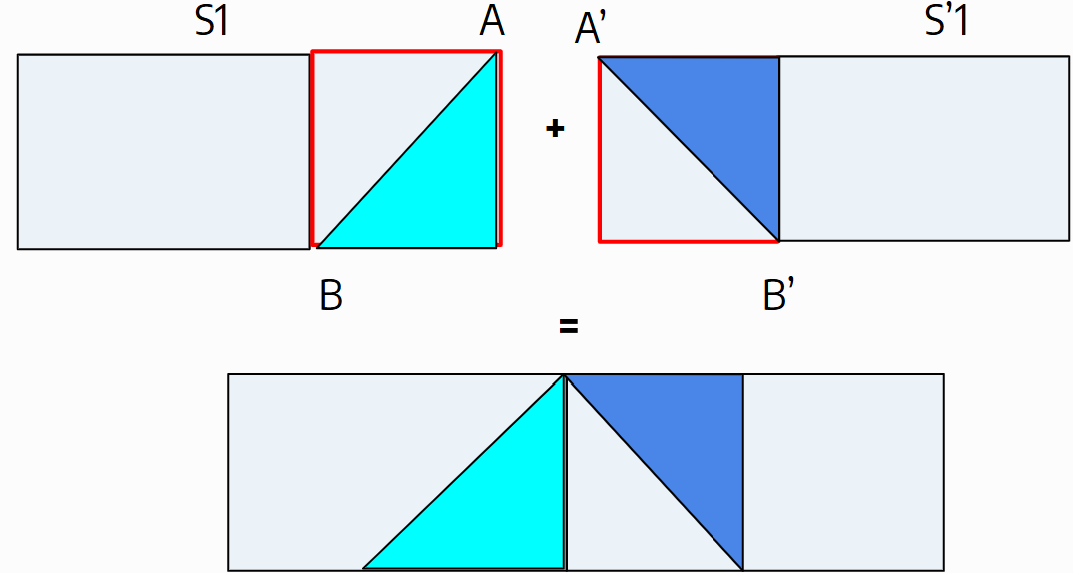}
        \caption{Invalid segment}
        \label{invalid}
    \end{minipage}
\label{fig:two_images}
\end{figure}

To ensure the resulting vertical slices are structurally complementary (meaning they maintain consistent relative order and direction between the two images), a directional validation is performed on each consecutive pair of anchors $(A,B)$ and their corresponding pair ($A', B'$) from the second image:

\begin{itemize}
  \item Valid Segments (Consistent Direction):
  \begin{itemize}
    \item If the x-coordinate of $A$ is greater than $B$ ($x_A > x_B$) AND $A’$ is greater than $B’$ ($x_A’ > x_B’$): The segments are valid, as both pairs move from right-to-left (see Figure \ref{case2}).
    \item If $x_A < x_B$ and $x_A’< x_B’$: The segments are valid, as both pairs move from left-to-right (see Figure \ref{case1}).
  \end{itemize}
  \item Invalid Segments (Inconsistent Direction):
  If the direction is inconsistent between the two images (e.g., $x_A > x_B$ and $x_A’ < x_B’$ or vice versa), the segments are rejected as they are not complementary (see Figure \ref{invalid}). This prevents structural inconsistencies from being introduced during the final reconstruction
\end{itemize}

\subsubsection{Reconstruction}

\begin{figure*}[h]
   \centering
   \includegraphics[width=1.2\textwidth]{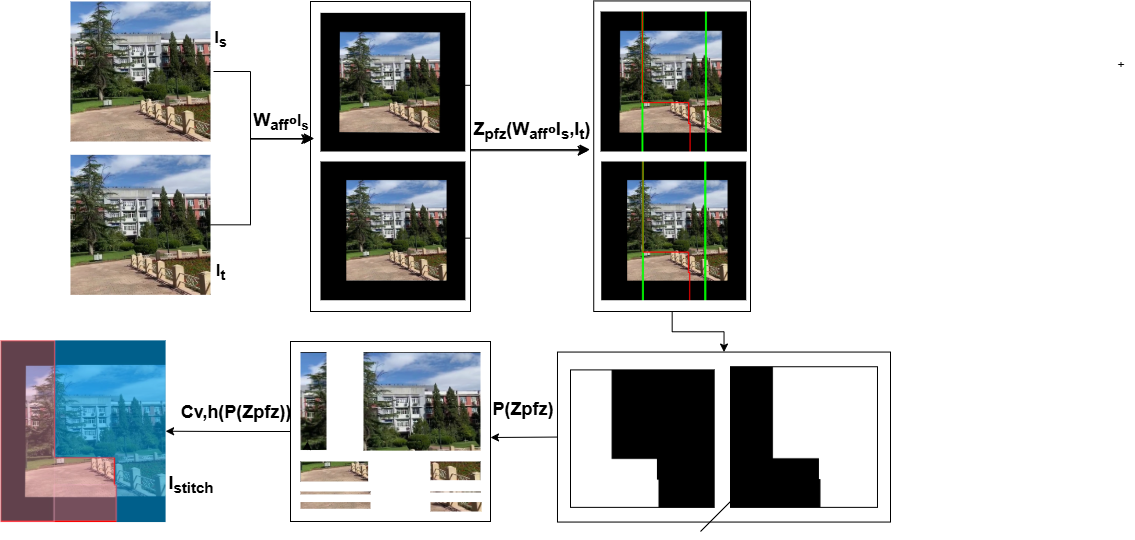} 
   \caption{We warp the source image $I_s$ into the domain of the target $I_t$ using a locally adaptive affine warping strategy ($W_{aff}\circ I_s$). A parallax-minimized zone $Z_{pfz}$ is then identified in the overlapping area, from which the optimal stitching line is extracted. Based on this line, both images are partitioned into corresponding vertical slices via operator $P$. Finally, slices are concatenated horizontally and vertically $(C_{v,h})$ to produce the final stitched image $I_{stitch}$}
   \label{detailed_flowchart}
\end{figure*}

After the segments have been validated, the process moves to the \textbf{concatenation phase}. Because both input images share the identical partitioning pattern, each segment of the first image corresponds directly to a segment of the second image. Corresponding segments are merged using a simple \textbf{linear alpha transition}: the contribution of the left segment decreases linearly from left to right, while the contribution of the right segment increases symmetrically. A final, light Gaussian smoothing is then applied across the seam area to suppress any residual visible artifacts. The resulting blended composites are concatenated horizontally, and the constructed rows are subsequently stacked vertically to form the final stitched image.\\
\\Figure \ref{detailed_flowchart} summarizes all the described processes.

\label{sec:formatting}

\section{Experimentation and results}

The experiments were conducted within a Google Colaboratory environment, leveraging 12.67GB of RAM and 107.72 GB of storage. We utilized publicly available and widely recognized datasets provided by Nie et al. \cite{nie2021unsupervised}, Du et al. \cite{du2022geometric}, and Hermann et al. \cite{herrmann2018object}. The code is available on  \href{https://github.com/Camertronix-Cm/seamlessly-natural-sena-/tree/main}{GitHub}.

\subsection{Quantitative evaluation} 

Table \ref{tab:results} This section summarizes quantitative results for recent algorithms on the UDIS-D dataset (1,106 images grouped by difficulty: easy, moderate, and hard). Alignment accuracy in the overlap region between stitched images is evaluated using \textbf{Peak Signal-to-Noise Ratio (PSNR)} and the Structural \textbf{Similarity Index Measure (SSIM)}.
\begin{table}[hbt]
\centering
\begin{tabular}{|l|llll|llll|}
\hline
Metrics                                                      & \multicolumn{4}{c|}{PSNR}                                                                                & \multicolumn{4}{c|}{SSIM}                                                                                \\ \hline
Datasets                                                     & \multicolumn{1}{l|}{Easy}  & \multicolumn{1}{l|}{Mod} & \multicolumn{1}{l|}{Hard}  & Avg        & \multicolumn{1}{l|}{Easy}  & \multicolumn{1}{l|}{Mod} & \multicolumn{1}{l|}{Hard}  & Avg        \\ \hline
$I_{3x3}$                                                         & \multicolumn{1}{l|}{15.87} & \multicolumn{1}{l|}{12.76}    & \multicolumn{1}{l|}{10.68} & 12.86          & \multicolumn{1}{l|}{0.530} & \multicolumn{1}{l|}{0.286}    & \multicolumn{1}{l|}{0.146} & 0.303          \\ \hline
\begin{tabular}[c]{@{}l@{}}SIFT+RANSAC\cite{fischler1981random}\end{tabular} & \multicolumn{1}{l|}{28.75} & \multicolumn{1}{l|}{24.08}    & \multicolumn{1}{l|}{18.55} & 23.27          & \multicolumn{1}{l|}{0.916} & \multicolumn{1}{l|}{0.833}    & \multicolumn{1}{l|}{0.636} & 0.779          \\ \hline
APAP \cite{zaragoza2013projective}                                                   & \multicolumn{1}{l|}{27.96} & \multicolumn{1}{l|}{24.39}    & \multicolumn{1}{l|}{20.21} & 23.79          & \multicolumn{1}{l|}{0.901} & \multicolumn{1}{l|}{0.837}    & \multicolumn{1}{l|}{0.682} & 0.794          \\ \hline
ELA \cite{li2017parallax}                                                    & \multicolumn{1}{l|}{29.36} & \multicolumn{1}{l|}{25.10}    & \multicolumn{1}{l|}{19.19} & 24.01          & \multicolumn{1}{l|}{0.917} & \multicolumn{1}{l|}{0.855}    & \multicolumn{1}{l|}{0.691} & 0.808          \\ \hline
SPW \cite{liao2019single}                                                    & \multicolumn{1}{l|}{26.98} & \multicolumn{1}{l|}{22.67}    & \multicolumn{1}{l|}{16.77} & 21.60          & \multicolumn{1}{l|}{0.880} & \multicolumn{1}{l|}{0.758}    & \multicolumn{1}{l|}{0.490} & 0.687          \\ \hline
LPC \cite{jia2021leveraging}                                                    & \multicolumn{1}{l|}{26.94} & \multicolumn{1}{l|}{22.63}    & \multicolumn{1}{l|}{19.31} & 22.59          & \multicolumn{1}{l|}{0.878} & \multicolumn{1}{l|}{0.764}    & \multicolumn{1}{l|}{0.610} & 0.736          \\ \hline
UDIS \cite{nie2021unsupervised}                                                   & \multicolumn{1}{l|}{25.16} & \multicolumn{1}{l|}{20.96}    & \multicolumn{1}{l|}{18.36} & 21.17          & \multicolumn{1}{l|}{0.834} & \multicolumn{1}{l|}{0.669}    & \multicolumn{1}{l|}{0.495} & 0.648          \\ \hline
UDIS++ \cite{nie2023parallax}                                                       & \multicolumn{1}{l|}{\textbf{30.19}} & \multicolumn{1}{l|}{25.84}    & \multicolumn{1}{l|}{21.57} & 25.43          & \multicolumn{1}{l|}{\textbf{0.933}} & \multicolumn{1}{l|}{\textbf{0.875}}    & \multicolumn{1}{l|}{0.739} & 0.838          \\ \hline
OURS                                                         & \multicolumn{1}{l|}{24.92} & \multicolumn{1}{l|}{\textbf{26.00}}    & \multicolumn{1}{l|}{\textbf{27.89}} & {\textbf{26.27}} & \multicolumn{1}{l|}{0.823} & \multicolumn{1}{l|}{0.839}    & \multicolumn{1}{l|}{\textbf{0.882}} & {\textbf{0.848}} \\ \hline
\end{tabular}
\caption{Quantitative results with PSNR and SSIM.}
\label{tab:results}
\end{table}\\
As shown in the table \ref{tab:results}, our method achieves the highest PSNR and SSIM values among all compared approaches, indicating superior stitching quality. This confirms that our geometry-driven strategy consistently reflects better structural preservation and visual fidelity across diverse scenes. 

\subsection{Qualitative evaluation} 
We compare our approach against several state-of-the-art methods, including  APAP~\cite{zaragoza2013projective}, ELA~\cite{li2017parallax}, UDIS~\cite{nie2021unsupervised}, UDIS++ ~\cite{nie2023parallax} and SEAMLESS~\cite{chen2024seamless}. Please feel free to zoom in on the images to better observe the highlighted elements. Additional data is given in the online resource and comprehensive visuals comparisons are provided on \href{https://drive.google.com/drive/folders/1CUV0PjbWwC7lh_VVOazLt5XnPYELgjfc?usp=drive_link}{Google Drive}.

\begin{figure}[H]
  \centering
  \setlength{\tabcolsep}{6pt} 
  \renewcommand{\arraystretch}{0.5} 
  \begin{tabular}{ccc}
    \begin{minipage}[b]{0.32\linewidth}
      \centering
      \includegraphics[width=\linewidth]{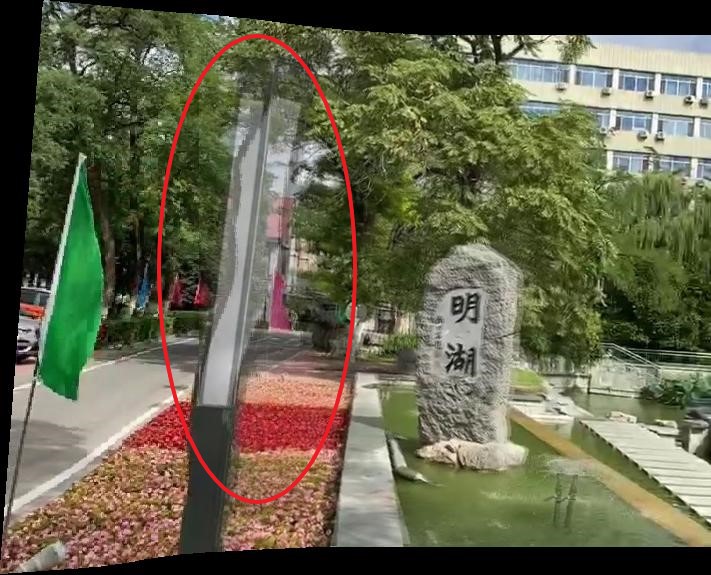}
      \\[4pt] APAP
    \end{minipage} &
    \begin{minipage}[b]{0.32\linewidth}
      \centering
      \includegraphics[width=\linewidth]{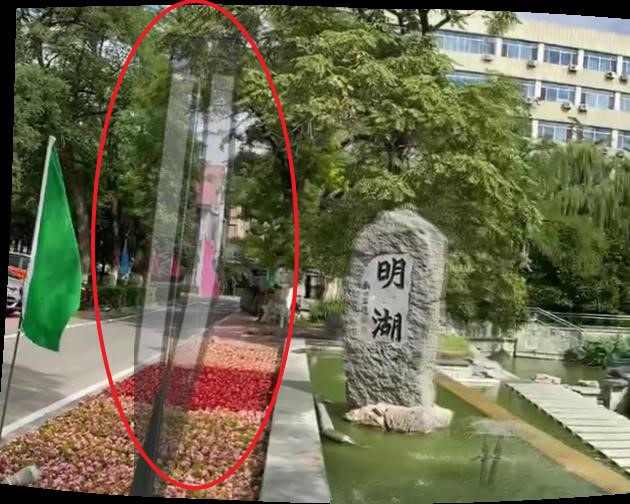}
      \\[4pt] ELA
    \end{minipage} &
    \begin{minipage}[b]{0.32\linewidth}
      \centering
      \includegraphics[width=\linewidth]{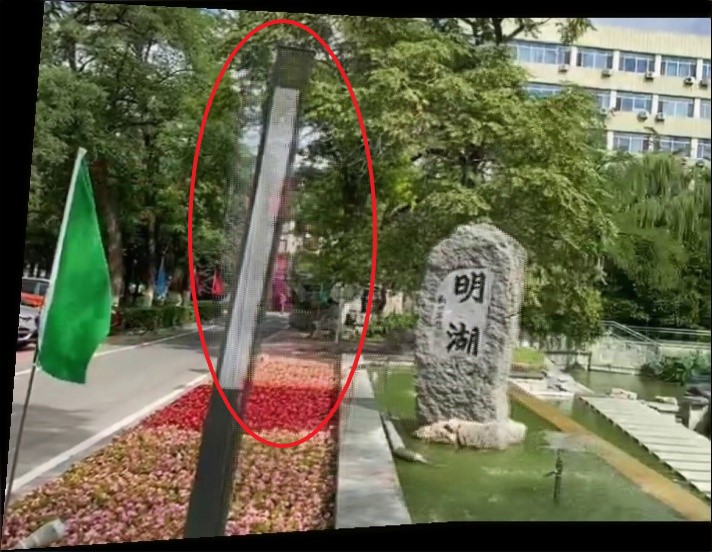}
      \\[4pt] UDIS
    \end{minipage} \\[6pt]
    \begin{minipage}[b]{0.32\linewidth}
      \centering
      \includegraphics[width=\linewidth]{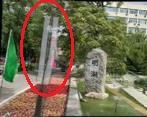}
      \\[4pt] UDIS++
    \end{minipage} &
    \begin{minipage}[b]{0.32\linewidth}
      \centering
      \includegraphics[width=\linewidth]{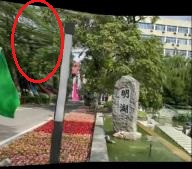}
      \\[4pt] SEAMLESS
    \end{minipage} &
    \begin{minipage}[b]{0.32\linewidth}
      \centering
      \includegraphics[width=\linewidth]{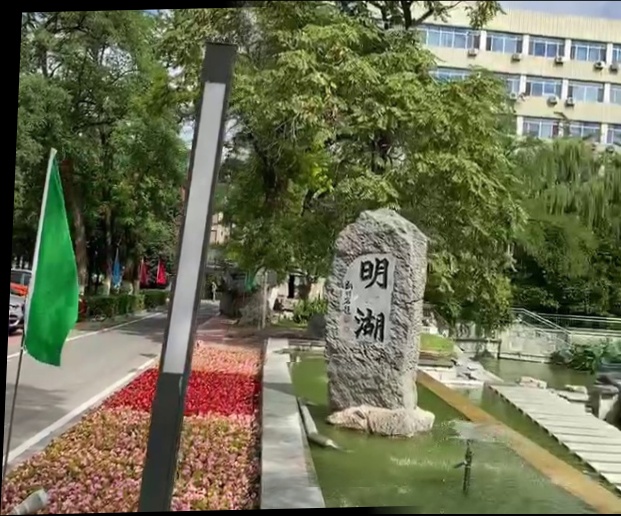}
      \\[4pt] OURS
    \end{minipage}
  \end{tabular}
 \caption{APAP\cite{zaragoza2013projective},ELA\cite{li2017parallax}, and UDIS++ \cite{nie2023parallax} exhibit pronounced ghosting artifacts, while UDIS \cite{nie2021unsupervised} produces noticeable blurring and SEAMLESS\cite{chen2024seamless} introduces stretching in certain regions. Moreover, most of these approaches display a generally blurred texture across the stitched image. In contrast, our method eliminates duplication and preserves the original sharp texture of the input images.}
  \label{qualitative1}
\end{figure}

\begin{figure}[h]
  \centering
  \setlength{\tabcolsep}{6pt} 
  \renewcommand{\arraystretch}{0.5} 
  \begin{tabular}{ccc}
    \begin{minipage}[b]{0.32\linewidth}
      \centering
      \includegraphics[width=\linewidth]{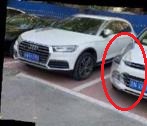}
      \\[4pt] APAP \cite{zaragoza2013projective}
    \end{minipage} &
    \begin{minipage}[b]{0.32\linewidth}
      \centering
      \includegraphics[width=\linewidth]{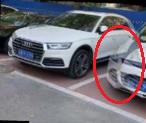}
      \\[4pt] ELA \cite{li2017parallax}
    \end{minipage} &
    \begin{minipage}[b]{0.32\linewidth}
      \centering
      \includegraphics[width=\linewidth]{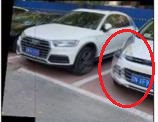}
      \\[4pt] UDIS \cite{nie2021unsupervised}
    \end{minipage} \\[6pt]
    \begin{minipage}[b]{0.32\linewidth}
      \centering
      \includegraphics[width=\linewidth]{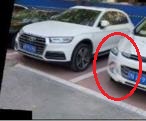}
      \\[4pt] UDIS++ \cite{nie2023parallax}
    \end{minipage} &
    \begin{minipage}[b]{0.32\linewidth}
      \centering
      \includegraphics[width=\linewidth]{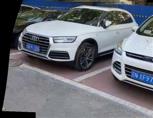}
      \\[4pt] SEAMLESS \cite{chen2024seamless}
    \end{minipage} &
    \begin{minipage}[b]{0.32\linewidth}
      \centering
      \includegraphics[width=\linewidth]{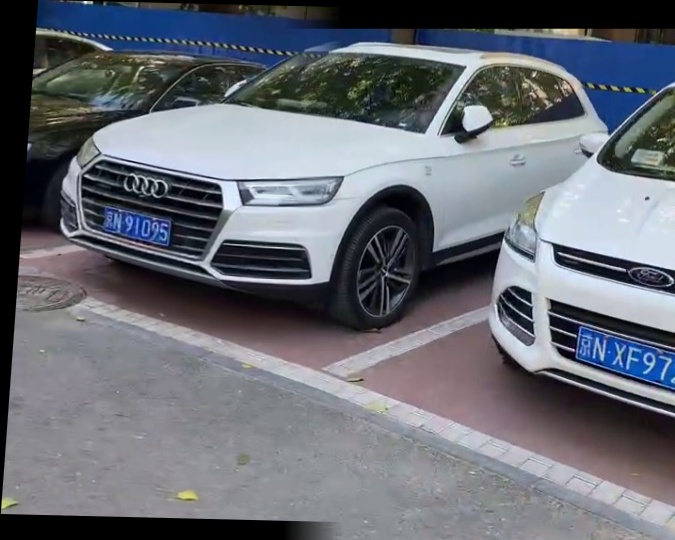}
      \\[4pt] OURS
    \end{minipage}
  \end{tabular}
 \caption{We observe ghosting artifacts in APAP, ELA, UDIS, UDIS++ and a blurred texture in SEAMLESS. Our result is free of these artifacts.}
  \label{qualitative2}
\end{figure}

\begin{figure}[H]
  \centering
  \setlength{\tabcolsep}{2pt} 
  \renewcommand{\arraystretch}{0.2} 
  \begin{tabular}{ccc}
    \begin{minipage}[b]{0.25\linewidth}
      \centering
      \includegraphics[width=\linewidth]{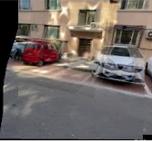}
      \\[2pt] APAP
    \end{minipage} &
    \begin{minipage}[b]{0.25\linewidth}
      \centering
      \includegraphics[width=\linewidth]{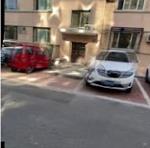}
      \\[2pt] ELA
    \end{minipage} &
    \begin{minipage}[b]{0.25\linewidth}
      \centering
      \includegraphics[width=\linewidth]{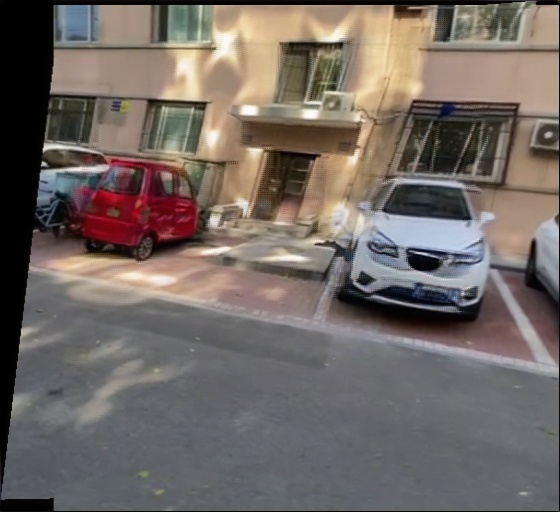}
      \\[2pt] UDIS
    \end{minipage} \\[2pt]
    \begin{minipage}[b]{0.25\linewidth}
      \centering
      \includegraphics[width=\linewidth]{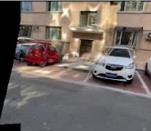}
      \\[2pt] UDIS++
    \end{minipage} &
    \begin{minipage}[b]{0.25\linewidth}
      \centering
      \includegraphics[width=\linewidth]{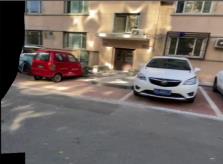}
      \\[2pt] SEAMLESS
    \end{minipage} &
    \begin{minipage}[b]{0.25\linewidth}
      \centering
      \includegraphics[width=\linewidth]{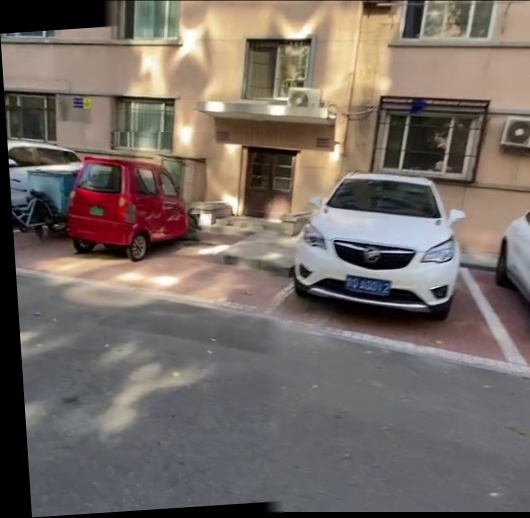}
      \\[2pt] OURS
    \end{minipage}
  \end{tabular}
  \caption{Ghosting artifacts, and blurring are observed in the state-of-the-art methods. Our stitched result is seamlessly natural.}
  \label{qualitative3}
\end{figure}

\newcommand{\PanelWidth}{0.45}  
\newcommand{\MainFrac}{0.60}   
\newcommand{\ThumbFrac}{0.4}   
\setlength{\fboxrule}{0.1pt}
\setlength{\fboxsep}{1pt}

\newcommand{\stitchPanelV}[5]{%
  \begin{minipage}[t]{\PanelWidth\linewidth}
    \begin{tabular}[t]{@{}p{\MainFrac\linewidth}@{\hspace{0.012\linewidth}}p{\ThumbFrac\linewidth}@{}}
      \vspace{0pt}\includegraphics[width=\linewidth]{#1} &
      \vspace{0pt}%
      \fbox{\includegraphics[width=\linewidth]{#2}}\par\vspace{4pt}%
      \fbox{\includegraphics[width=\linewidth]{#3}}\par\vspace{4pt}%
      \fbox{\includegraphics[width=\linewidth]{#4}}%
    \end{tabular}
    \par\vspace{4pt}\centering\textbf{#5}%
  \end{minipage}%
}

\begin{figure}[H]
  \centering
  \stitchPanelV{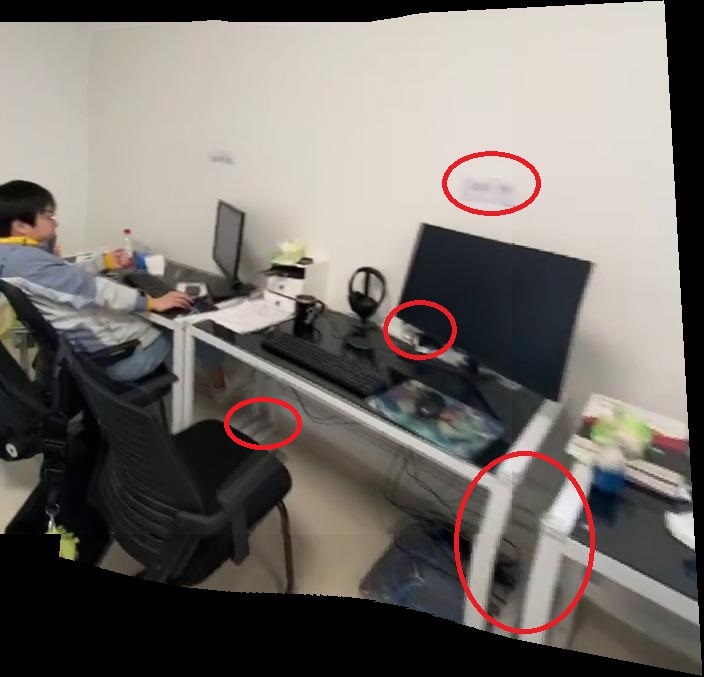}{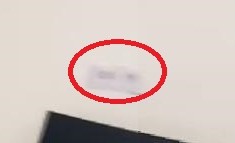}{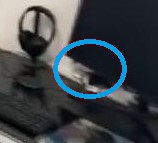}{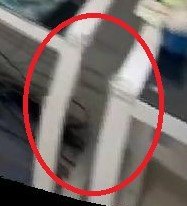}{APAP}
  \hfill
  \stitchPanelV{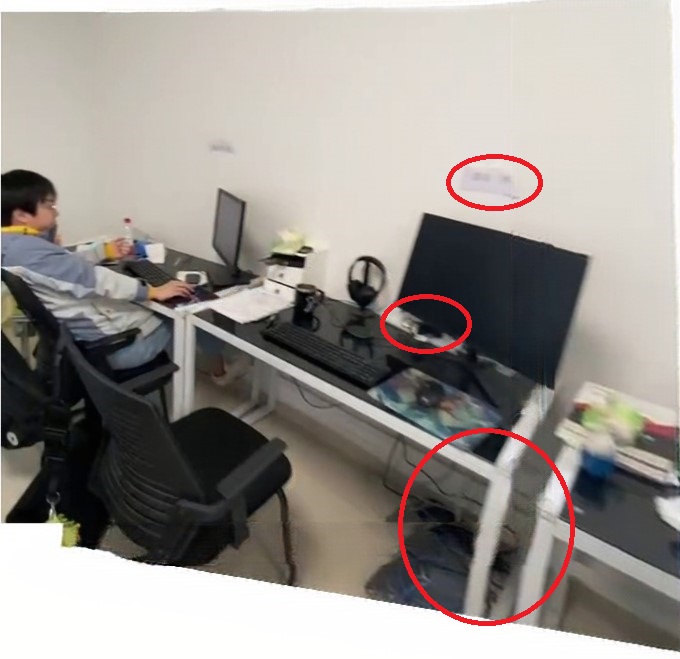}{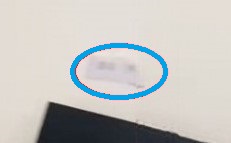}{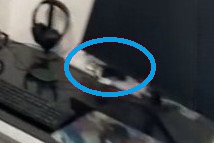}{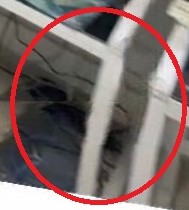}{EPISNET}
  \hfill
  \stitchPanelV{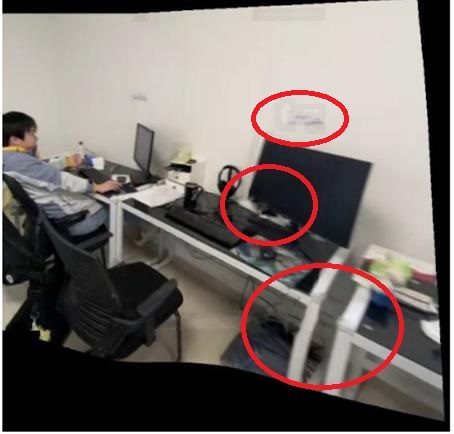}{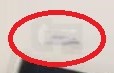}{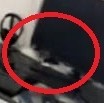}{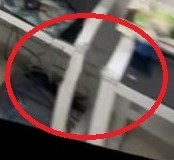}{SEAMLESS}
  \hfill
  \stitchPanelV{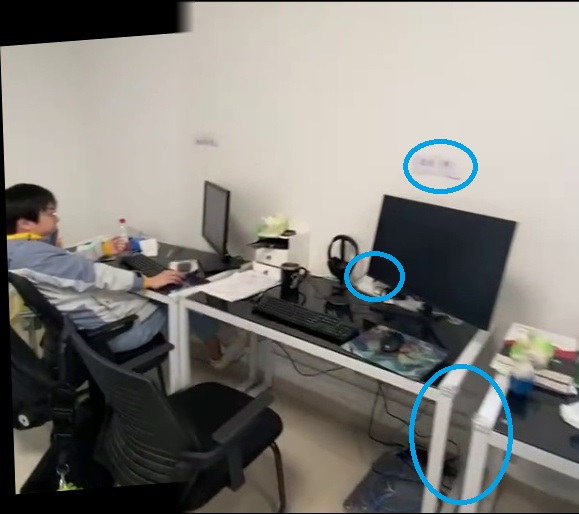}{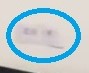}{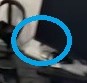}{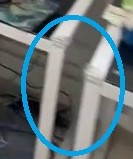}{OURS}
\caption{Similar artifacts as in \ref{qualitative2} for the state of the art.}
  \label{fig:stitch_two_panels}
\end{figure}

\begin{figure}[H]
\centering
  \begin{minipage}[b]{0.30\linewidth}
    \centering
    \includegraphics[width=\linewidth]{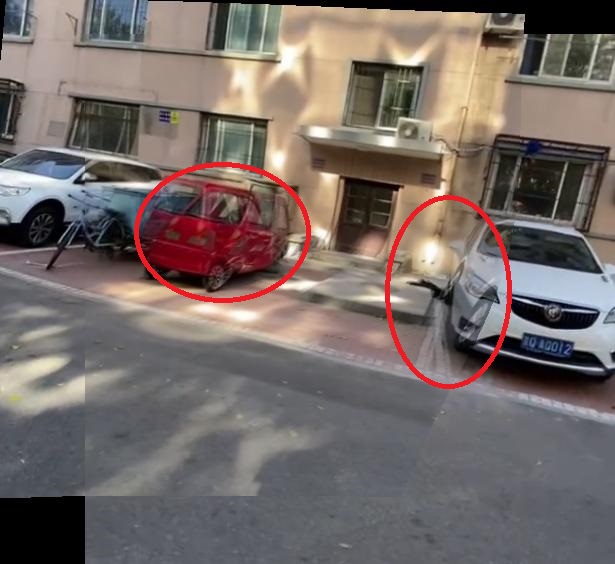}\\[1pt]
    \small ELA
  \end{minipage}
  \begin{minipage}[b]{0.30\linewidth}
    \centering
    \includegraphics[width=\linewidth]{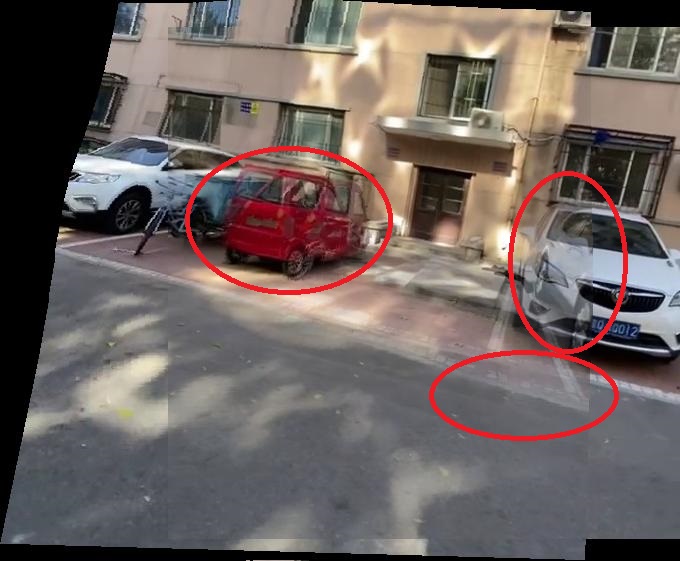}\\[1pt]
    \small APAP
  \end{minipage}

 \begin{minipage}[b]{0.30\linewidth}
    \centering
    \includegraphics[width=\linewidth]{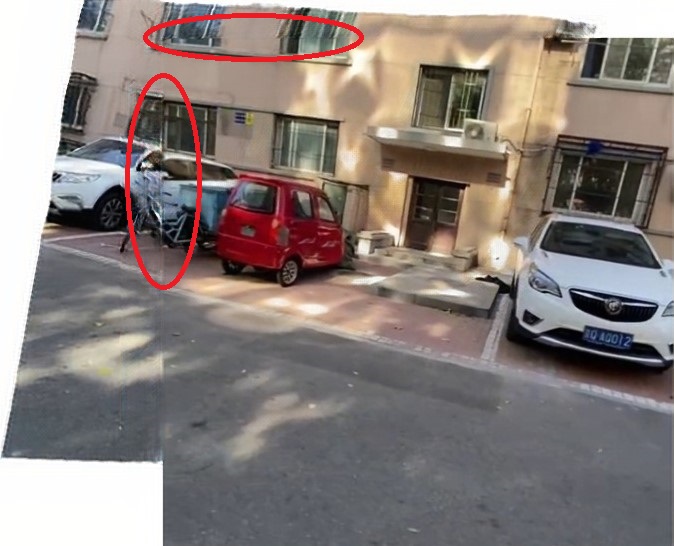}\\[1pt]
    \small EPISNET
  \end{minipage}
\begin{minipage}[b]{0.33\linewidth}
    \centering
    \includegraphics[width=\linewidth]{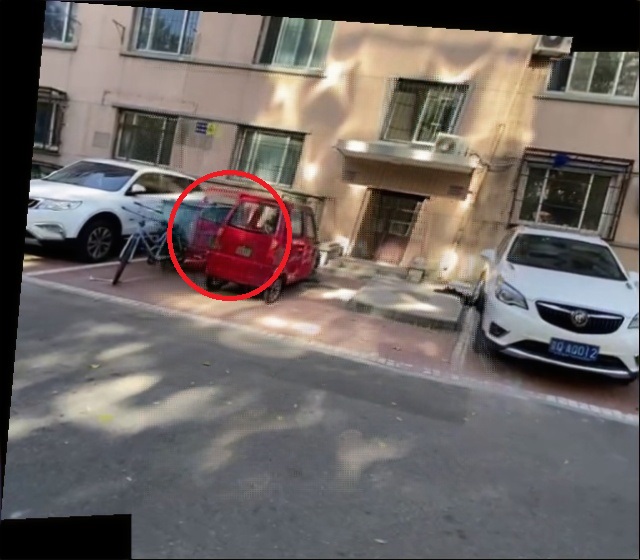}\\[1pt]
    \small UDIS
  \end{minipage}
\begin{minipage}[b]{0.33\linewidth}
    \centering
    \includegraphics[width=\linewidth]{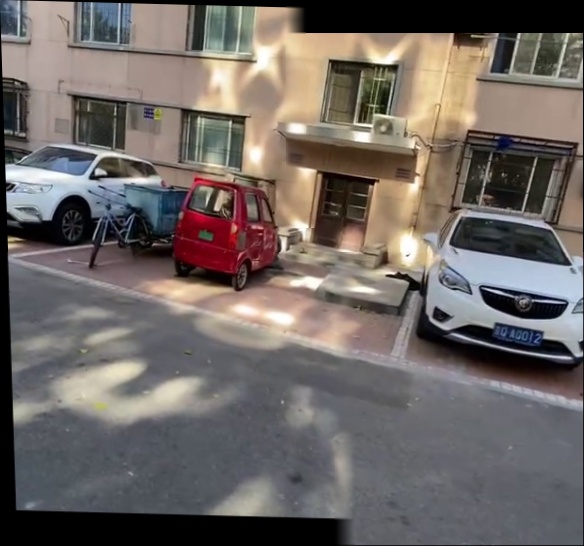}\\[1pt]
    \small OURS
  \end{minipage}
 
\caption{Similar artifacts as in \ref{qualitative2} for the state of the art.}
  \label{qualitative4}
\end{figure}

\begin{figure}[H]
\centering
  \begin{minipage}[b]{0.30\linewidth}
    \centering
    \includegraphics[width=\linewidth]{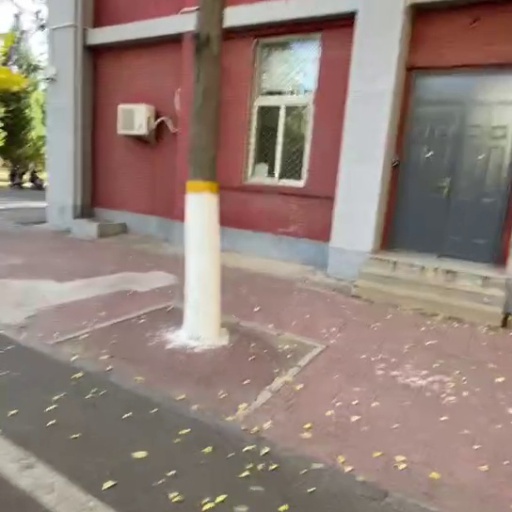}\\[2pt]
    \small SOURCE
  \end{minipage}
  \begin{minipage}[b]{0.30\linewidth}
    \centering
    \includegraphics[width=\linewidth]{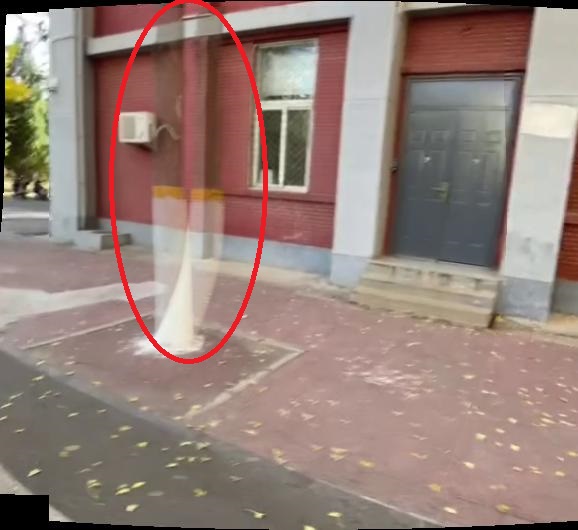}\\[2pt]
    \small ELA
  \end{minipage}
  \begin{minipage}[b]{0.30\linewidth}
    \centering
    \includegraphics[width=\linewidth]{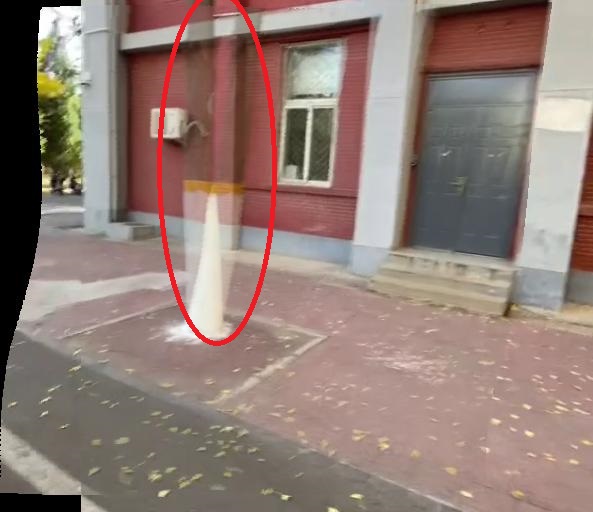}\\[2pt]
    \small APAP
  \end{minipage}
\begin{minipage}[b]{0.30\linewidth}
    \centering
    \includegraphics[width=\linewidth]{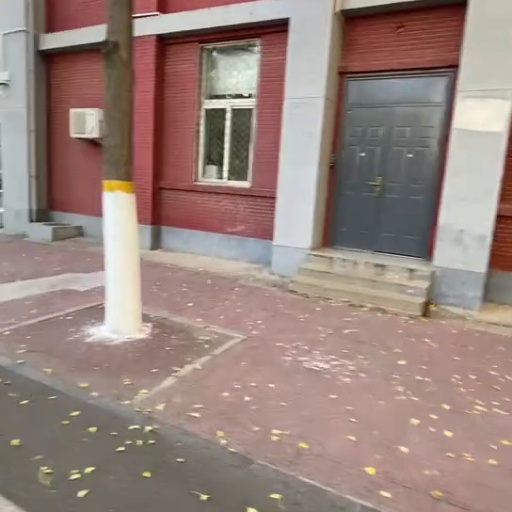}\\[2pt]
    \small TARGET
  \end{minipage}
\begin{minipage}[b]{0.30\linewidth}
    \centering
    \includegraphics[width=\linewidth]{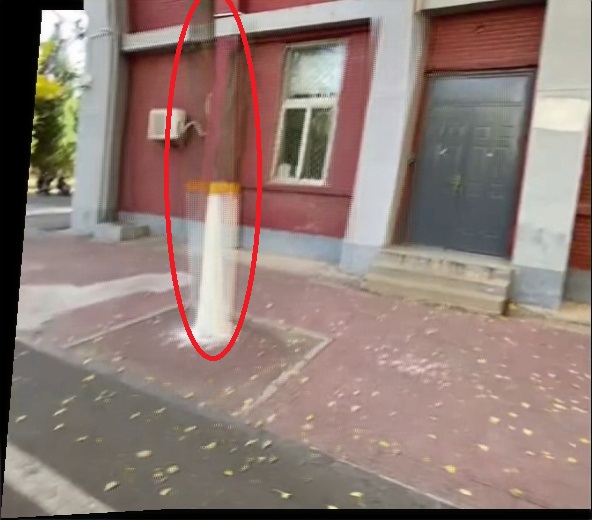}\\[2pt]
    \small UDIS
  \end{minipage}
\begin{minipage}[b]{0.30\linewidth}
    \centering
    \includegraphics[width=\linewidth]{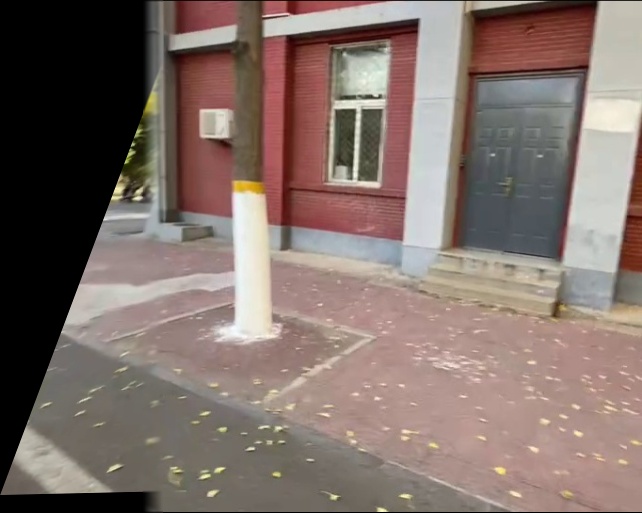}\\[2pt]
    \small OURS
  \end{minipage}
 
\caption{Similar artifacts as in \ref{qualitative2} for the state of the art.}
  \label{qualitative5}
\end{figure}

\begin{figure}[H]
\centering
  \begin{minipage}[b]{0.20\linewidth}
    \centering
    \includegraphics[width=\linewidth]{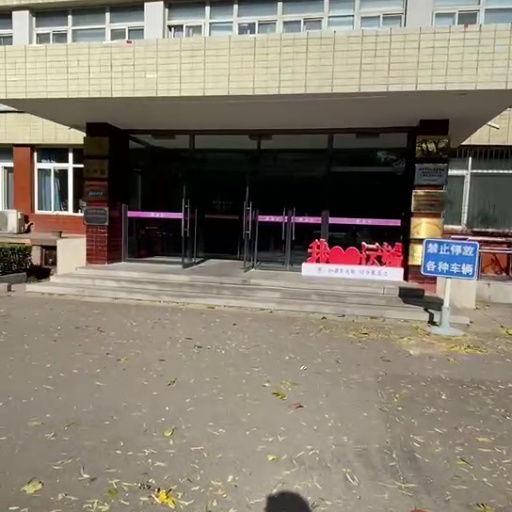}\\[2pt]
    \small SOURCE
  \end{minipage}
  \begin{minipage}[b]{0.30\linewidth}
    \centering
    \includegraphics[width=\linewidth]{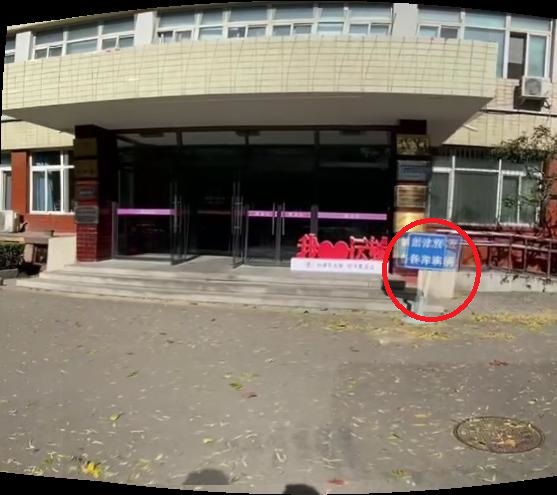}\\[2pt]
    \small ELA
  \end{minipage}
  \begin{minipage}[b]{0.30\linewidth}
    \centering
    \includegraphics[width=\linewidth]{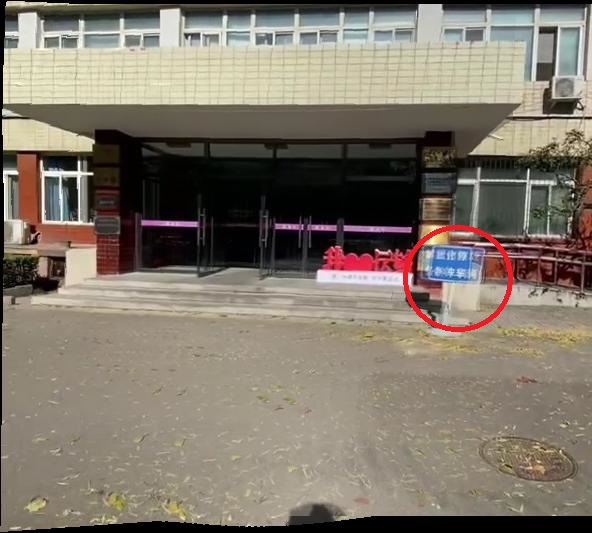}\\[2pt]
    \small APAP
  \end{minipage}
\begin{minipage}[b]{0.20\linewidth}
    \centering
    \includegraphics[width=\linewidth]{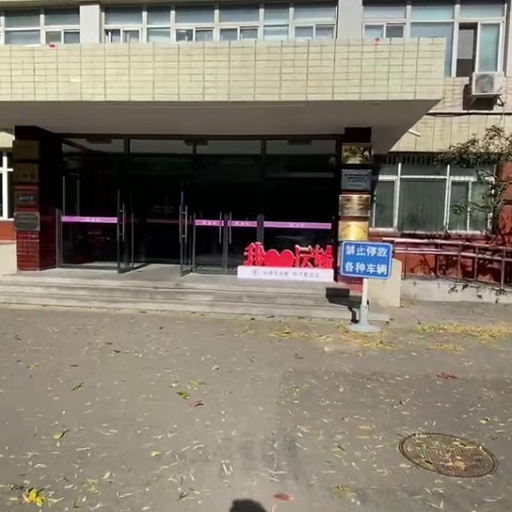}\\[2pt]
    \small TARGET
  \end{minipage}
\begin{minipage}[b]{0.30\linewidth}
    \centering
    \includegraphics[width=\linewidth]{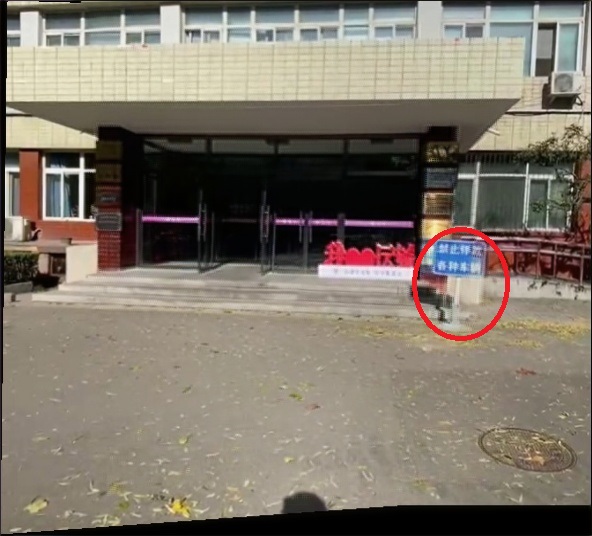}\\[2pt]
    \small UDIS
  \end{minipage}
\begin{minipage}[b]{0.30\linewidth}
    \centering
    \includegraphics[width=\linewidth]{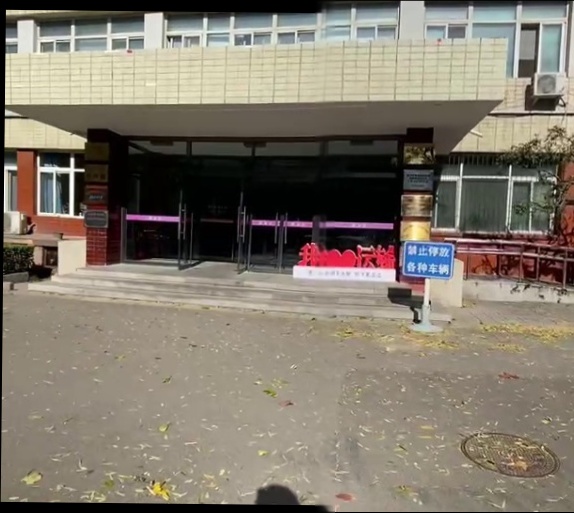}\\[2pt]
    \small OURS
  \end{minipage}
 
\caption{Similar artifacts as in \ref{qualitative2} for the state of the art.}
  \label{qualitative6}
\end{figure}

\begin{figure}[H]
\centering
  \begin{minipage}[b]{0.30\linewidth}
    \centering
    \includegraphics[width=\linewidth]{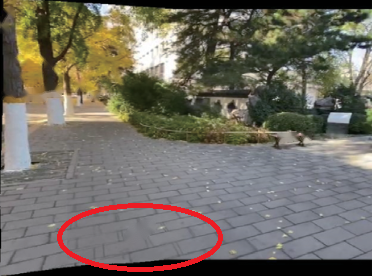}\\[2pt]
    \small APAP
  \end{minipage}
  \begin{minipage}[b]{0.30\linewidth}
    \centering
    \includegraphics[width=\linewidth]{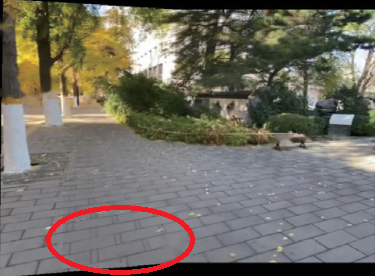}\\[2pt]
    \small LPC
  \end{minipage}

 \begin{minipage}[b]{0.30\linewidth}
    \centering
    \includegraphics[width=\linewidth]{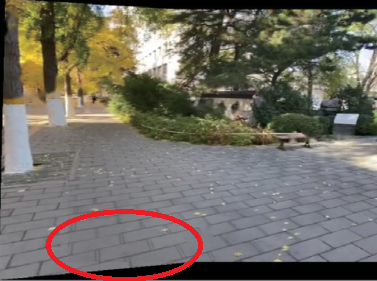}\\[2pt]
    \small SPW
  \end{minipage}
\begin{minipage}[b]{0.30\linewidth}
    \centering
    \includegraphics[width=\linewidth]{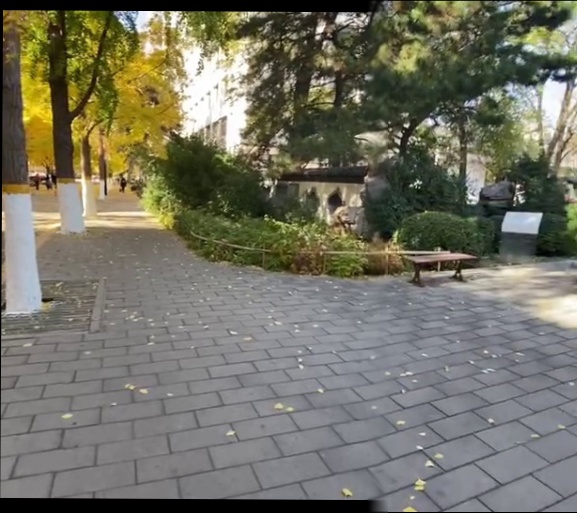}\\[2pt]
    \small OURS
  \end{minipage}
 
\caption{Misalignments (tiles in the floor) in APAP, LPC, SPW, and correct alignment in SENA.}
  \label{qualitative7}
\end{figure}

\newcommand{\MainColW}{0.30\linewidth}
\newcommand{\ThumbColW}{0.10\linewidth}
\newcommand{\ColSep}{0.010\linewidth}

\newcommand{\RowSeven}[8]{%
  \noindent
  \begin{tabular}{@{}p{\MainColW}
                  @{\hspace{\ColSep}}p{\ThumbColW}
                  @{\hspace{\ColSep}}p{\ThumbColW}
                  @{\hspace{\ColSep}}p{\ThumbColW}
                  @{\hspace{\ColSep}}p{\ThumbColW}
                  @{\hspace{\ColSep}}p{\ThumbColW}
                  @{\hspace{\ColSep}}p{\ThumbColW}@{}}
    \includegraphics[width=\linewidth]{#1} &
    \includegraphics[width=\linewidth]{#2} &
    \includegraphics[width=\linewidth]{#3} &
    \includegraphics[width=\linewidth]{#4} &
    \includegraphics[width=\linewidth]{#5} &
    \includegraphics[width=\linewidth]{#6} &
    \includegraphics[width=\linewidth]{#7}
  \end{tabular}\\[2pt]\textbf{#8}\par
}

\begin{figure}[H]
  \centering

  \RowSeven{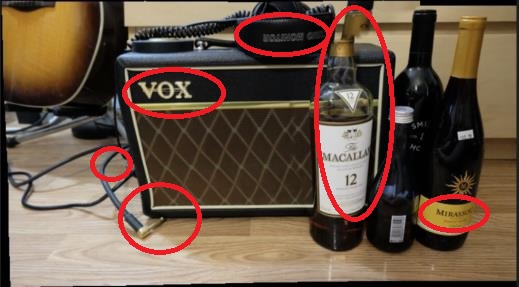}
           {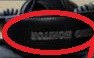}
           {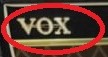}
           {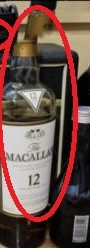}
           {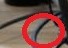}
           {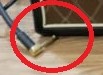}
           {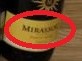}
           {UDIS++}

  \vspace{0.2em}

  \RowSeven{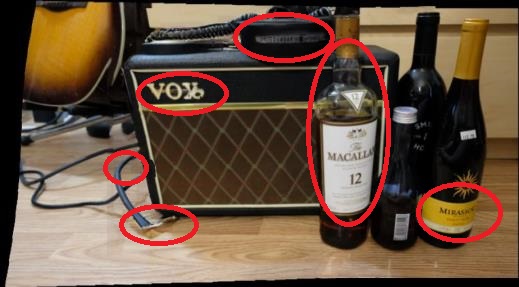}
           {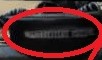}
           {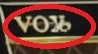}
           {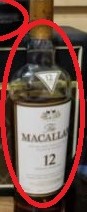}
           {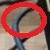}
           {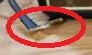}
           {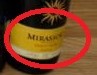}
           {SEAMLESS}

  \vspace{0.2em}

  \RowSeven{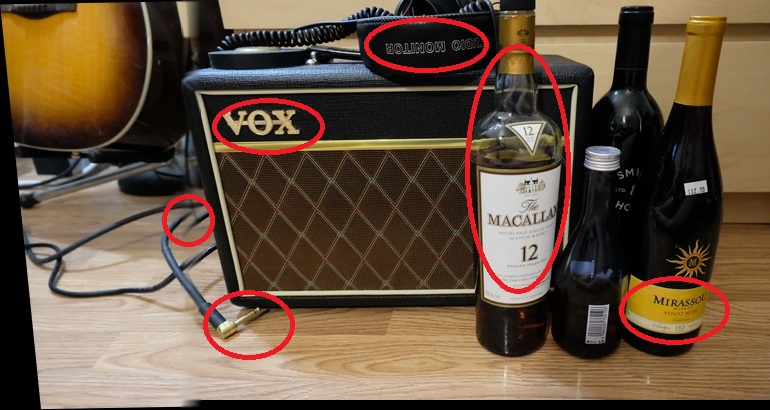}
           {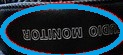}
           {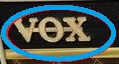}
           {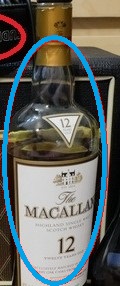}
           {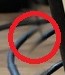}
           {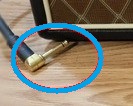}
           {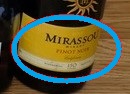}
           {OURS}

  \caption{UDIS++ \cite{nie2023parallax} and SEAMLESS \cite{chen2024seamless} present undistinguishable writings due to the blurred texture of the image, and misalignments. In SENA, the writings are clear as in the input images and a single slight misalignment is observed. }
  \label{fig:stitch_three_rows_six_zooms}
\end{figure}

Qualitative results show that state-of-the-art methods often produce noticeable artifacts, including duplication, texture loss, and stretching of image structures. In contrast, SENA generates sharper, higher-resolution stitched images, while preserving structural integrity and visual realism.

\subsection{Ablation studies}
We evaluate here the effectiveness of our three main components — the locally adaptive image warping, the adequate parallax-free zone identification, and the segment-based reconstruction — in improving alignment and removing artifacts.
\begin{figure}[H]
    \centering
    \begin{minipage}{0.48\textwidth}
        \centering
        \includegraphics[width=\linewidth]{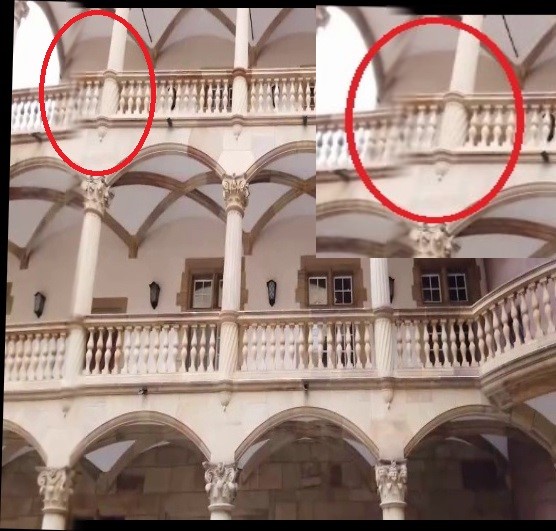}
        \caption{Without our warping strategy. This leads to alignment errors and geometric distortions.}
        \label{fig:image1}
    \end{minipage}
    \hfill
    \begin{minipage}{0.48\textwidth}
        \centering
        \includegraphics[width=\linewidth]{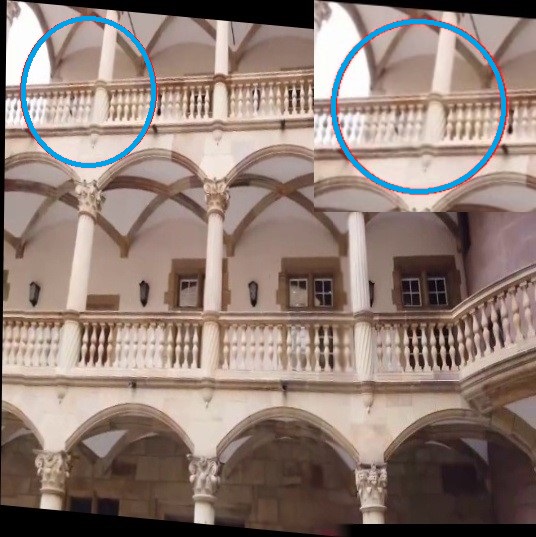}
        \caption{With our locally-adaptive image warping}
        \label{fig:image2}
    \end{minipage}
\end{figure}
\begin{figure}[H]
    \centering
    \begin{minipage}{0.48\textwidth}
        \centering
        \includegraphics[width=\linewidth]{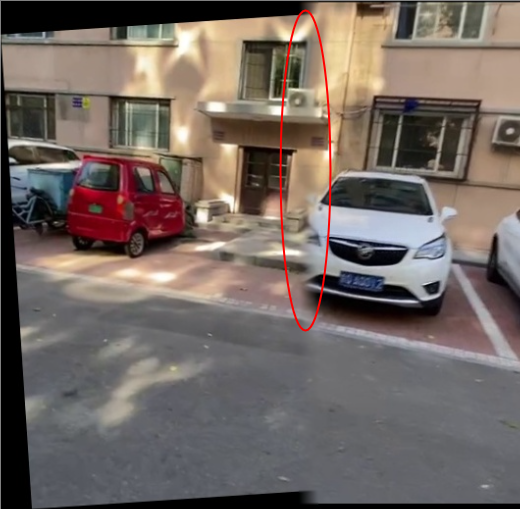}
        \caption{Without the identification of a parallax-free zone, the stitching might be performed in an unstable region, leading to visual artifacts in the result}
        \label{fig:image1}
    \end{minipage}
    \hfill
    \begin{minipage}{0.48\textwidth}
        \centering
        \includegraphics[width=\linewidth]{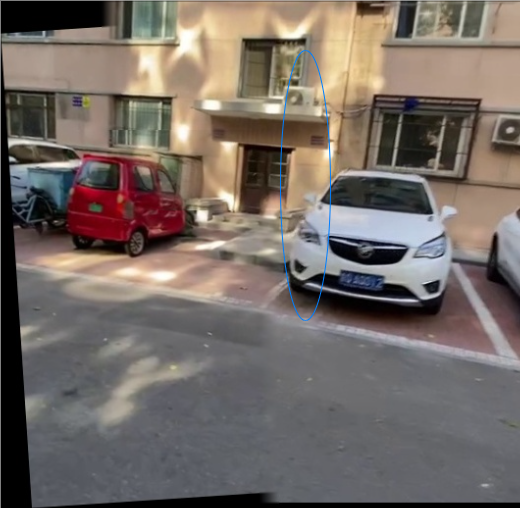}
        \caption{When identifying a parallax-free zone, the stitching is performed in a more stable region with consistent keypoints, resulting to an improved alignment.}
        \label{fig:image2}
    \end{minipage}
\end{figure}
\begin{figure}[H]
    \centering
    \begin{minipage}{0.48\textwidth}
        \centering
        \includegraphics[width=\linewidth]{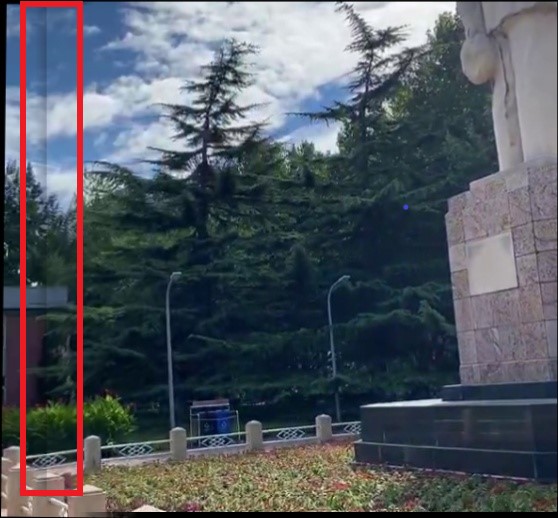}
        \caption{Without blending}
        \label{fig:image1}
    \end{minipage}
    \hfill
    \begin{minipage}{0.48\textwidth}
        \centering
        \includegraphics[width=\linewidth]{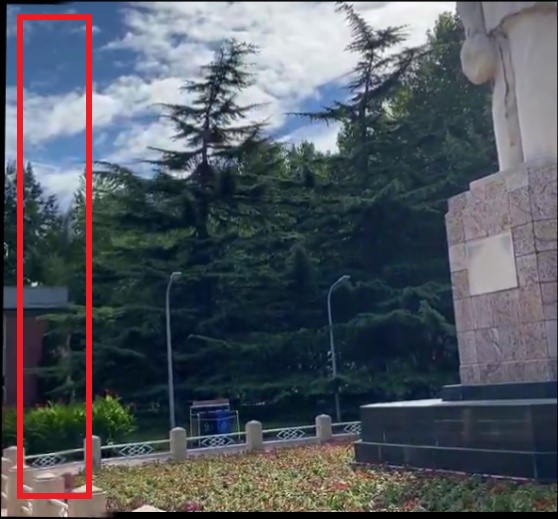}
        \caption{with our blending (linear alpha transition + light Gaussian smoothing)}
        \label{fig:image2}
    \end{minipage}
\end{figure}

\subsection{Limitations} 

As demonstrated earlier, SENA effectively stitches images while preserving their natural appearance and geometric structure. However, certain limitations remain. The alignment between images is achieved through a multi-stage local deformation process: a global affine transformation estimated via RANSAC provides coarse alignment, which is then refined within the overlap region using locally adaptive affine models interpolated into a smooth Free-Form Deformation (FFD) field. This design allows accurate alignment while mitigating excessive distortion and maintaining shape and texture consistency. Nevertheless, in scenes with complex depth variations, the estimation of local models may become unstable, leading to locally inconsistent deformations or slight geometric discontinuities. Furthermore, the quality of the detected keypoints directly influences the identification of the adequate parallax-minimized zone and the subsequent stitching line extraction. When feature detection or matching is unreliable, artifacts such as minor ghosting, blurring, or local misalignment may still appear in the final panorama.
\begin{figure}[H]
    \centering
    \subfigure{
        \includegraphics[width=0.50\textwidth]{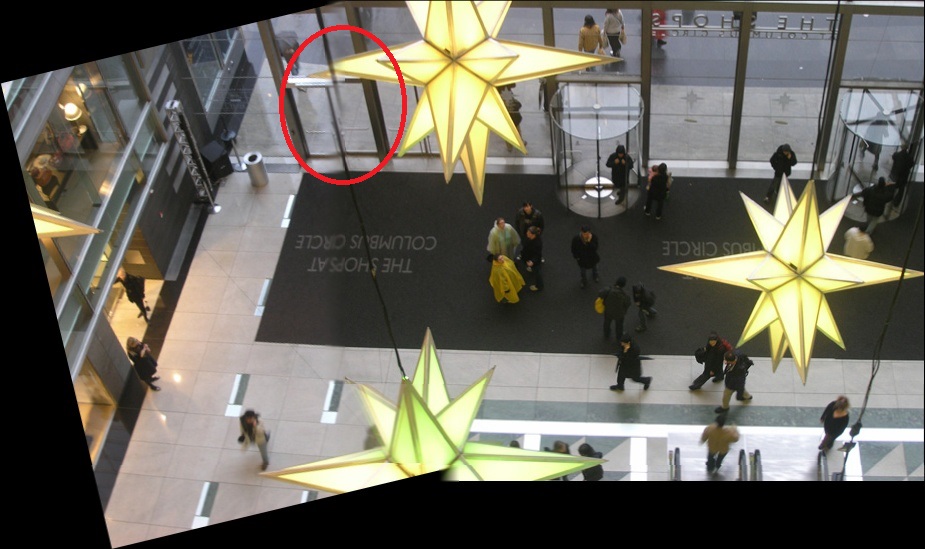}
}
    \hfill
    \subfigure{
        \includegraphics[width=0.40\textwidth]{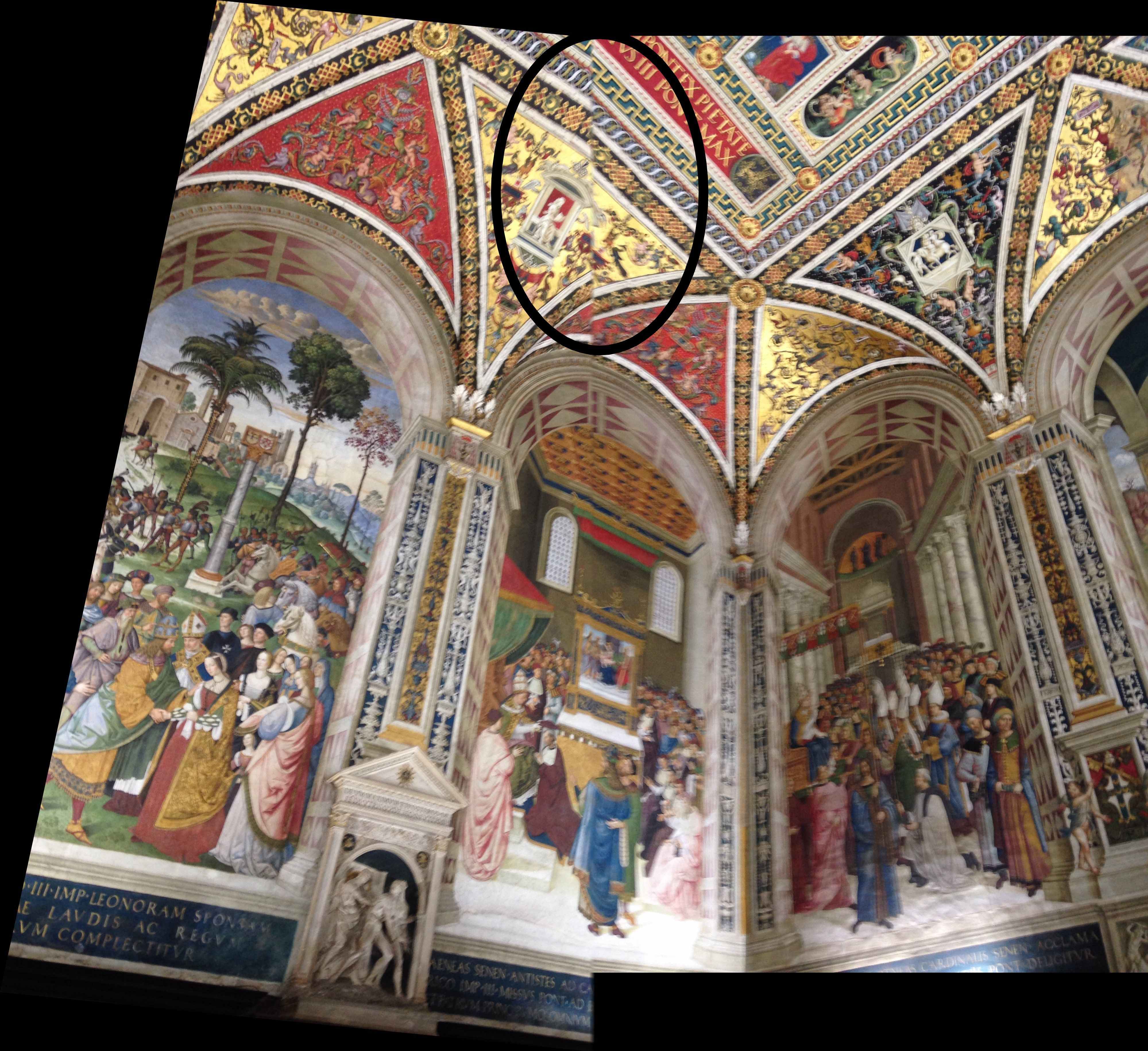} 
    }
    \caption{Some failure cases}
\end{figure}

\section{Conclusion}
This work introduced SENA, a structure-preserving image stitching framework explicitly designed to address the geometric limitations of homography-based and over-flexible warping models. Rather than compensating for depth induced parallax through non-physical distortions, SENA adopts a hierarchical affine deformation strategy, combining global affine initialization, local affine refinement within the overlap region, and a smoothly regularized free-form deformation field. This multi-scale design preserves local shape, parallelism, and aspect ratios while providing sufficient flexibility to ensure accurate alignment.
To further mitigate parallax-related artifacts, SENA incorporates a geometry-driven stitching mechanism that identifies a parallax minimized adequate zone directly from the consistency of RANSAC filtered correspondences, avoiding reliance on semantic segmentation or learning-based preprocessing. Based on this zone, an anchor-based segmentation strategy is applied, partitioning the overlapping images into corresponding vertical slices aligned by refined keypoints. This guarantees one-to-one geometric correspondence across segments and enforces structural continuity across the stitching boundary, effectively reducing ghosting, duplication, and texture stretching.

Extensive experiments demonstrate that SENA consistently achieves visually coherent panoramas with improved structural fidelity and reduced distortion compared to representative homography-based and spline-based methods. While the method remains robust across a wide range of scenes, performance may degrade under extreme viewpoint changes or in severely texture-deprived environments, where feature extraction becomes unreliable. Future work will explore adaptive local deformation models and hybrid learning-based components to further enhance robustness under such challenging conditions.
\bibliography{sn-bibliography}

@article{fischler1981random,
  title={Random sample consensus: a paradigm for model fitting with applications to image analysis and automated cartography},
  author={Fischler, Martin A and Bolles, Robert C},
  journal={Communications of the ACM},
  volume={24},
  number={6},
  pages={381--395},
  year={1981},
  publisher={ACM New York, NY, USA}
}

@inproceedings{du2022geometric,
  title={Geometric structure preserving warp for natural image stitching},
  author={Du, Peng and Ning, Jifeng and Cui, Jiguang and Huang, Shaoli and Wang, Xinchao and Wang, Jiaxin},
  booktitle={Proceedings of the IEEE/CVF conference on computer vision and pattern recognition},
  pages={3688--3696},
  year={2022}
}

@inproceedings{herrmann2018object,
  title={Object-centered image stitching},
  author={Herrmann, Charles and Wang, Chen and Bowen, Richard Strong and Keyder, Emil and Zabih, Ramin},
  booktitle={Proceedings of the European Conference on Computer Vision (ECCV)},
  pages={821--835},
  year={2018}
}

@inproceedings{bay2006surf,
  title={Surf: Speeded up robust features},
  author={Bay, Herbert and Tuytelaars, Tinne and Van Gool, Luc},
  booktitle={Computer Vision--ECCV 2006: 9th European Conference on Computer Vision, Graz, Austria, May 7-13, 2006. Proceedings, Part I 9},
  pages={404--417},
  year={2006},
  organization={Springer}
}

@article{lowe2004distinctive,
  title={Distinctive image features from scale-invariant keypoints},
  author={Lowe, David G},
  journal={International journal of computer vision},
  volume={60},
  pages={91--110},
  year={2004},
  publisher={Springer}
}

@inproceedings{zaragoza2013projective,
  title={As-projective-as-possible image stitching with moving DLT},
  author={Zaragoza, Julio and Chin, Tat-Jun and Brown, Michael S and Suter, David},
  booktitle={Proceedings of the IEEE conference on computer vision and pattern recognition},
  pages={2339--2346},
  year={2013}
}

@inproceedings{yadav2018selfie,
  title={Selfie Stitch: Dual Homography Based Image Stitching for Wide-Angle Selfie},
  author={Yadav, Sourabh and Choudhary, Pradeep and Goel, Srishti and Parameswaran, Sankaranarayanan and Bajpai, Pankaj and Kim, Jaehyun},
  booktitle={2018 IEEE International Conference on Multimedia \& Expo Workshops (ICMEW)},
  pages={1--4},
  year={2018},
  organization={IEEE}
}

@inproceedings{jia2021leveraging,
  title={Leveraging line-point consistence to preserve structures for wide parallax image stitching},
  author={Jia, Qi and Li, ZhengJun and Fan, Xin and Zhao, Haotian and Teng, Shiyu and Ye, Xinchen and Latecki, Longin Jan},
  booktitle={Proceedings of the IEEE/CVF conference on computer vision and pattern recognition},
  pages={12186--12195},
  year={2021}
}

@inproceedings{nie2023parallax,
  title={Parallax-tolerant unsupervised deep image stitching},
  author={Nie, Lang and Lin, Chunyu and Liao, Kang and Liu, Shuaicheng and Zhao, Yao},
  booktitle={Proceedings of the IEEE/CVF international conference on computer vision},
  pages={7399--7408},
  year={2023}
}

@article{huang2021semantic,
  title={Semantic segmentation guided feature point classification and seam fusion for image stitching},
  author={Huang, Huafeng and Chen, Fei and Cheng, Hang and Li, Liyao and Wang, Meiqing},
  journal={Journal of Algorithms \& Computational Technology},
  volume={15},
  pages={17483026211065399},
  year={2021},
  publisher={SAGE Publications Sage UK: London, England}
}

@inproceedings{li2023combined,
  title={Combined regional homography-affine warp for image stitching},
  author={Li, Xinyi and He, Lin and He, Xinguo},
  booktitle={Fourteenth International Conference on Graphics and Image Processing (ICGIP 2022)},
  volume={12705},
  pages={258--264},
  year={2023},
  organization={SPIE}
}

@article{liao2019single,
  title={Single-perspective warps in natural image stitching},
  author={Liao, Tianli and Li, Nan},
  journal={IEEE transactions on image processing},
  volume={29},
  pages={724--735},
  year={2019},
  publisher={IEEE}
}

@article{wen2022structure,
  title={Structure preservation and seam optimization for parallax-tolerant image stitching},
  author={Wen, Shaoping and Wang, Xiaolei and Zhang, Weichao and Wang, Guanjun and Huang, Mengxing and Yu, Benguo},
  journal={IEEE Access},
  volume={10},
  pages={78713--78725},
  year={2022},
  publisher={IEEE}
}

@article{nie2021unsupervised,
  title={Unsupervised deep image stitching: Reconstructing stitched features to images},
  author={Nie, Lang and Lin, Chunyu and Liao, Kang and Liu, Shuaicheng and Zhao, Yao},
  journal={IEEE Transactions on Image Processing},
  volume={30},
  pages={6184--6197},
  year={2021},
  publisher={IEEE}
}

@article{ji2022research,
  title={Research on image stitching method based on improved ORB and stitching line calculation},
  author={Ji, Xiu and Yang, Huamin and Han, Cheng},
  journal={Journal of Electronic Imaging},
  volume={31},
  number={5},
  pages={051404--051404},
  year={2022},
  publisher={Society of Photo-Optical Instrumentation Engineers}
}

@article{liu2023utilization,
  title={Utilization of merge-sorting method to improve stitching efficiency in multi-scene image stitching},
  author={Liu, Wei and Zhang, Kunhao and Zhang, Yan and He, Jian and Sun, Bo},
  journal={Applied Sciences},
  volume={13},
  number={5},
  pages={2791},
  year={2023},
  publisher={MDPI}
}

@inproceedings{potje2024xfeat,
  title={XFeat: Accelerated Features for Lightweight Image Matching},
  author={Potje, Guilherme and Cadar, Felipe and Araujo, Andr{\'e} and Martins, Renato and Nascimento, Erickson R},
  booktitle={Proceedings of the IEEE/CVF Conference on Computer Vision and Pattern Recognition},
  pages={2682--2691},
  year={2024}
}

@book{nghonda2023enable,
  title={Enable 360-Degree Immersion for Outside-In Camera Systems for Live Events},
  author={Nghonda, Erman},
  year={2023},
  publisher={University of Florida}
}

@inproceedings{chen2024seamless,
  title={Seamless-Through-Breaking: Rethinking Image Stitching for Optimal Alignment},
  author={Chen, KuanYan and Garg, Atik and Wang, Yu-Shuen},
  booktitle={Proceedings of the Asian Conference on Computer Vision},
  pages={4352--4367},
  year={2024}
}

@article{qin2021image,
  title={Image stitching by feature positioning and seam elimination},
  author={Qin, Yunbai and Li, Jialiang and Jiang, Pinqun and Jiang, Frank},
  journal={Multimedia Tools and Applications},
  volume={80},
  number={14},
  pages={20869--20881},
  year={2021},
  publisher={Springer}
}

@article{chai2022upscaling,
  title={An Upscaling--Downscaling Optimal Seamline Detection Algorithm for Very Large Remote Sensing Image Mosaicking},
  author={Chai, Xuchao and Chen, Jianyu and Mao, Zhihua and Zhu, Qiankun},
  journal={Remote Sensing},
  volume={15},
  number={1},
  pages={89},
  year={2022},
  publisher={MDPI}
}

@inproceedings{garg2020stitching,
  title={Stitching Strip Determination for Optimal Seamline Search},
  author={Garg, Atik and Dung, Lan-Rong},
  booktitle={2020 4th International Conference on Imaging, Signal Processing and Communications (ICISPC)},
  pages={29--33},
  year={2020},
  organization={IEEE}
}

@article{li2017parallax,
  title={Parallax-tolerant image stitching based on robust elastic warping},
  author={Li, Jing and Wang, Zhengming and Lai, Shiming and Zhai, Yongping and Zhang, Maojun},
  journal={IEEE Transactions on multimedia},
  volume={20},
  number={7},
  pages={1672--1687},
  year={2017},
  publisher={IEEE}
}

@article{zheng2019novel,
  title={A novel projective-consistent plane based image stitching method},
  author={Zheng, Jin and Wang, Yue and Wang, Hanzi and Li, Bo and Hu, Hai-Miao},
  journal={IEEE Transactions on Multimedia},
  volume={21},
  number={10},
  pages={2561--2575},
  year={2019},
  publisher={IEEE}
}

@article{nie2020view,
  title={A view-free image stitching network based on global homography},
  author={Nie, Lang and Lin, Chunyu and Liao, Kang and Liu, Meiqin and Zhao, Yao},
  journal={Journal of Visual Communication and Image Representation},
  volume={73},
  pages={102950},
  year={2020},
  publisher={Elsevier}
}

\end{document}


\title{\centering \textit{International Journal of Computer Vision (IJCV)}\\
\textbf{Supplementary Material for "Seamlessly Natural: Image Stitching with Natural Appearance Preservation"}}

\author[1]{\fnm{Gaetane Lorna N.} \sur{TCHANA}}\email{gaetane.tchana@facsciences-uy1.cm}

\author[1]{\fnm{Damaris Belle M.} \sur{Fotso}}\email{makwate.fotso@facsciences-uy1.cm}
\author[2]{\fnm{Antonio} \sur{Hendricks}}\email{a.hendricks1@ufl.edu}

\author*[2]{\fnm{Christophe} \sur{Bobda}}\email{cbobda@ece.ufl.edu}

\affil[1]{\orgname{University of Yaounde I}, \orgaddress{\city{Yaoundé}, \postcode{812}, \country{Cameroon}}}

\affil[2]{\orgname{University of Florida, ECE}, \orgaddress{\city{Gainesville}, \postcode{32611}, \state{FL}, \country{US}}}
\maketitle

\section{Results}

We compare our approach against several state-of-the-art methods, including LPC~\cite{jia2021leveraging}, APAP~\cite{zaragoza2013projective}, ELA~\cite{li2017parallax}, SPW~\cite{liao2019single}, UDIS~\cite{nie2021unsupervised}, EPISNET\cite{nie2020view}, UDIS++ ~\cite{nie2023parallax} and Seamless \cite{chen2024seamless}.  We utilized publicly available and widely recognized datasets provided by Nie et al. \cite{nie2021unsupervised}, and Jia et al. \cite{jia2021leveraging}. 
Please feel free to zoom in on the images to better observe the highlighted elements. Rectangles indicate areas of excessive distortion, while circles mark other types of artifacts or defects that can be localized in the images.

\begin{figure}[H]
\centering

\newcommand{\ColW}{0.48\linewidth}     
\newcommand{\SmallGap}{2pt}
\newcommand{\BigGap}{6pt}

\begin{minipage}[t]{\ColW}
  \centering

  \begin{minipage}{\linewidth}
    \centering
    \begin{minipage}[t]{0.45\linewidth}
      \centering
      \includegraphics[width=\linewidth]{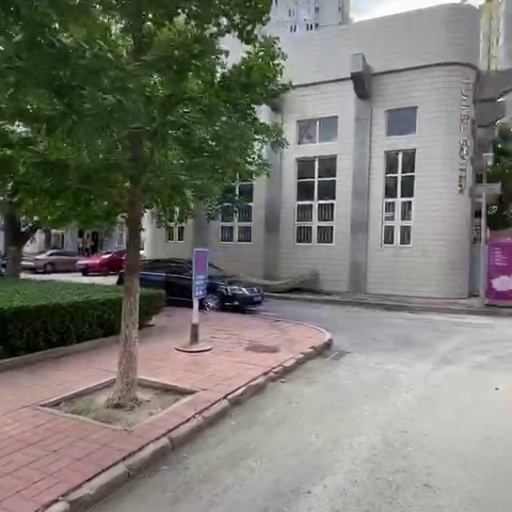}
    \end{minipage}\hfill
    \begin{minipage}[t]{0.45\linewidth}
      \centering
      \includegraphics[width=\linewidth]{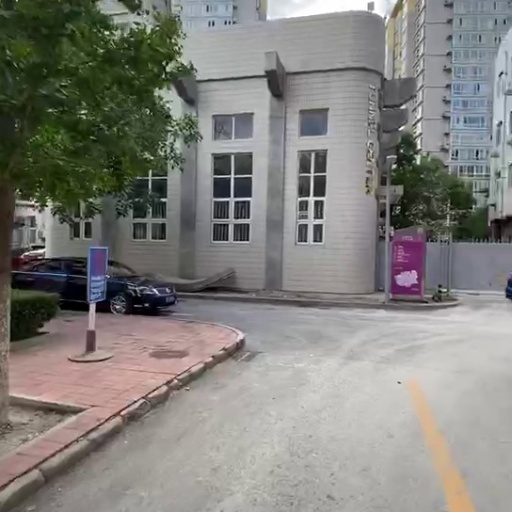}
    \end{minipage}\\[4pt]
    \small Input images
  \end{minipage}

  \vspace{\BigGap}

  \newcommand{\MethodBlock}[2]{%
    \begin{minipage}{\linewidth}
      \centering
      \includegraphics[width=0.95\linewidth]{#1}\\[2pt]
      \small #2
    \end{minipage}\par\vspace{\BigGap}
  }

  \MethodBlock{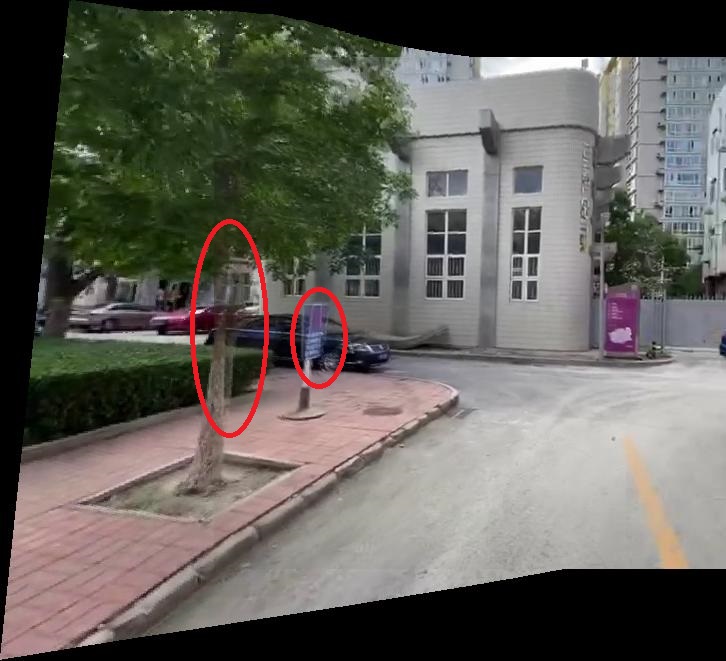}{APAP}
  \MethodBlock{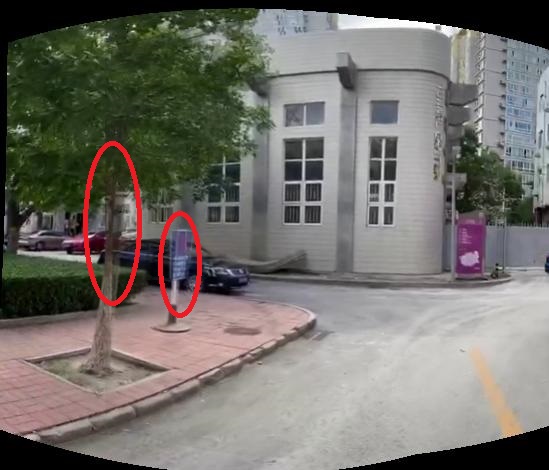}{ELA}
  \MethodBlock{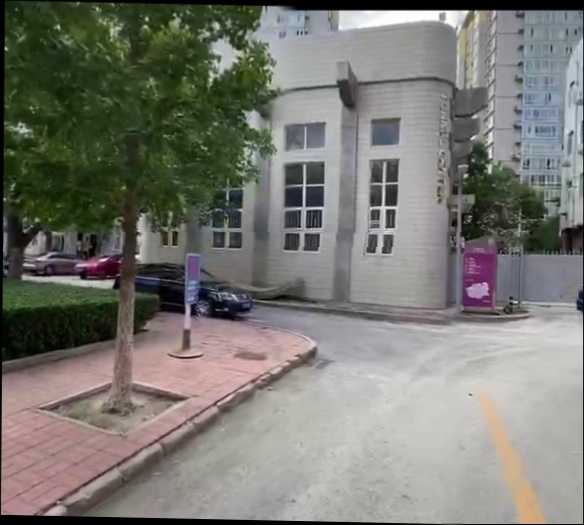}{OURS}

\end{minipage}%
\hfill%
\begin{minipage}[t]{\ColW}
  \centering

  \begin{minipage}{\linewidth}
    \centering
    \begin{minipage}[t]{0.45\linewidth}
      \centering
      \includegraphics[width=\linewidth]{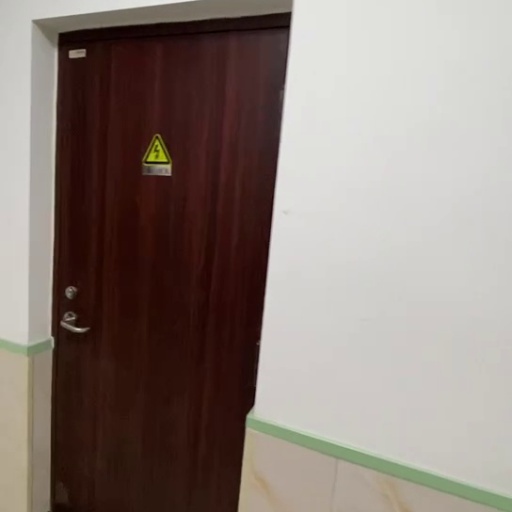}
    \end{minipage}\hfill
    \begin{minipage}[t]{0.45\linewidth}
      \centering
      \includegraphics[width=\linewidth]{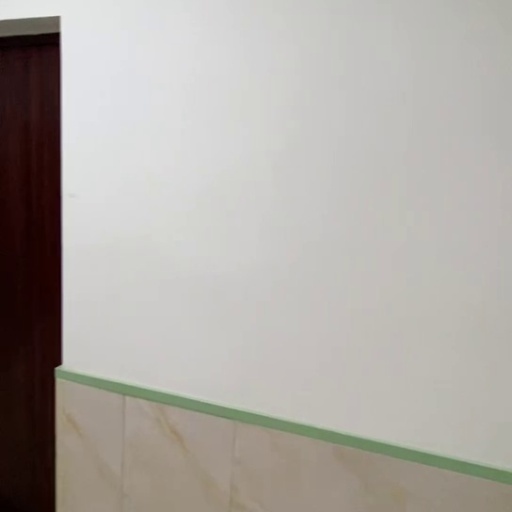}
    \end{minipage}\\[4pt]
    \small input images
  \end{minipage}

  \vspace{\BigGap}

  \begin{minipage}{\linewidth}
    \centering
    \includegraphics[width=0.40\linewidth]{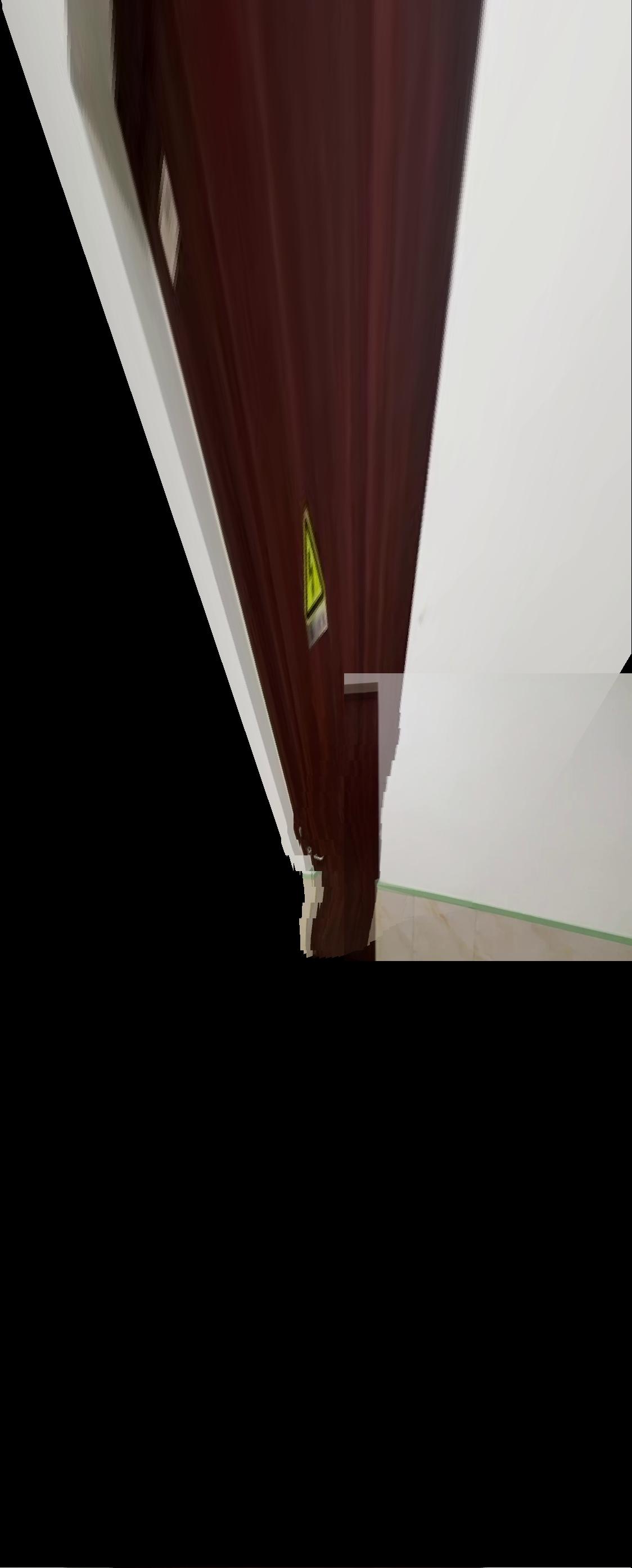}\\[2pt]
    \small APAP
  \end{minipage}\par\vspace{\BigGap}

  \begin{minipage}{\linewidth}
    \centering
    \includegraphics[width=0.80\linewidth]{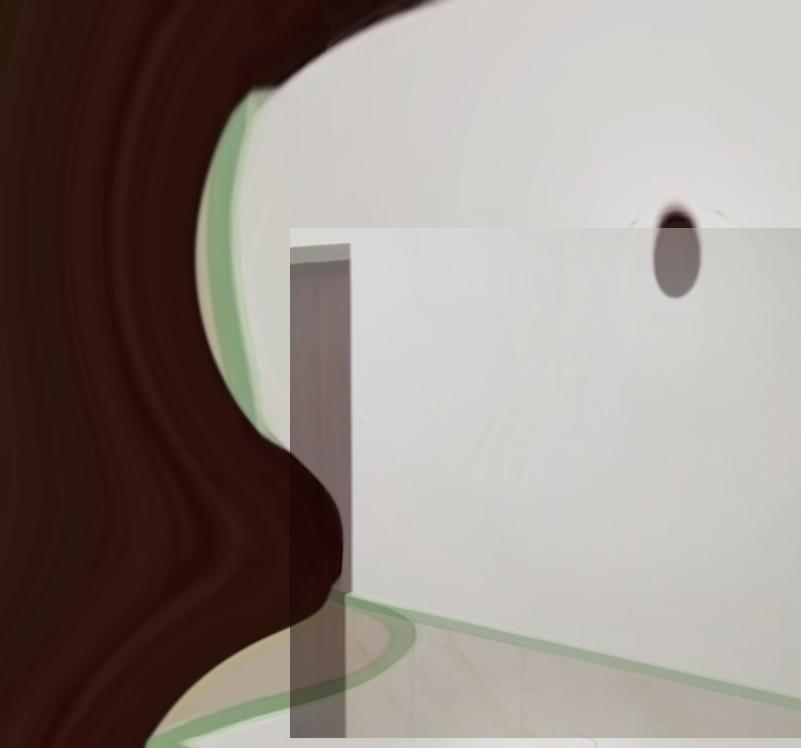}\\[2pt]
    \small ELA
  \end{minipage}\par\vspace{\BigGap}

  \begin{minipage}{\linewidth}
    \centering
    \includegraphics[width=0.80\linewidth]{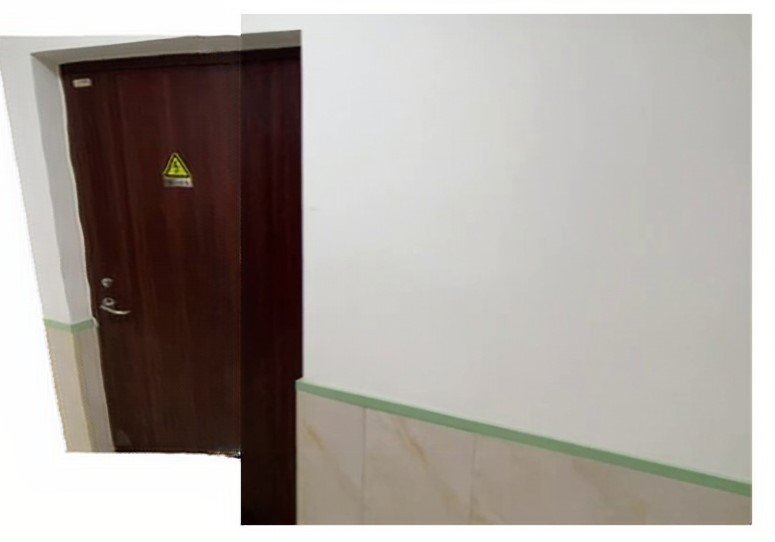}\\[2pt]
    \small EPISNET
  \end{minipage}\par\vspace{\BigGap}

  \begin{minipage}{\linewidth}
    \centering
    \includegraphics[width=0.80\linewidth]{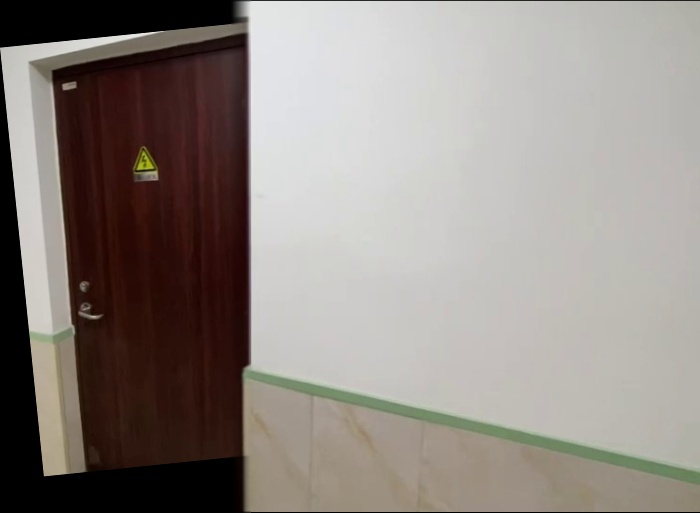}\\[2pt]
    \small OURS
  \end{minipage}

\end{minipage}
\caption{The results obtained with ELA and APAP (In the right-hand column) clearly illustrate the artifacts caused by an inaccurate homography estimation: the resulting image appears severely distorted. In contrast, our method (SENA) demonstrates robustness even in such extreme cases, successfully aligning the images while preserving their natural appearance.}
\label{fig:example1_and_more}
\end{figure}

\begin{figure}[H]
\centering
\setlength{\columnsep}{1em}
\begin{multicols}{2}

\centering
\includegraphics[width=0.48\linewidth]{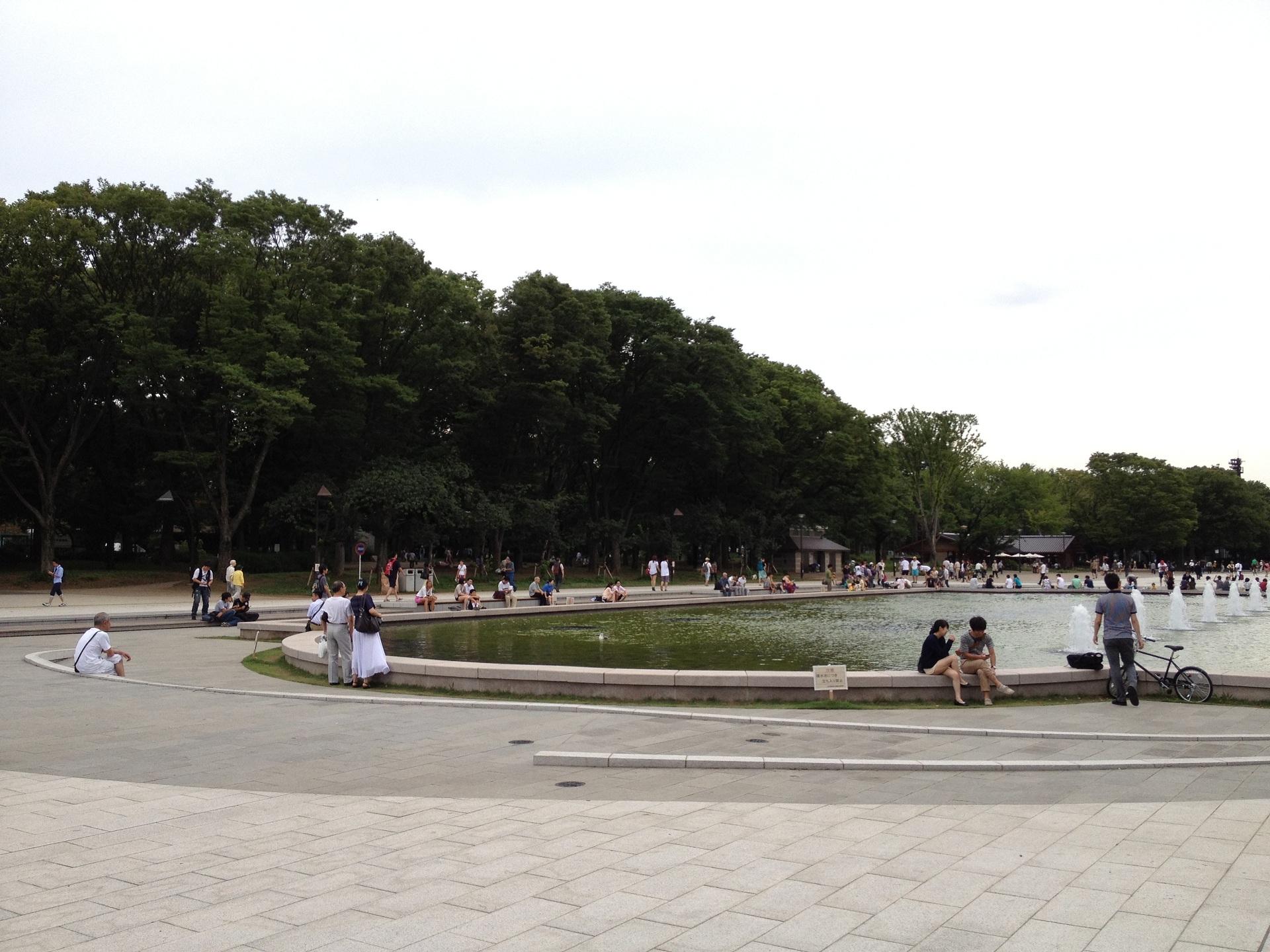}%
\hfill
\includegraphics[width=0.48\linewidth]{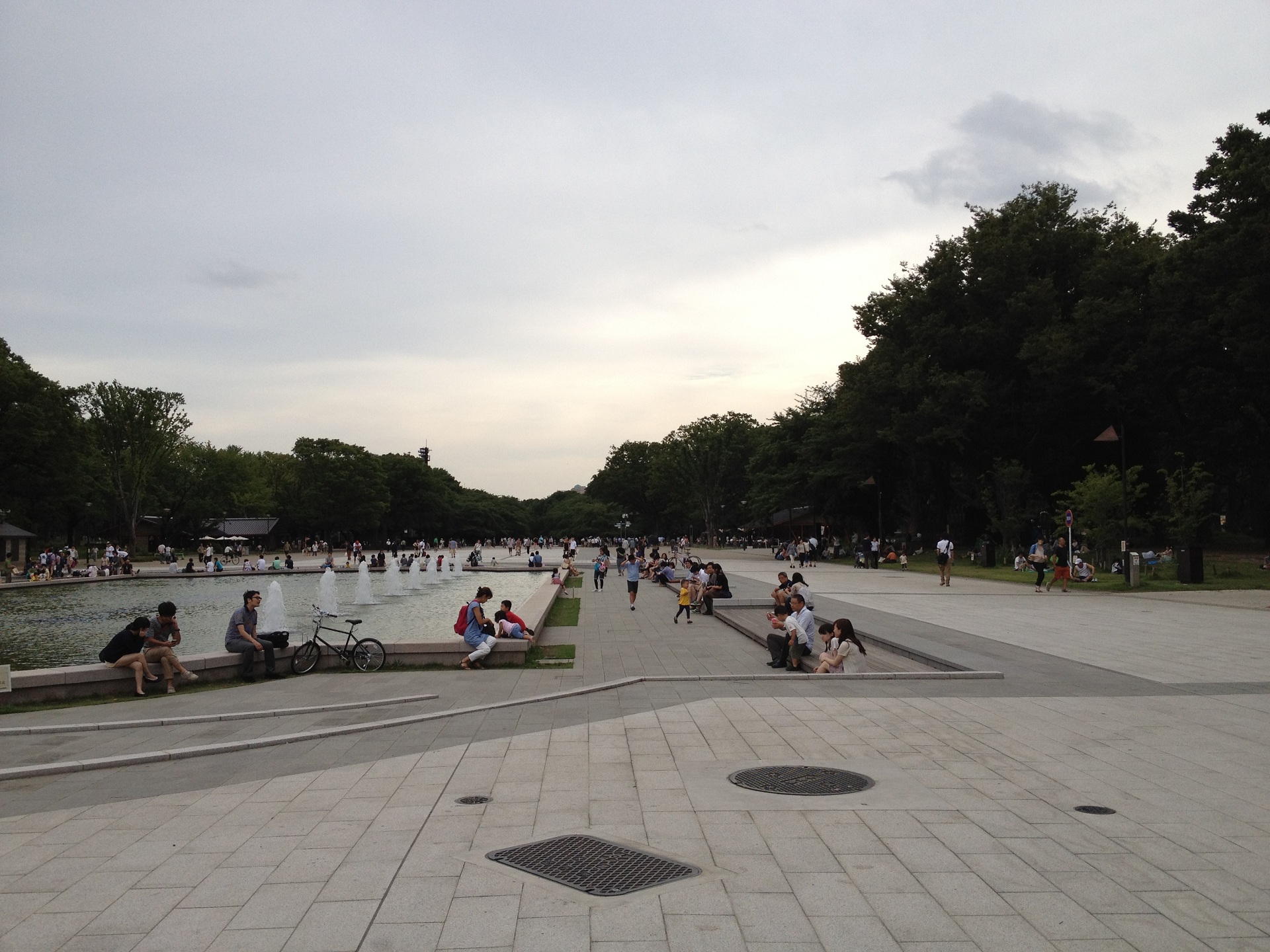}\\[0.4em]
{\small Input images}\\[0.4em]

\includegraphics[width=\linewidth]{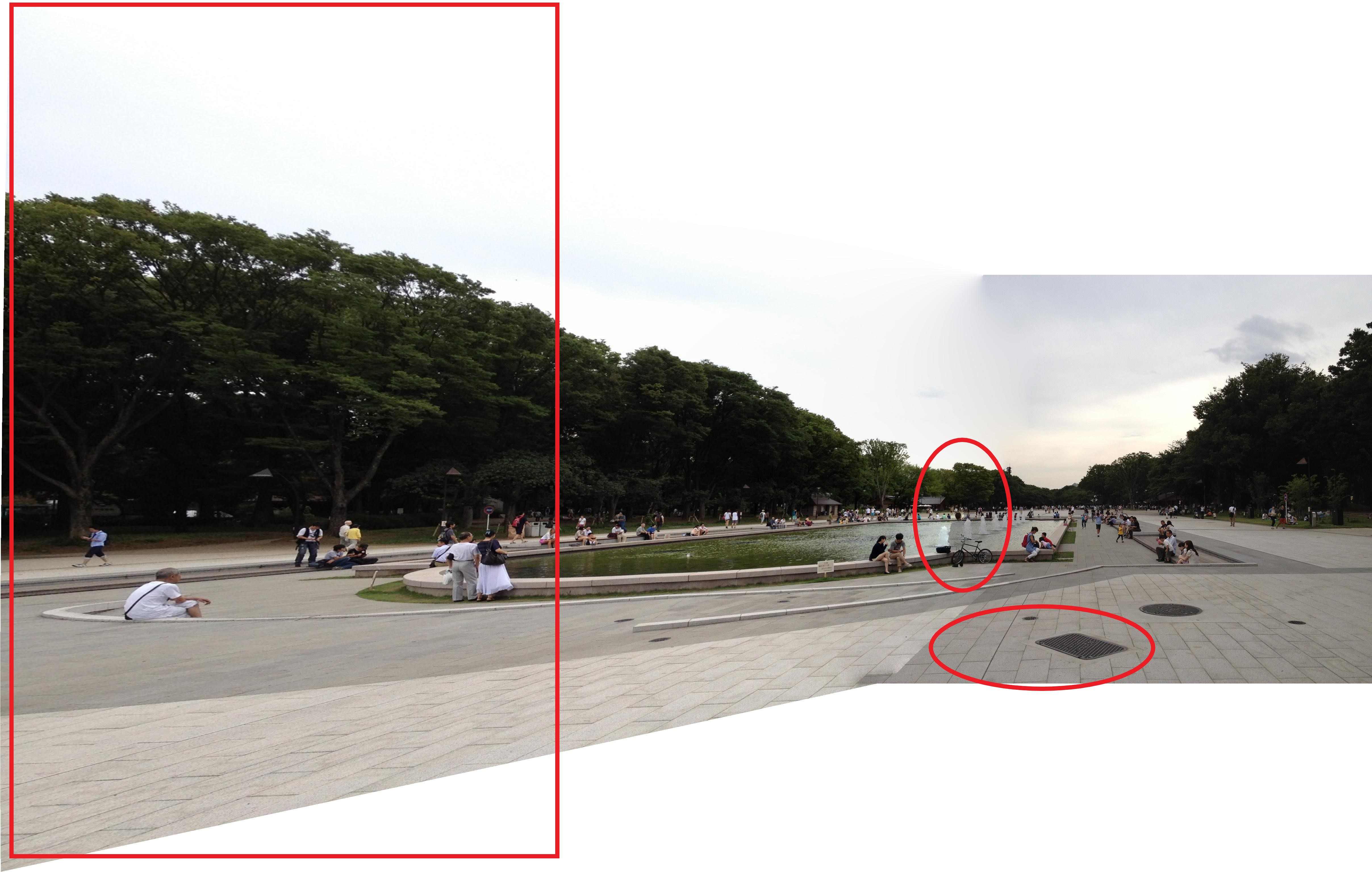}\\[-0.2em]
{\small LPC \cite{jia2021leveraging}}\\[0.4em]

\includegraphics[width=\linewidth]{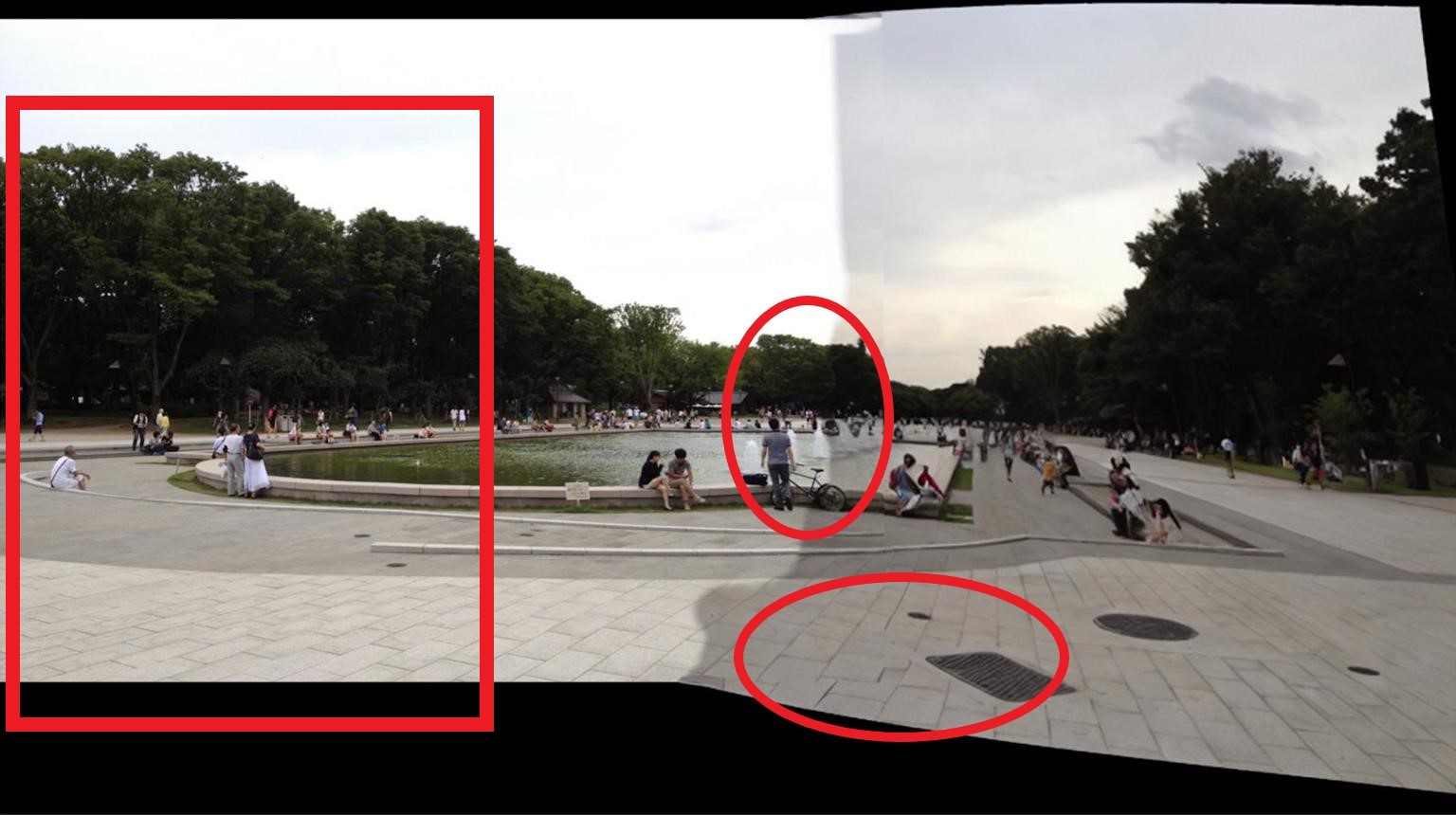}\\[-0.2em]
{\small Seamless \cite{chen2024seamless}}\\[0.4em]

\includegraphics[width=\linewidth]{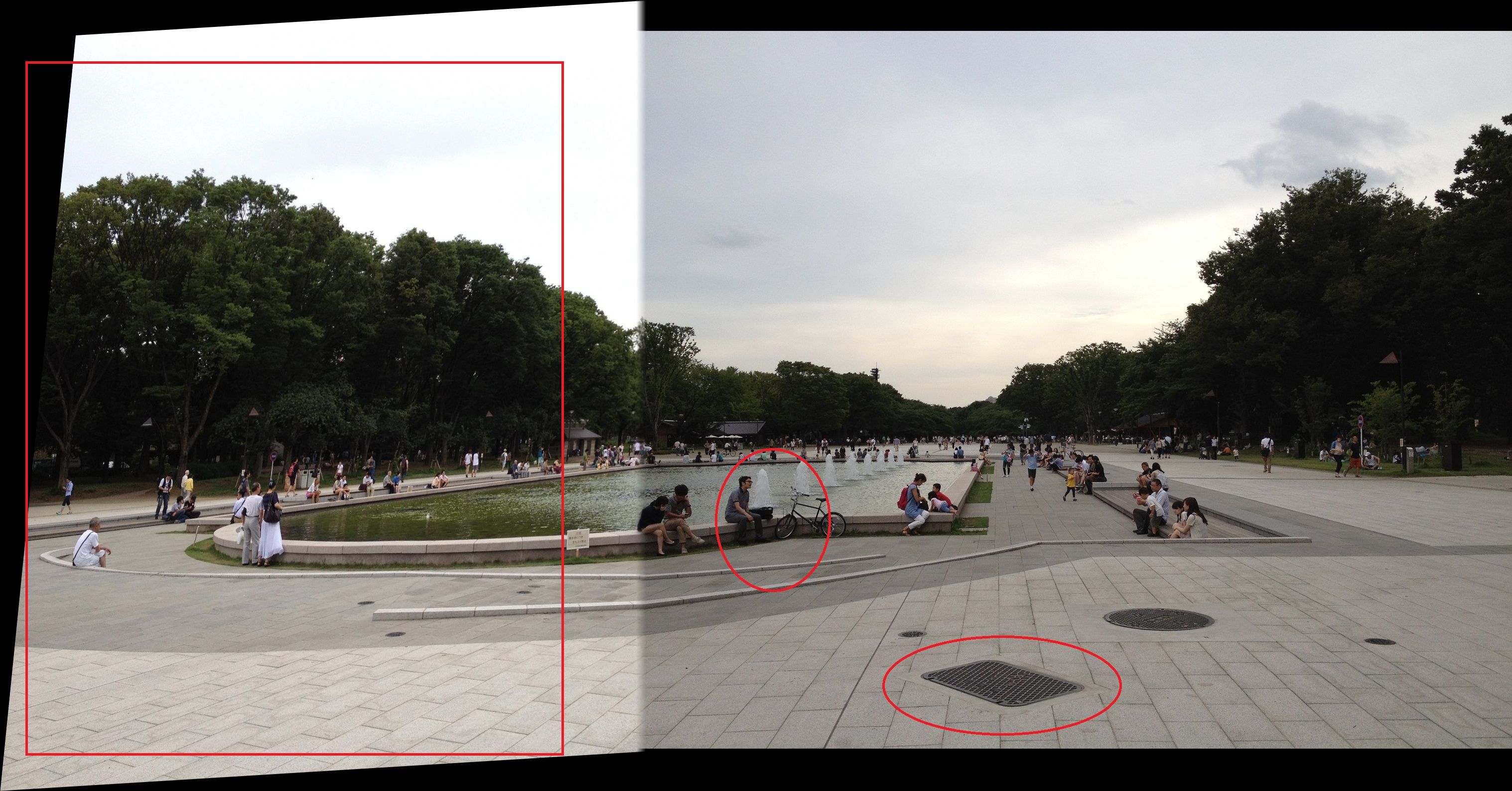}\\[-0.2em]

\centering
\includegraphics[width=0.48\linewidth]{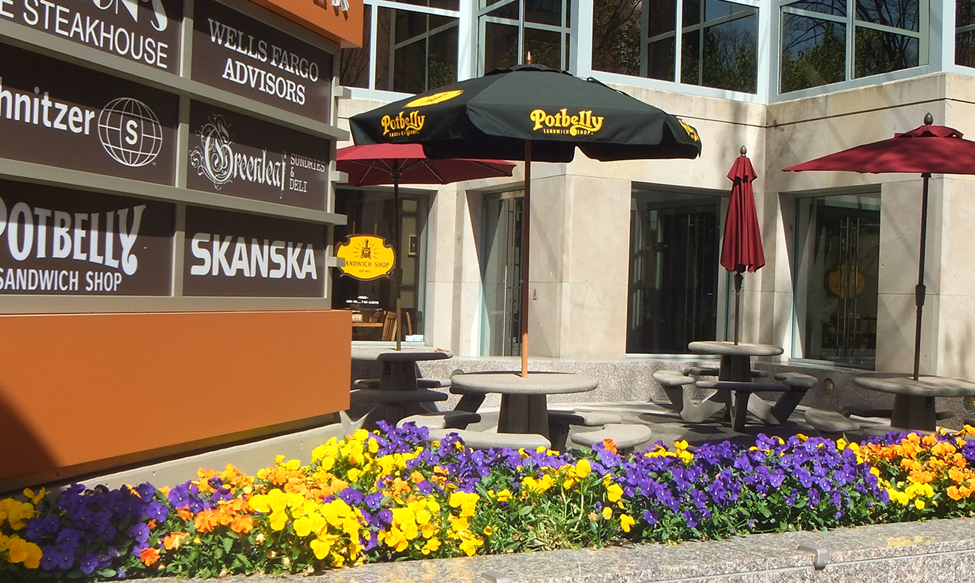}%
\hfill
\includegraphics[width=0.48\linewidth]{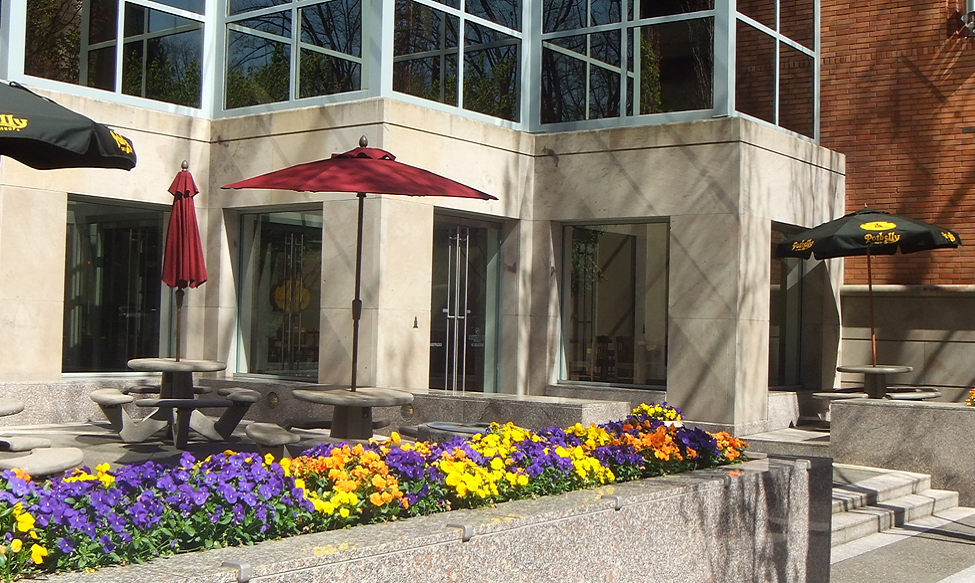}\\[0.4em]
{\small Input images}\\[0.4em]

\includegraphics[width=\linewidth]{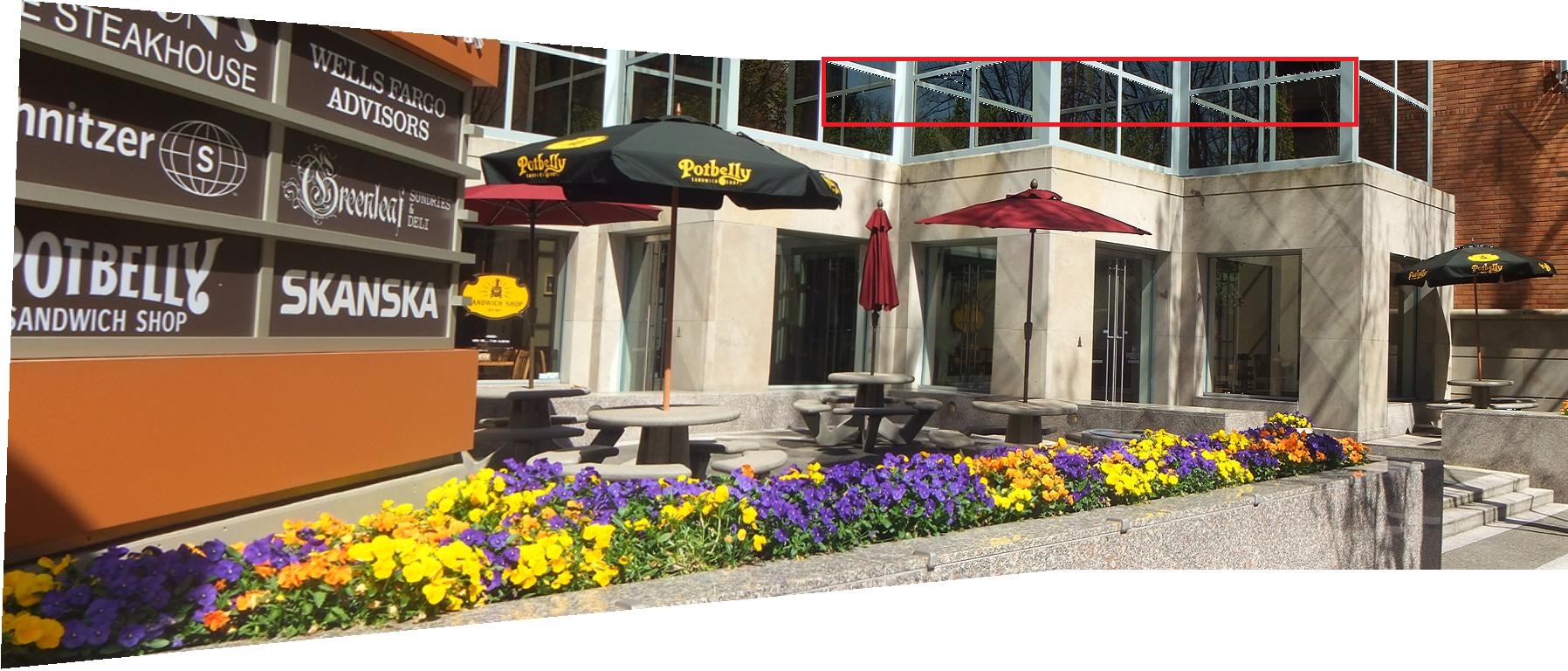}\\[-0.2em]
{\small LPC}\\[0.4em]

\includegraphics[width=\linewidth]{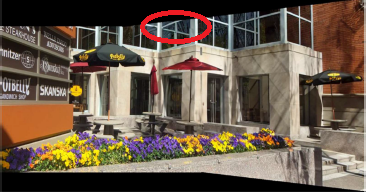}\\[-0.2em]
{\small Seamless}\\[0.4em]

\includegraphics[width=\linewidth]{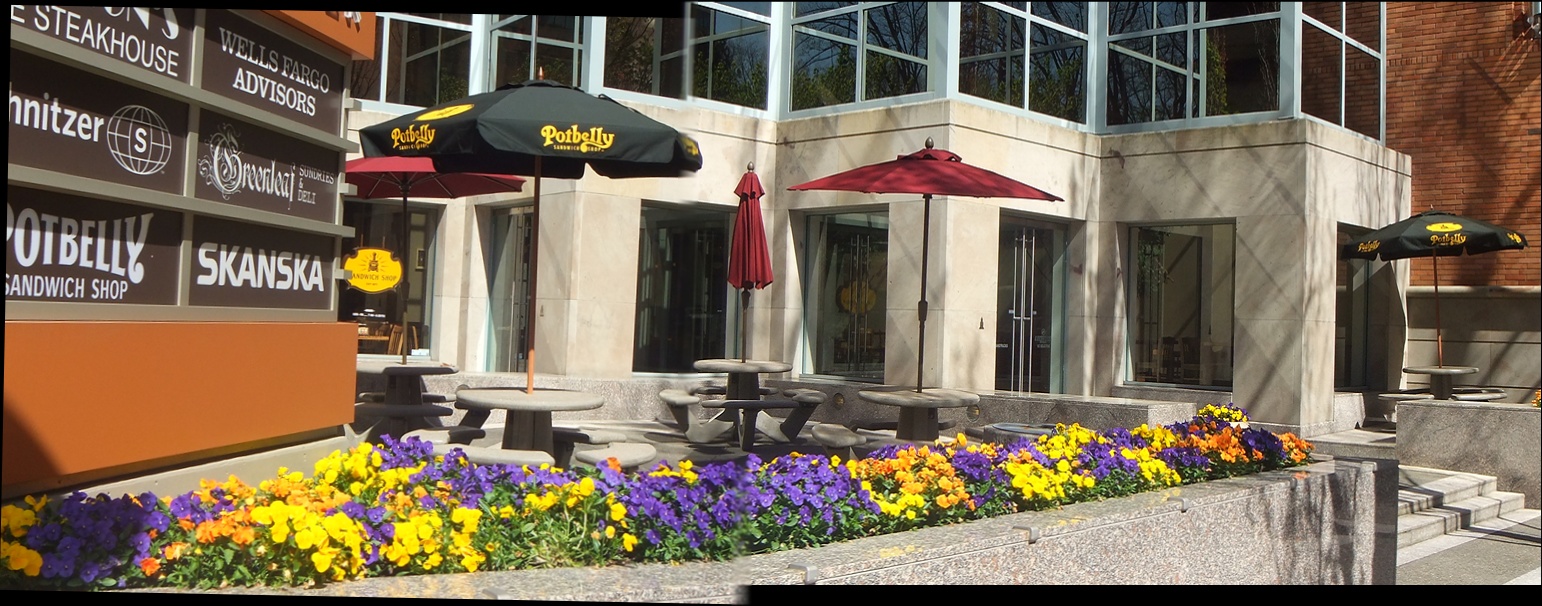}\\[-0.2em]
{\small \textbf{Ours (SENA)}}\\

\end{multicols}

\caption{Our method (SENA) preserves geometry and texture consistency across parallax variations.}
\label{fig:two_column_results}
\end{figure}

\begin{figure}[H]
\centering
\setlength{\columnsep}{1em}
\begin{multicols}{2}

\centering
\includegraphics[width=0.45\linewidth]{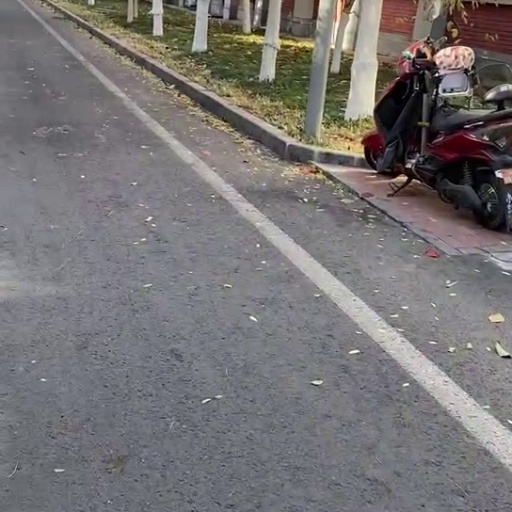}%
\hfill
\includegraphics[width=0.45\linewidth]{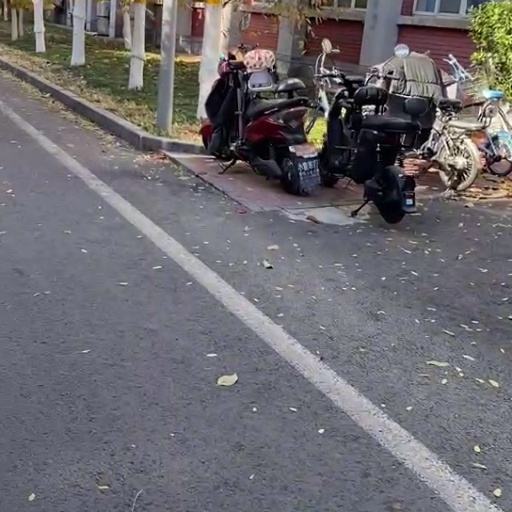}\\[0.3em]
{\small Input images}\\[0.3em]

\includegraphics[width=\linewidth]{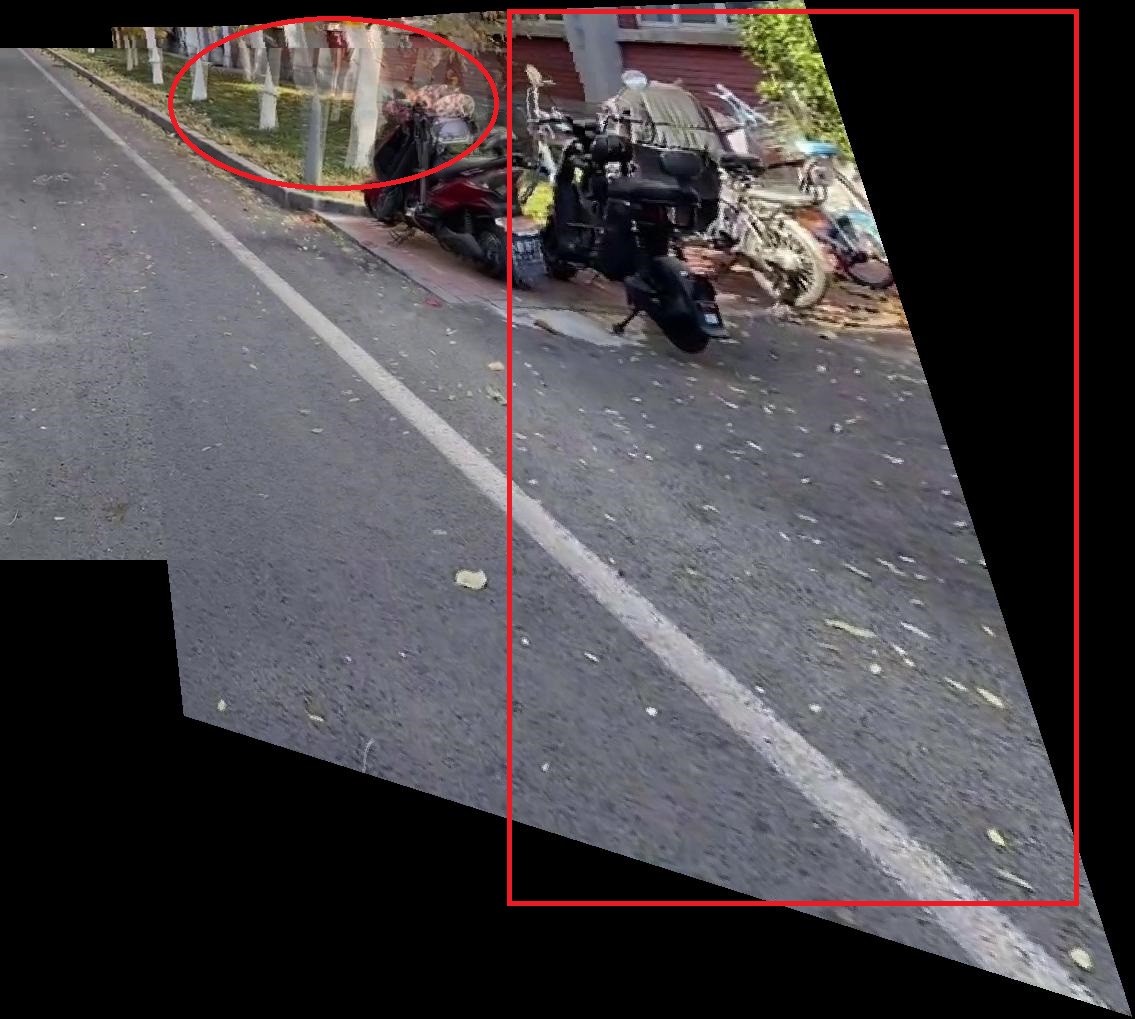}\\[-0.2em]
{\small APAP}\\[0.3em]
\includegraphics[width=\linewidth]{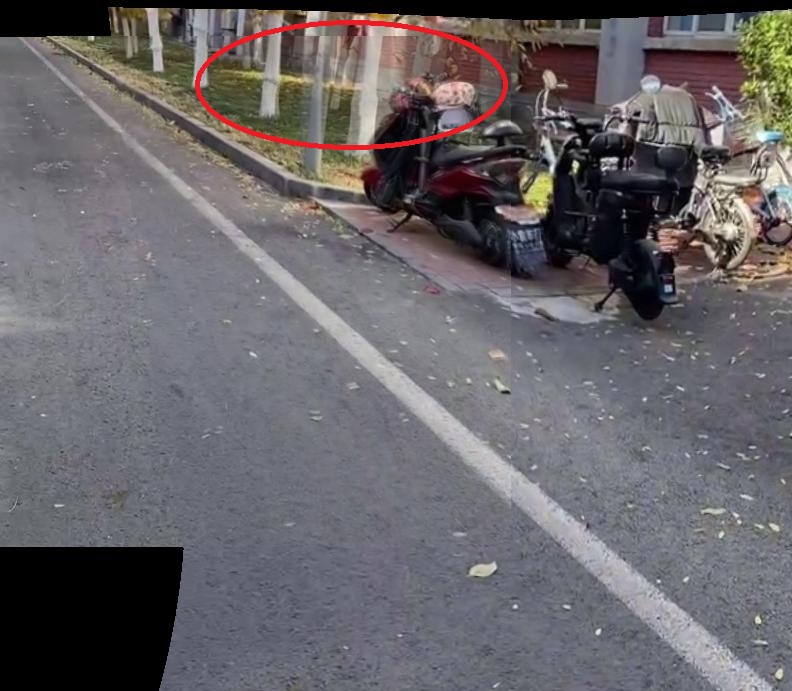}\\[-0.2em]
{\small ELA}\\[0.3em]
\includegraphics[width=\linewidth]{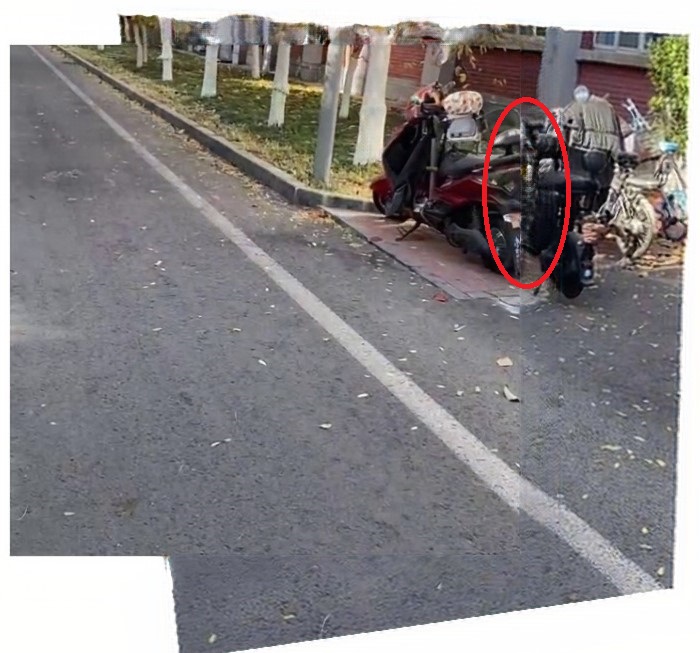}\\[-0.2em]
{\small EPISNET}\\[0.3em]
\includegraphics[width=\linewidth]{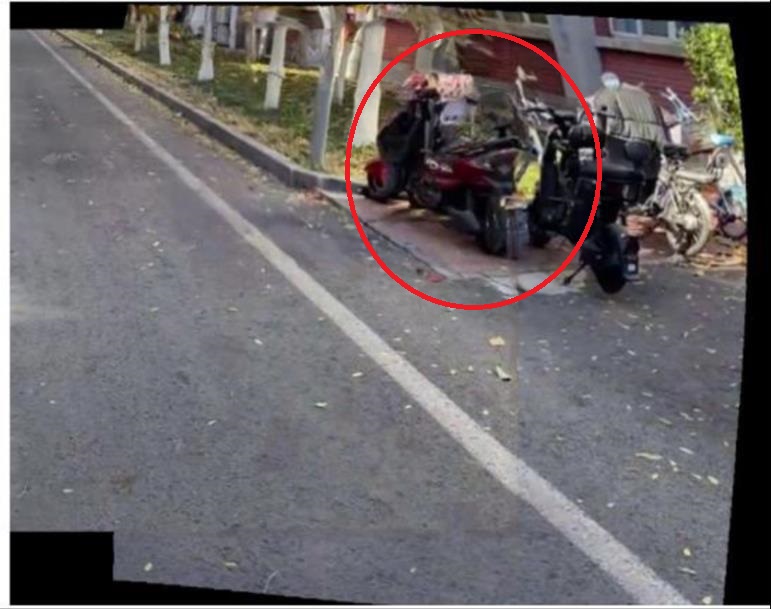}\\[-0.2em]
{\small Seamless }\\[0.3em]

\includegraphics[width=\linewidth]{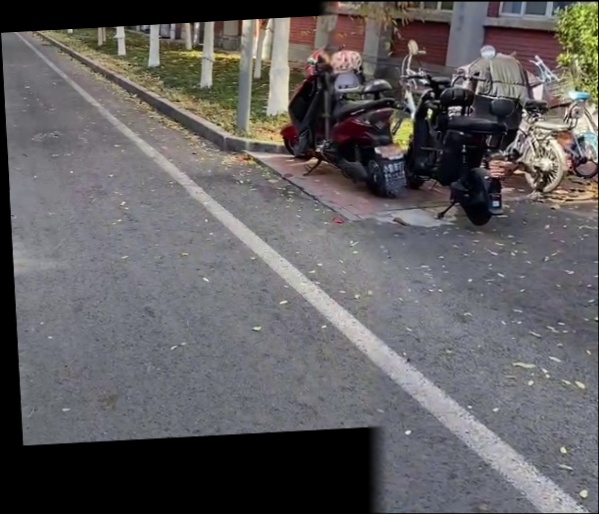}\\[-0.2em]
{\small Ours(SENA) }\\[0.3em]
\centering
\includegraphics[width=0.45\linewidth]{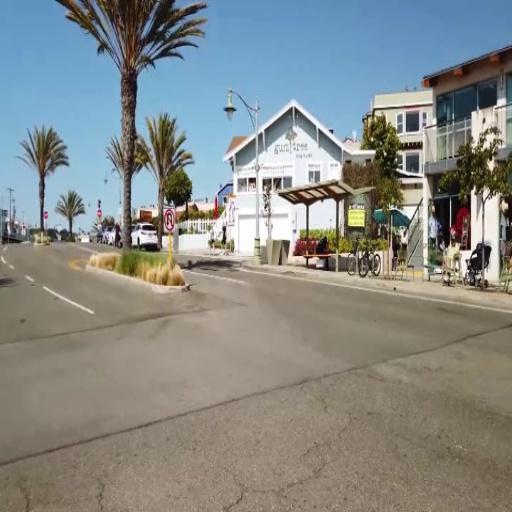}%
\hfill
\includegraphics[width=0.45\linewidth]{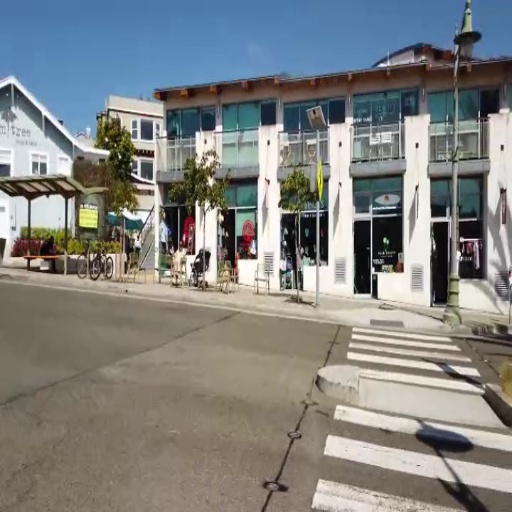}\\[0.3em]
{\small Input images}\\[0.3em]

\includegraphics[width=\linewidth]{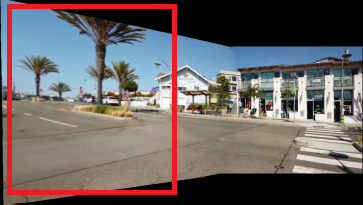}\\[-0.2em]
{\small SEAMLESS}\\[0.3em]

\includegraphics[width=\linewidth]{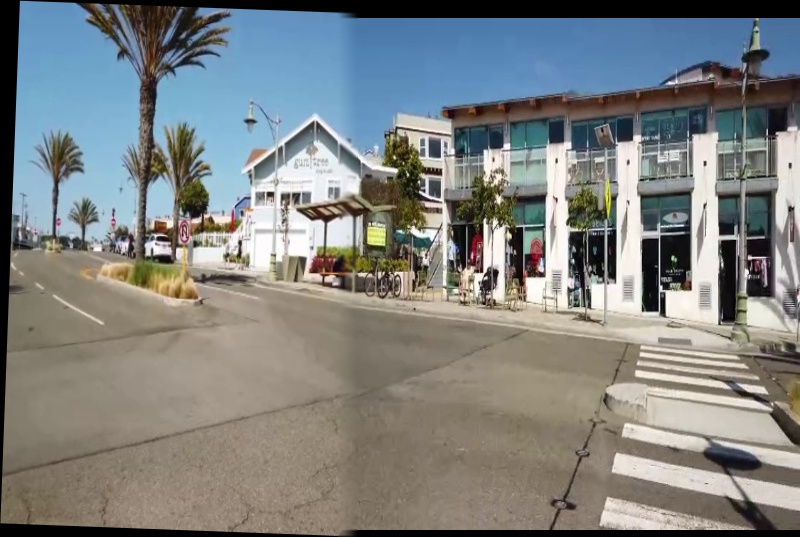}\\[-0.2em]
{\small \textbf{Ours (SENA)}}\\

\end{multicols}

\caption{Our method (SENA) preserves geometry and texture consistency across parallax variations.}
\label{fig:two_column_results}
\end{figure}

\begin{figure}[H]
\centering
\setlength{\columnsep}{1em}
\begin{multicols}{2}

\centering
\includegraphics[width=0.48\linewidth]{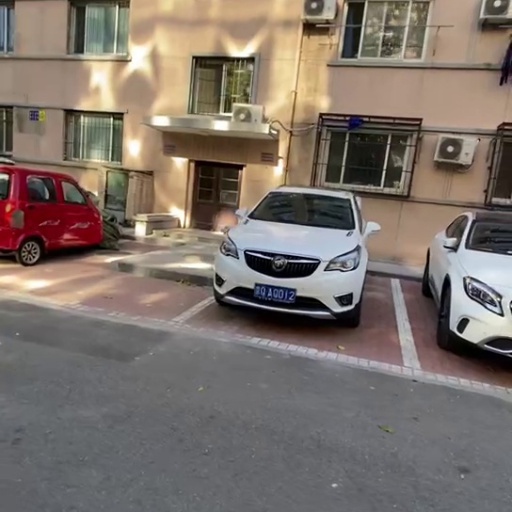}%
\hfill
\includegraphics[width=0.48\linewidth]{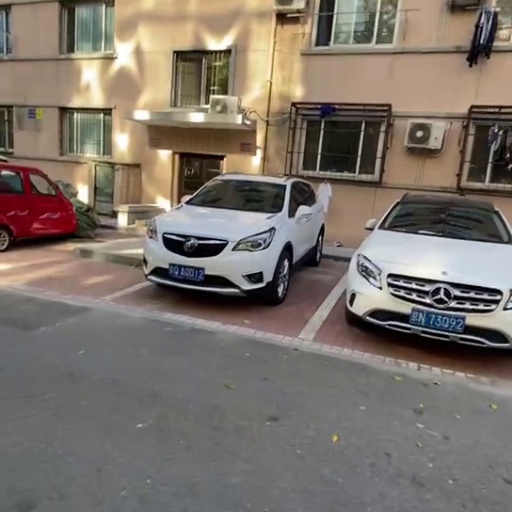}\\[0.4em]
{\small Input images}\\[0.4em]

\includegraphics[width=0.75\linewidth]{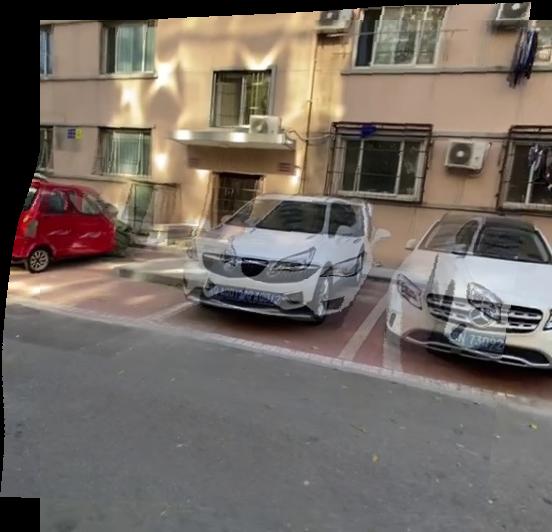}\\[-0.2em]
{\small APAP}\\[0.4em]

\includegraphics[width=0.75\linewidth]{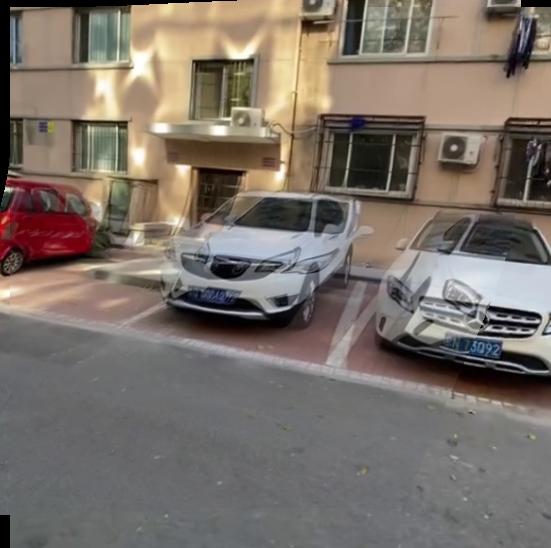}\\[-0.2em]
{\small ELA}\\[0.4em]
\includegraphics[width=0.75\linewidth]{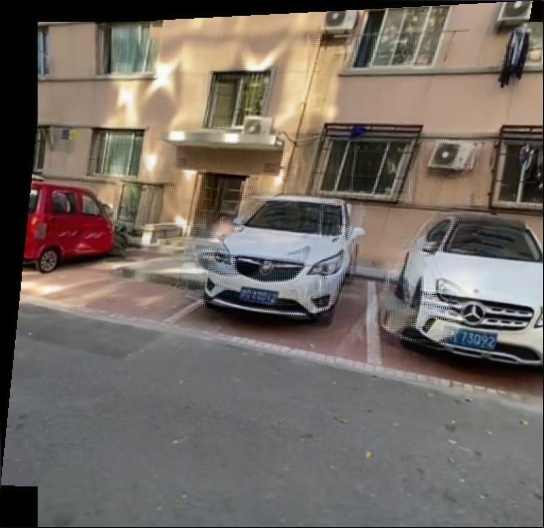}\\[-0.2em]
{\small UDIS}\\[0.4em]
\includegraphics[width=0.75\linewidth]{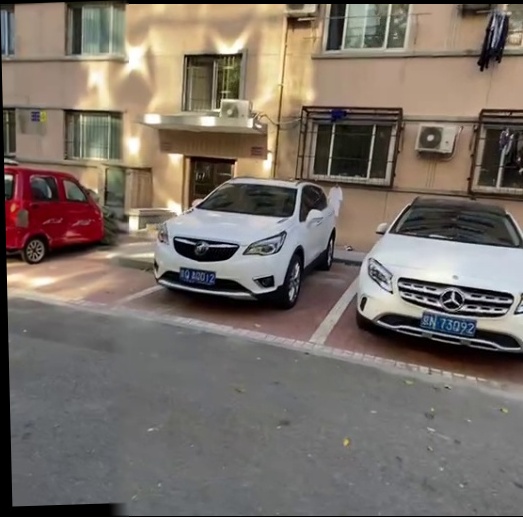}\\[-0.2em]
{\small OURS}\\[0.4em]

\centering
\includegraphics[width=0.48\linewidth]{001033_a.jpg}%
\hfill
\includegraphics[width=0.48\linewidth]{001033_b.jpg}\\[0.4em]
{\small Input images}\\[0.4em]

\includegraphics[width=0.85\linewidth]{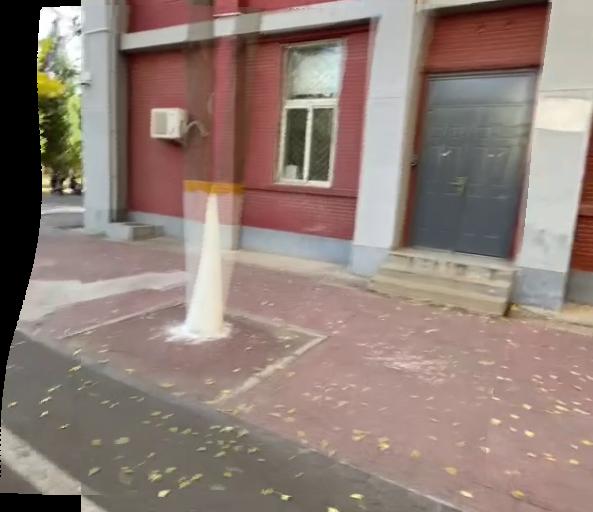}\\[-0.2em]
{\small APAP}\\[0.4em]
\includegraphics[width=0.85\linewidth]{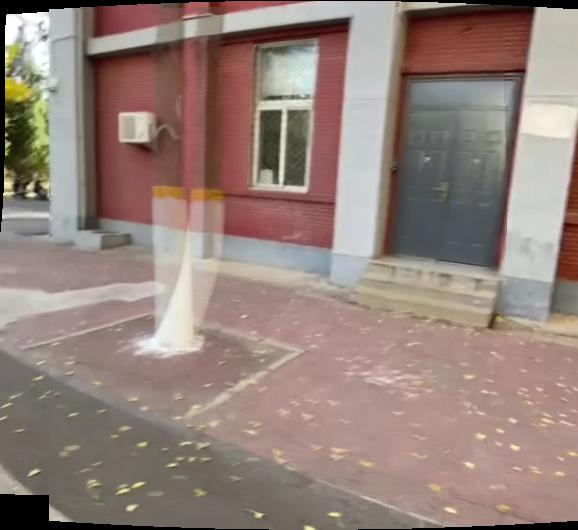}\\[-0.2em]
{\small ELA}\\[0.4em]
\includegraphics[width=0.85\linewidth]{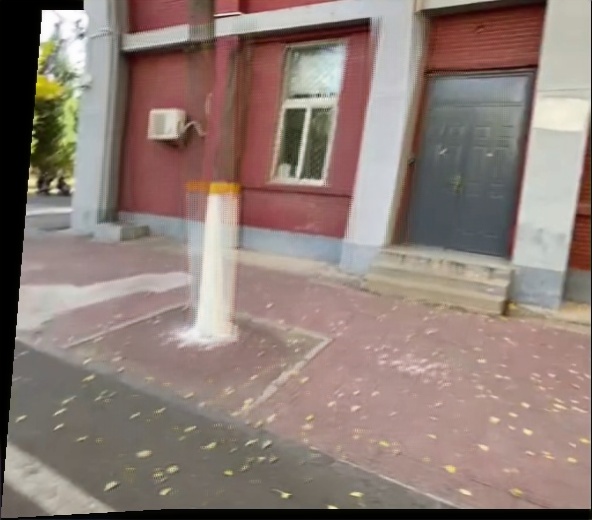}\\[-0.2em]
{\small UDIS}\\[0.4em]

\includegraphics[width=0.85\linewidth]{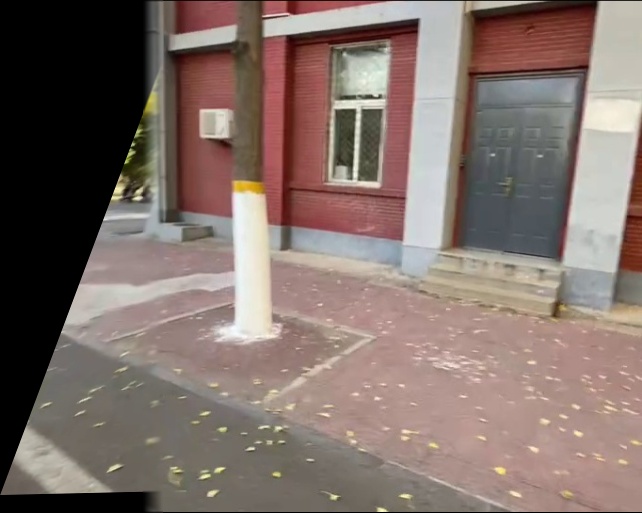}\\[-0.2em]
{\small \textbf{Ours (SENA)}}\\

\end{multicols}

\caption{Our method (SENA) preserves geometry and texture consistency across parallax variations.}
\label{fig:two_column_results}
\end{figure}

\begin{figure}[H]
\centering
\setlength{\columnsep}{1em}
\begin{multicols}{2}

\centering
\includegraphics[width=0.48\linewidth]{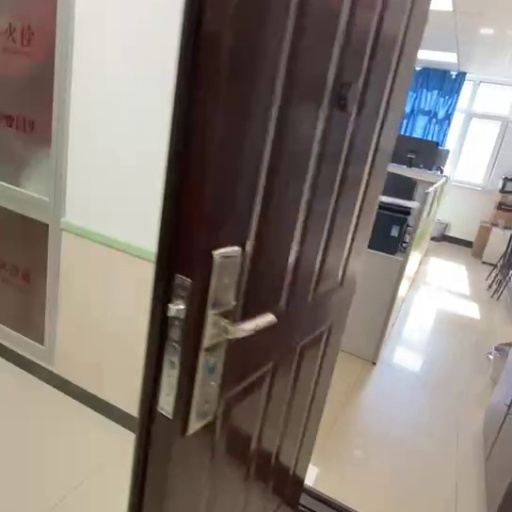}%
\hfill
\includegraphics[width=0.48\linewidth]{001046_b.jpg}\\[0.4em]
{\small Input images}\\[0.4em]

\includegraphics[width=0.75\linewidth]{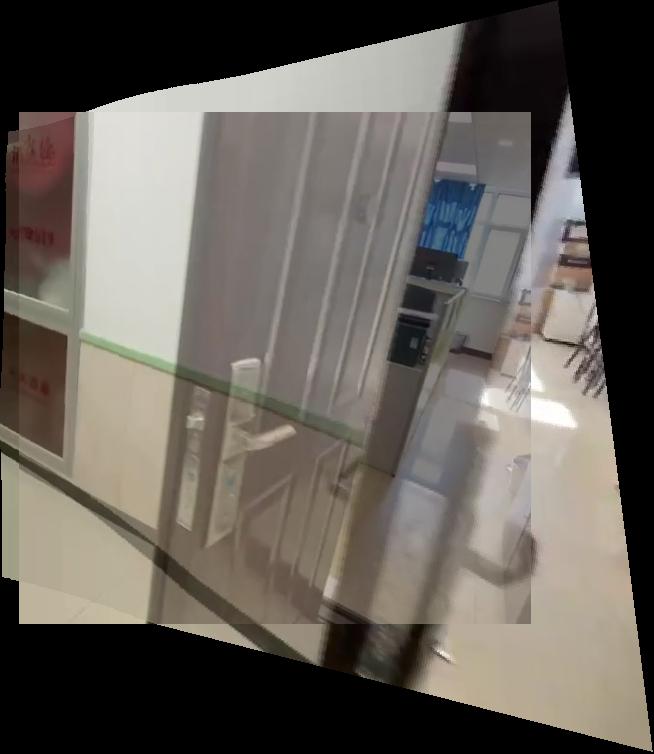}\\[-0.2em]
{\small APAP}\\[0.4em]

\includegraphics[width=0.75\linewidth]{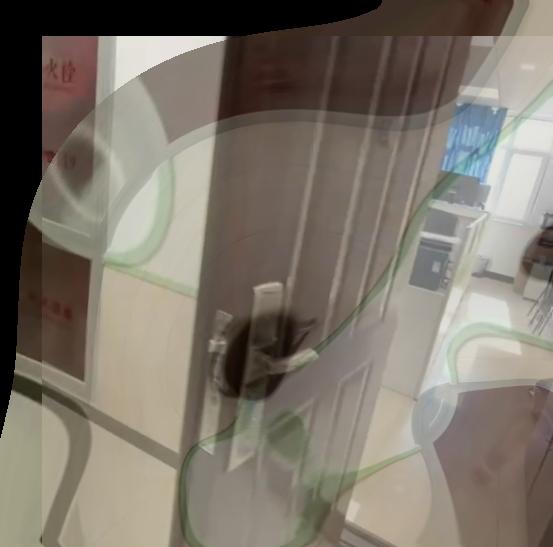}\\[-0.2em]
{\small ELA}\\[0.4em]
\includegraphics[width=0.80\linewidth]{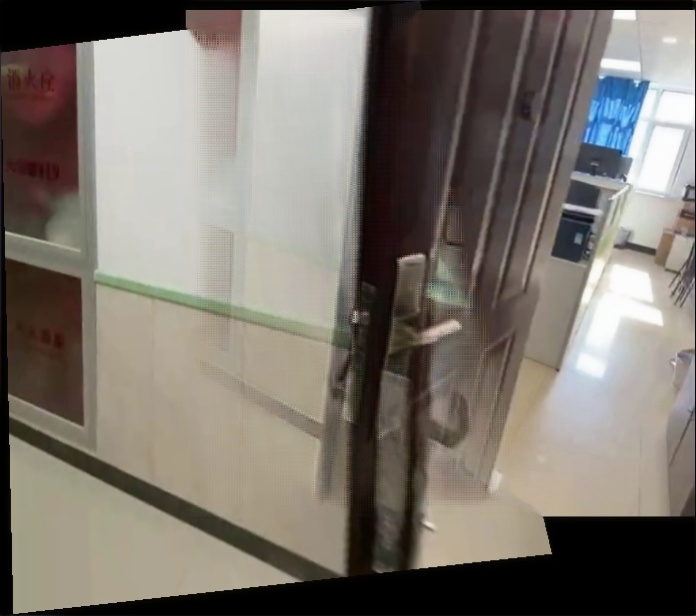}\\[-0.2em]
{\small UDIS}\\[0.4em]
\includegraphics[width=0.75\linewidth]{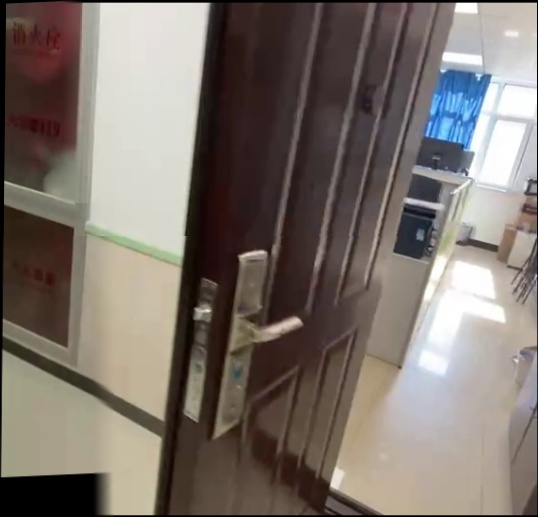}\\[-0.2em]
{\small OURS}\\[0.4em]

\centering
\includegraphics[width=0.48\linewidth]{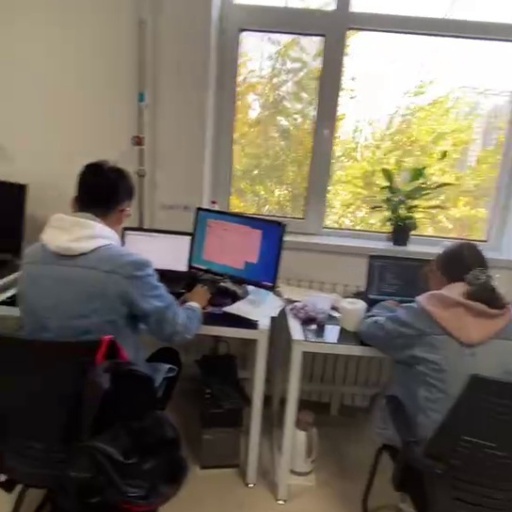}%
\hfill
\includegraphics[width=0.48\linewidth]{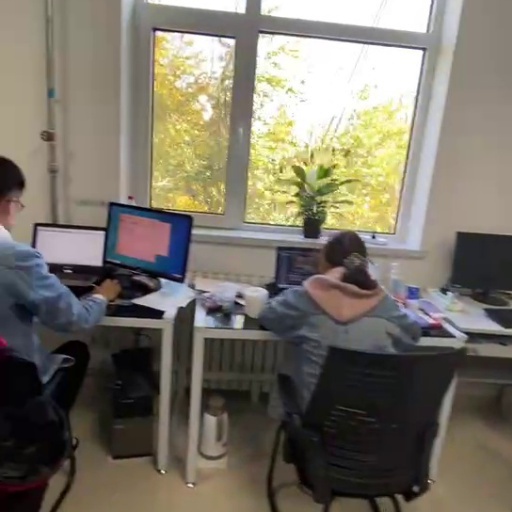}\\[0.4em]
{\small Input images}\\[0.4em]

\includegraphics[width=0.85\linewidth]{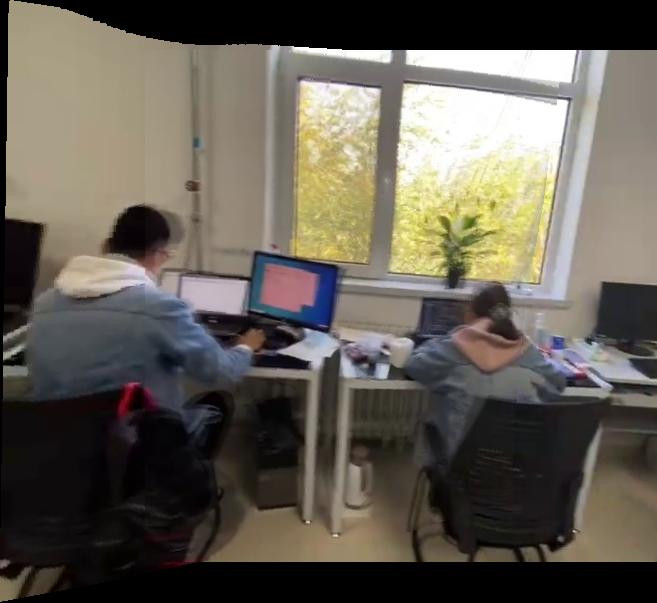}\\[-0.2em]
{\small APAP}\\[0.4em]
\includegraphics[width=0.85\linewidth]{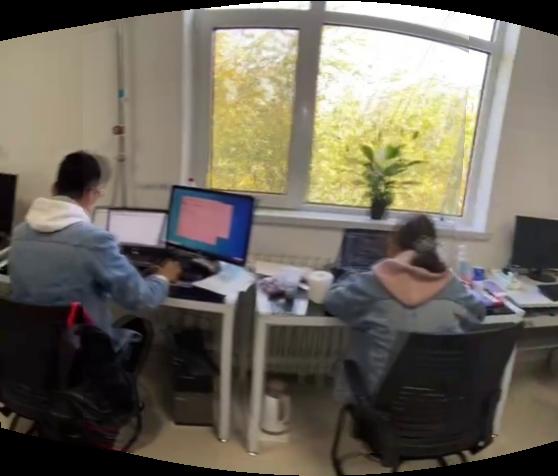}\\[-0.2em]
{\small ELA}\\[0.4em]
\includegraphics[width=0.85\linewidth]{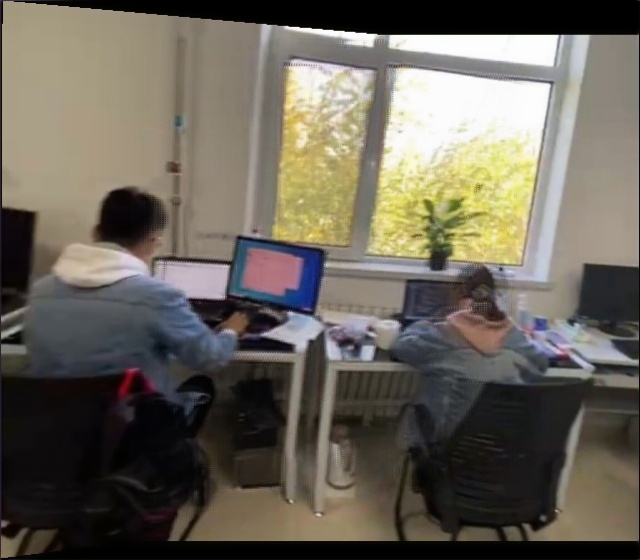}\\[-0.2em]
{\small UDIS}\\[0.4em]

\includegraphics[width=0.85\linewidth]{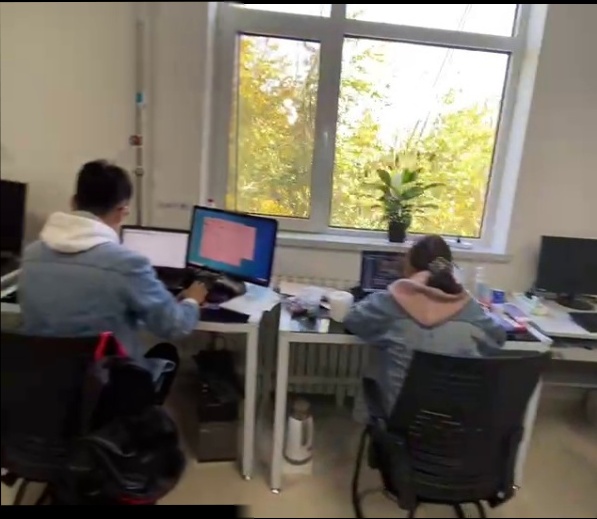}\\[-0.2em]
{\small \textbf{Ours (SENA)}}\\

\end{multicols}

\caption{Our method (SENA) preserves geometry and texture consistency across parallax variations.}
\label{fig:two_column_results}
\end{figure}

\begin{figure}[H]
\centering
\setlength{\columnsep}{1em}
\begin{multicols}{2}

\centering
\includegraphics[width=0.48\linewidth]{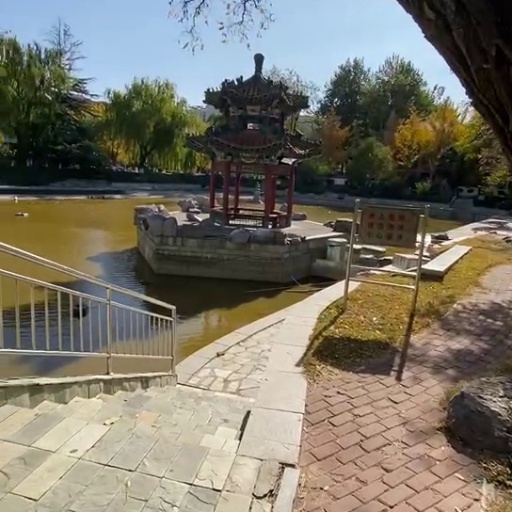}%
\hfill
\includegraphics[width=0.48\linewidth]{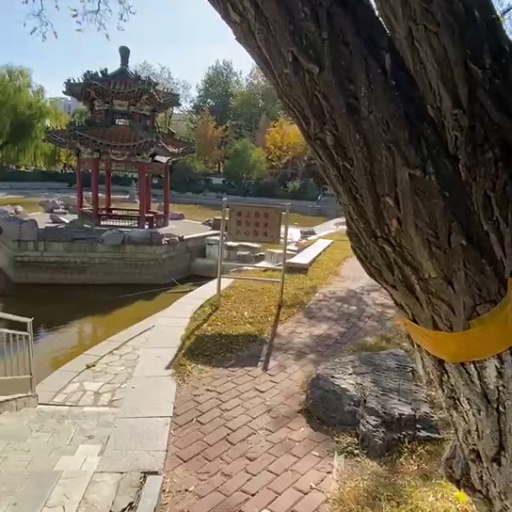}\\[0.4em]
{\small Input images}\\[0.4em]

\includegraphics[width=0.83\linewidth]{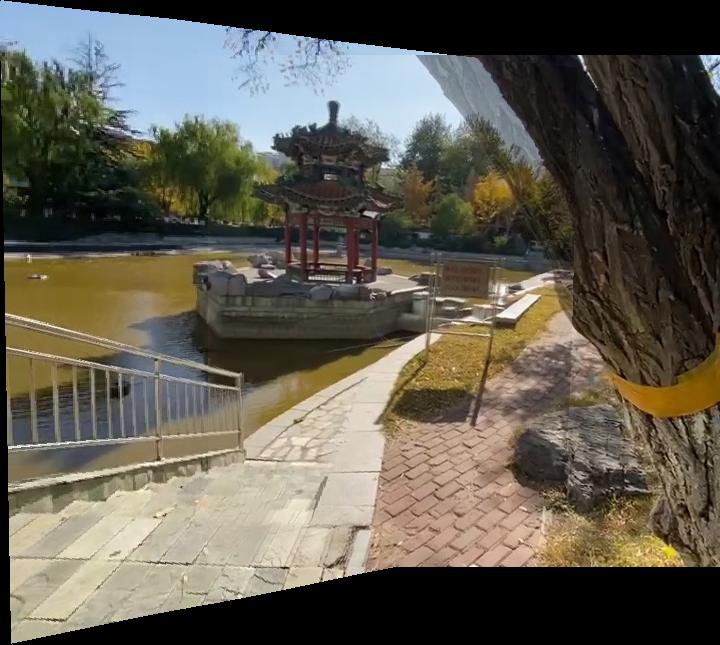}\\[-0.2em]
{\small APAP}\\[0.4em]

\includegraphics[width=0.83\linewidth]{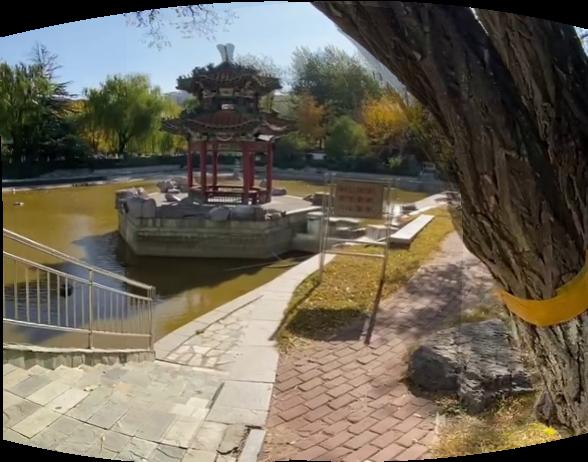}\\[-0.2em]
{\small ELA}\\[0.4em]
\includegraphics[width=0.83\linewidth]{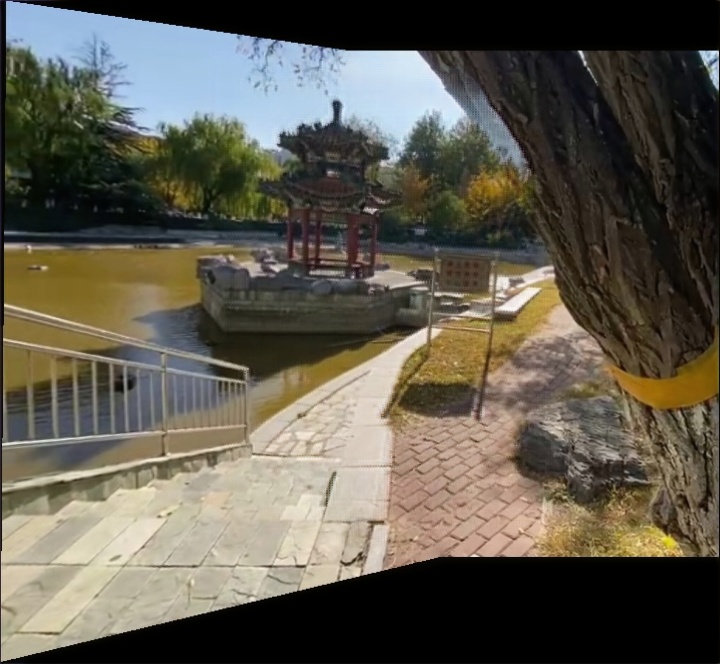}\\[-0.2em]
{\small UDIS}\\[0.4em]
\includegraphics[width=0.83\linewidth]{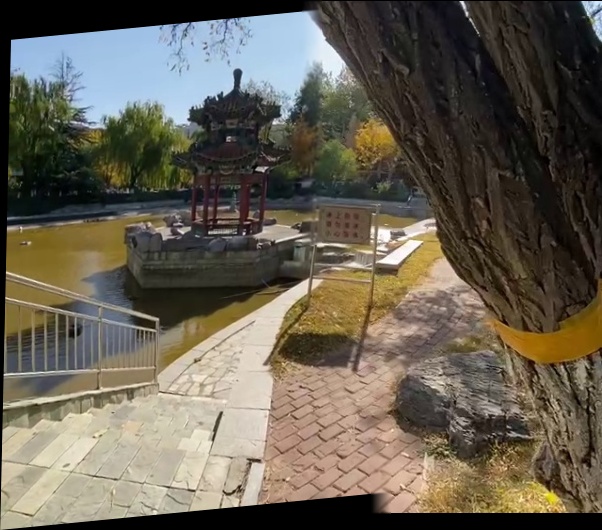}\\[-0.2em]
{\small OURS}\\[0.4em]

\centering
\includegraphics[width=0.48\linewidth]{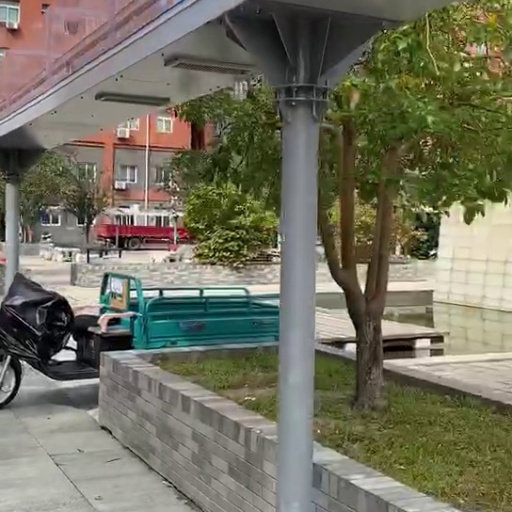}%
\hfill
\includegraphics[width=0.48\linewidth]{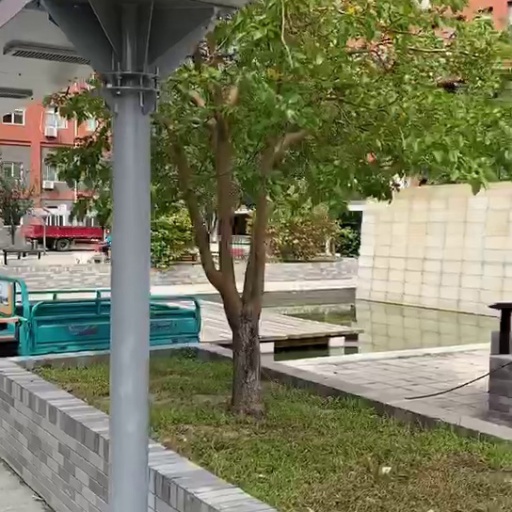}\\[0.4em]
{\small Input images}\\[0.4em]

\includegraphics[width=0.85\linewidth]{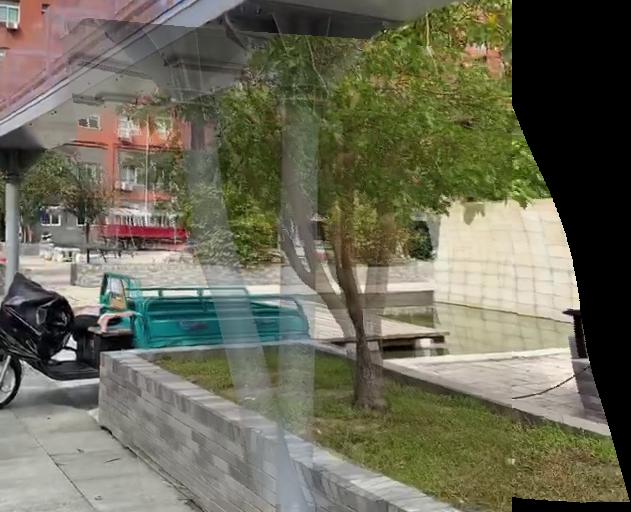}\\[-0.2em]
{\small APAP}\\[0.4em]
\includegraphics[width=0.85\linewidth]{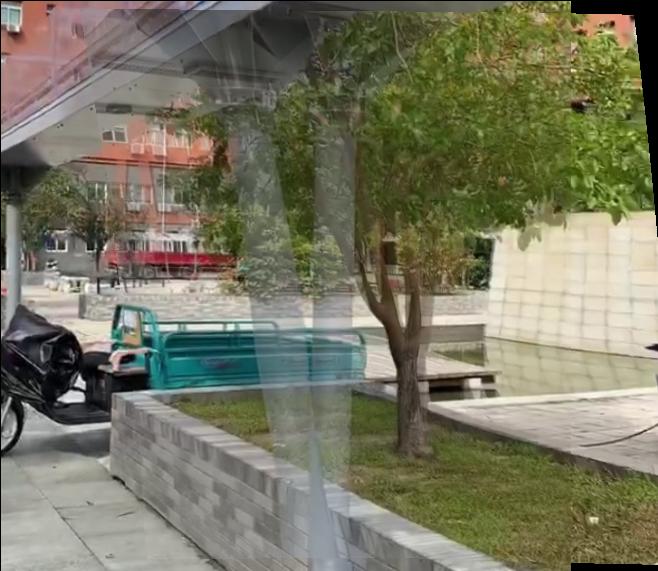}\\[-0.2em]
{\small ELA}\\[0.4em]
\includegraphics[width=0.85\linewidth]{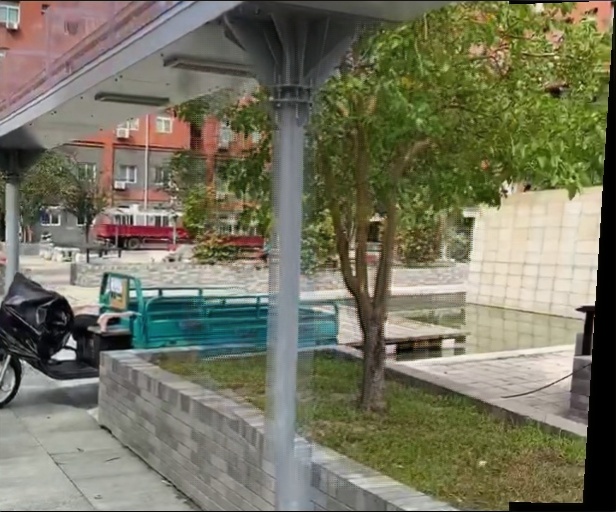}\\[-0.2em]
{\small UDIS}\\[0.4em]

\includegraphics[width=0.85\linewidth]{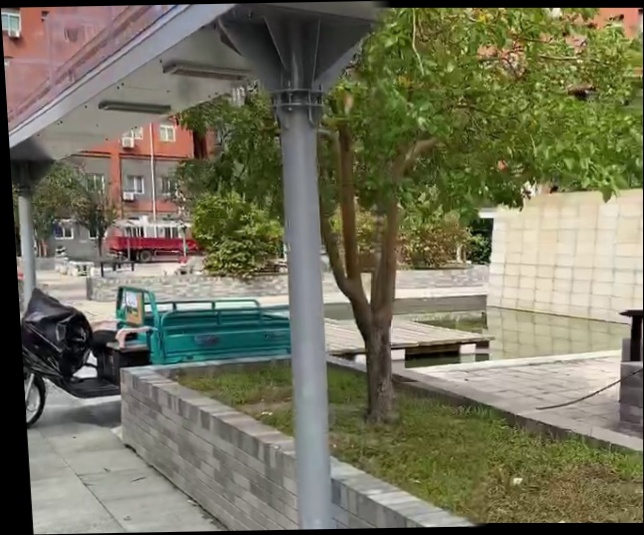}\\[-0.2em]
{\small \textbf{Ours (SENA)}}\\

\end{multicols}

\caption{Our method (SENA) preserves geometry and texture consistency across parallax variations.}
\label{fig:two_column_results}
\end{figure}

\begin{figure}[H]
\centering

\newcommand{\ColW}{0.48\linewidth}     
\newcommand{\SmallGap}{2pt}
\newcommand{\BigGap}{6pt}

\begin{minipage}[t]{\ColW}
  \centering

  \begin{minipage}{\linewidth}
    \centering
    \begin{minipage}[t]{0.45\linewidth}
      \centering
      \includegraphics[width=\linewidth]{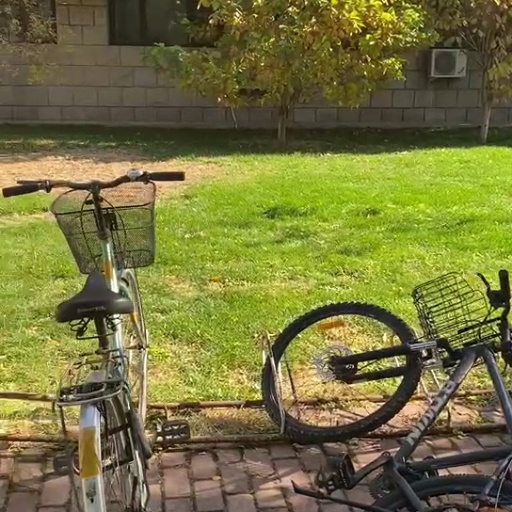}
    \end{minipage}\hfill
    \begin{minipage}[t]{0.45\linewidth}
      \centering
      \includegraphics[width=\linewidth]{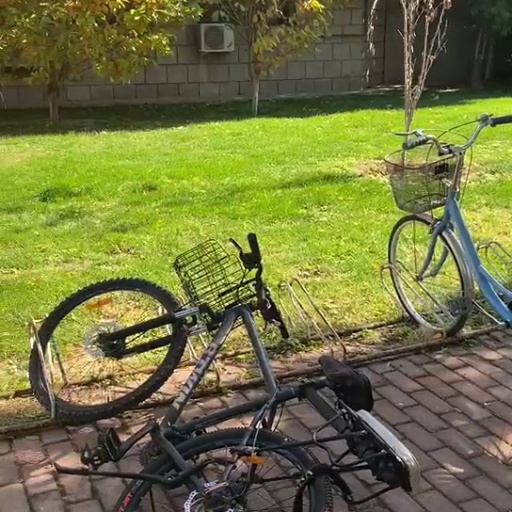}
    \end{minipage}\\[4pt]
    \small Input images
  \end{minipage}

  \vspace{\BigGap}

  \newcommand{\MethodBlock}[2]{%
    \begin{minipage}{\linewidth}
      \centering
      \includegraphics[width=0.95\linewidth]{#1}\\[2pt]
      \small #2
    \end{minipage}\par\vspace{\BigGap}
  }

  \MethodBlock{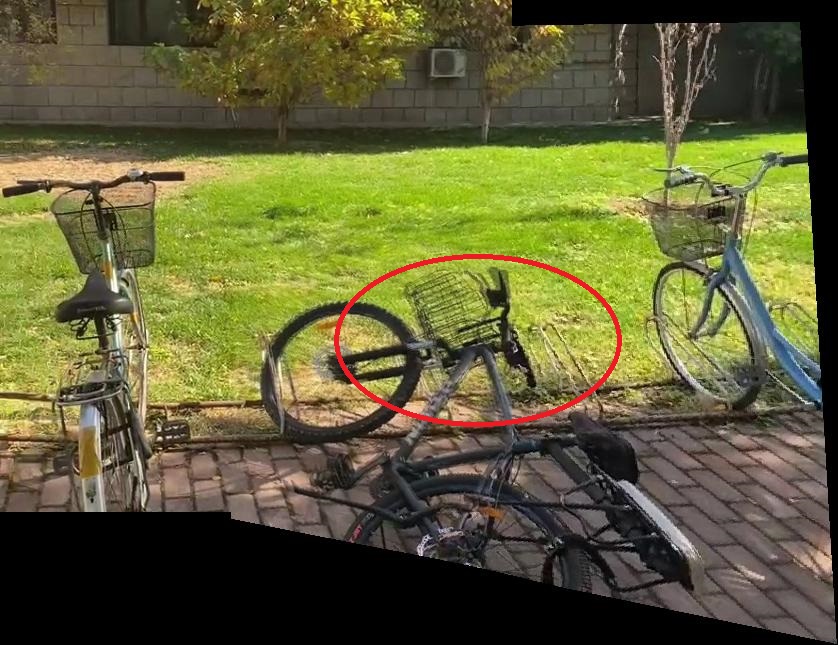}{APAP}
  \MethodBlock{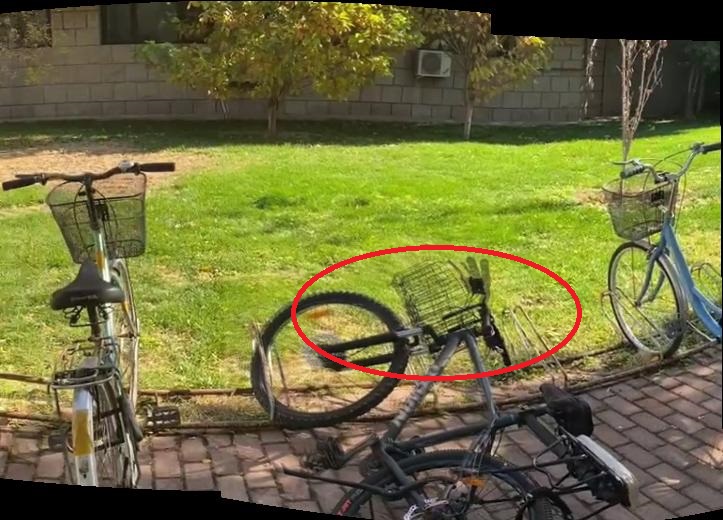}{ELA}
  \MethodBlock{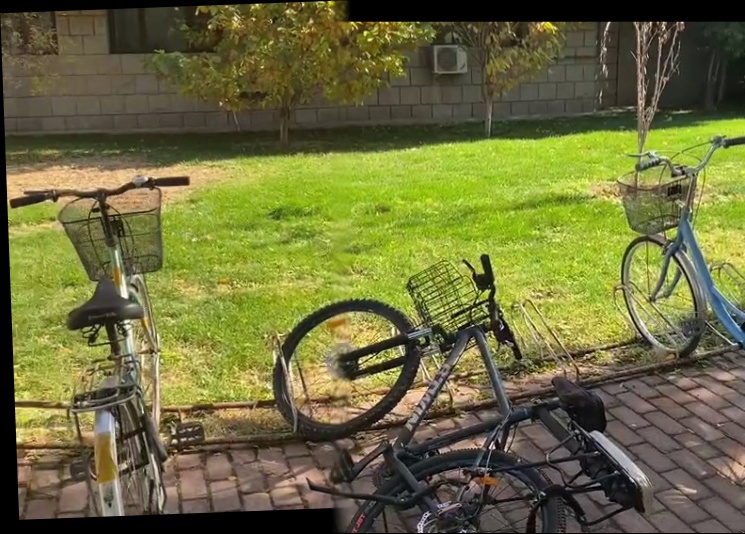}{OURS}

\end{minipage}%
\hfill%
\begin{minipage}[t]{\ColW}
  \centering

  \begin{minipage}{\linewidth}
    \centering
    \begin{minipage}[t]{0.45\linewidth}
      \centering
      \includegraphics[width=\linewidth]{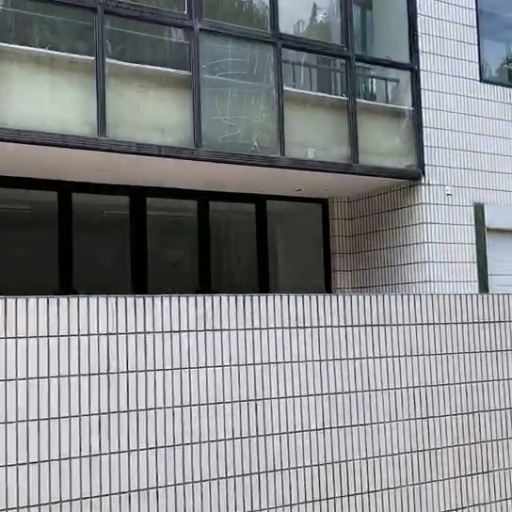}
    \end{minipage}\hfill
    \begin{minipage}[t]{0.45\linewidth}
      \centering
      \includegraphics[width=\linewidth]{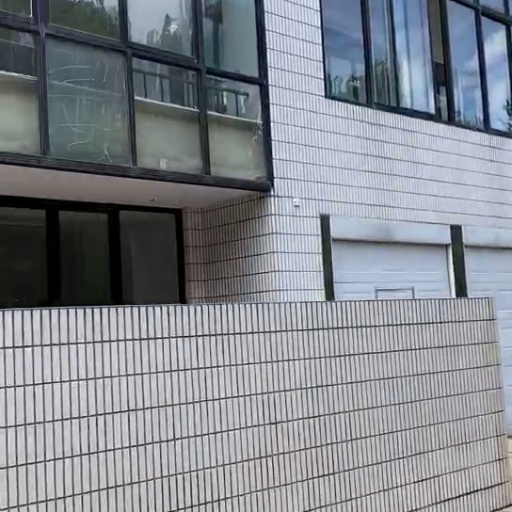}
    \end{minipage}\\[4pt]
    \small input images
  \end{minipage}

  \vspace{\BigGap}

  \begin{minipage}{\linewidth}
    \centering
    \includegraphics[width=0.95\linewidth]{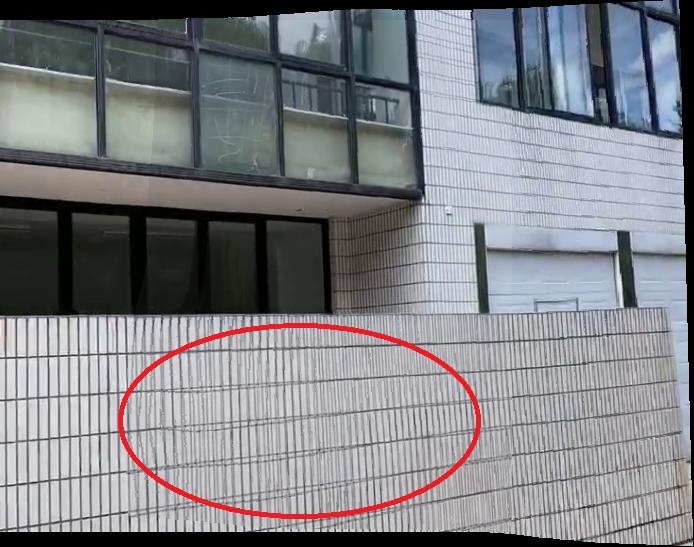}\\[2pt]
    \small APAP
  \end{minipage}\par\vspace{\BigGap}

  \begin{minipage}{\linewidth}
    \centering
    \includegraphics[width=0.95\linewidth]{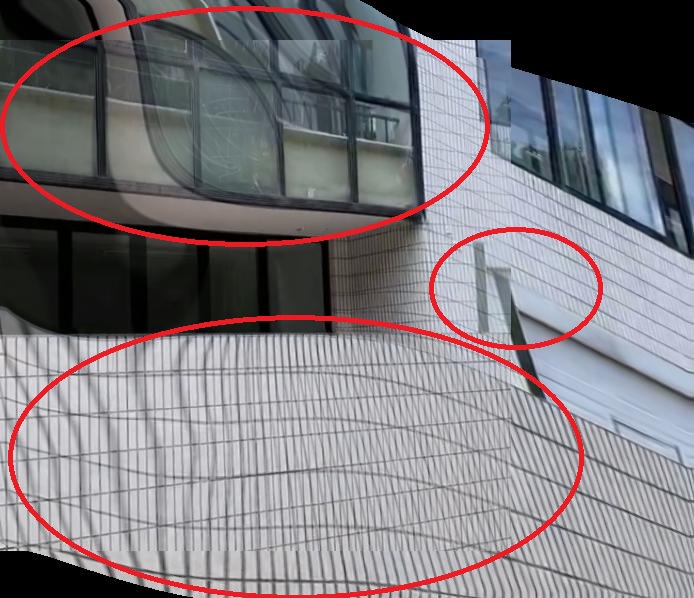}\\[2pt]
    \small ELA
  \end{minipage}\par\vspace{\BigGap}

  \begin{minipage}{\linewidth}
    \centering
    \includegraphics[width=0.95\linewidth]{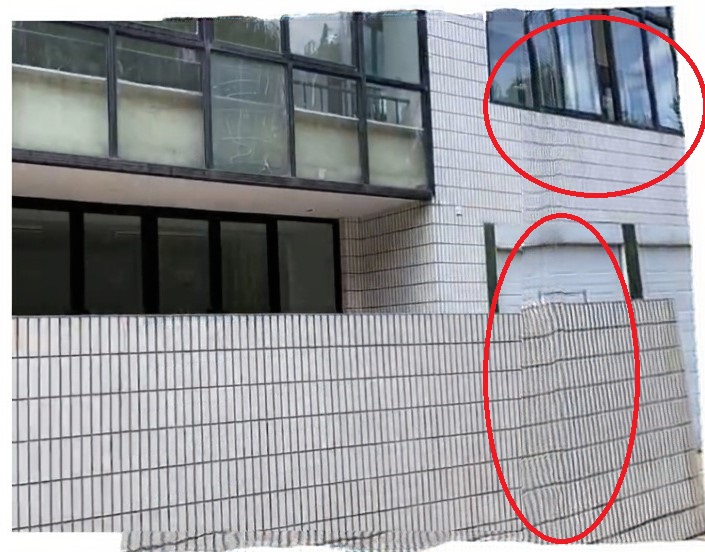}\\[2pt]
    \small EPISNET
  \end{minipage}\par\vspace{\BigGap}

  \begin{minipage}{\linewidth}
    \centering
    \includegraphics[width=0.95\linewidth]{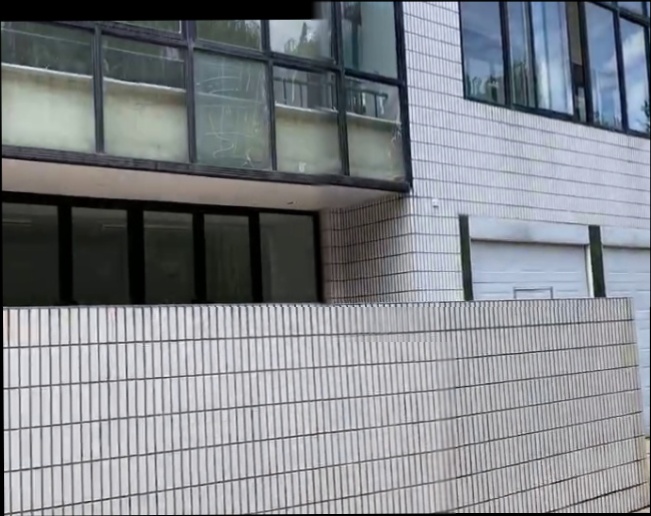}\\[2pt]
    \small OURS
  \end{minipage}

\end{minipage}

\label{fig:example1_and_more}
\end{figure}
\begin{figure}[H]
  \centering

  \begin{minipage}[b]{0.30\linewidth}
    \centering
    \includegraphics[width=\linewidth]{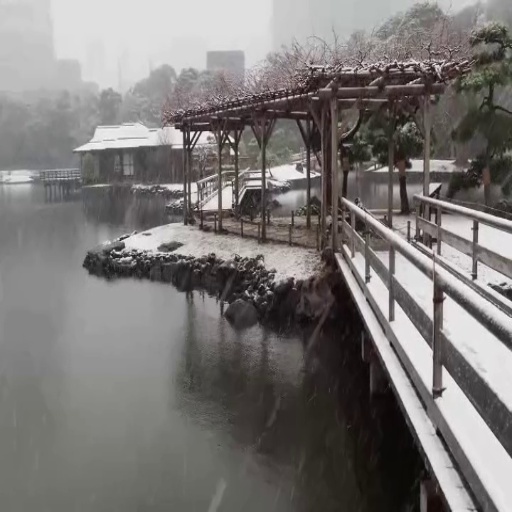}\\[2pt]
    \small source
  \end{minipage}\hfill
  \begin{minipage}[b]{0.30\linewidth}
    \centering
    \includegraphics[width=\linewidth]{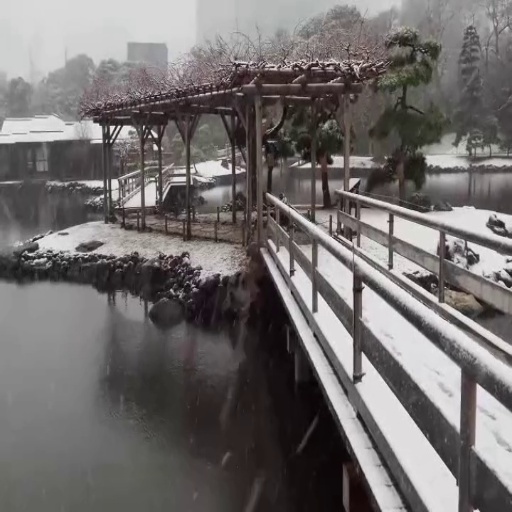}\\[2pt]
    \small target
  \end{minipage}

  \vspace{0.6em}

  \begin{minipage}[b]{0.68\linewidth}
    \centering
    \includegraphics[width=\linewidth]{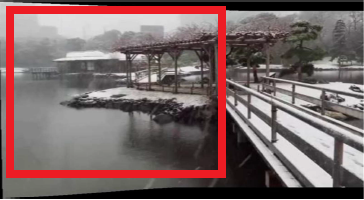}\\[2pt]
    \small SEAMLESS
  \end{minipage}

  \vspace{0.6em}

  \begin{minipage}[b]{0.60\linewidth}
    \centering
    \includegraphics[width=\linewidth]{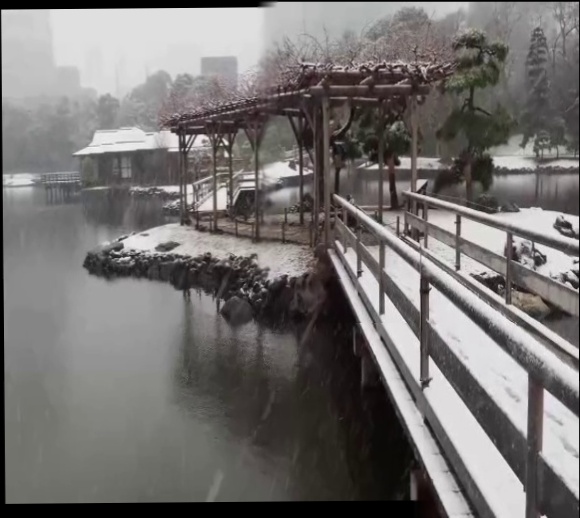}\\[2pt]
    \small OURS
  \end{minipage}
\caption{SEAMLESS and Our Approach.}
  \label{fig:comparison}
\end{figure}

\begin{figure}[H]
\centering
\setlength{\columnsep}{1em}
\begin{multicols}{2}

\centering

\includegraphics[width=0.48\linewidth]{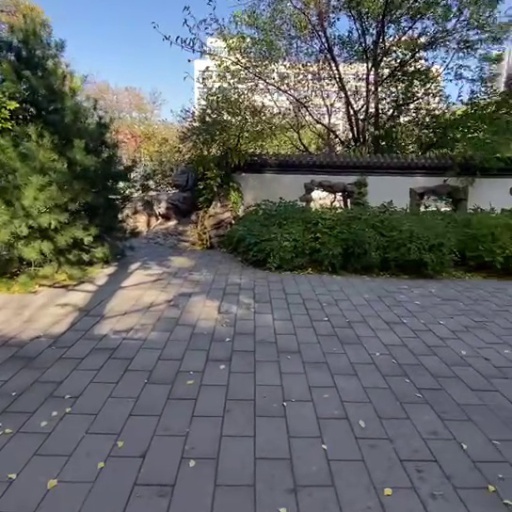}%
\hfill
\includegraphics[width=0.48\linewidth]{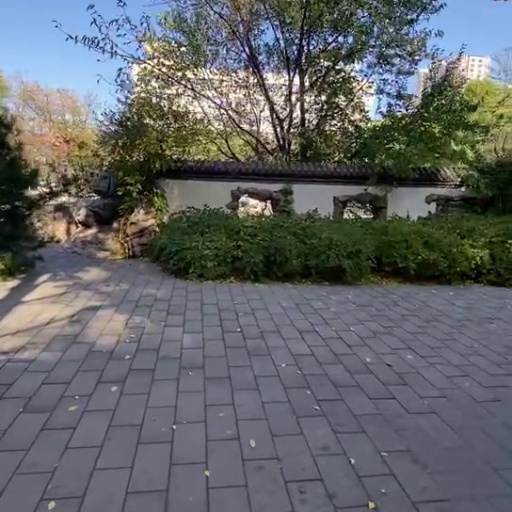}\\[0.4em]
{\small Input images}\\[0.4em]

\includegraphics[width=\linewidth]{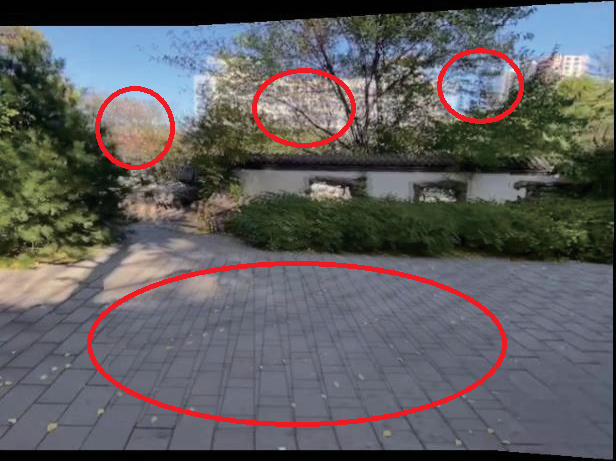}\\[-0.2em]
{\small APAP}\\[0.4em]

\includegraphics[width=\linewidth]{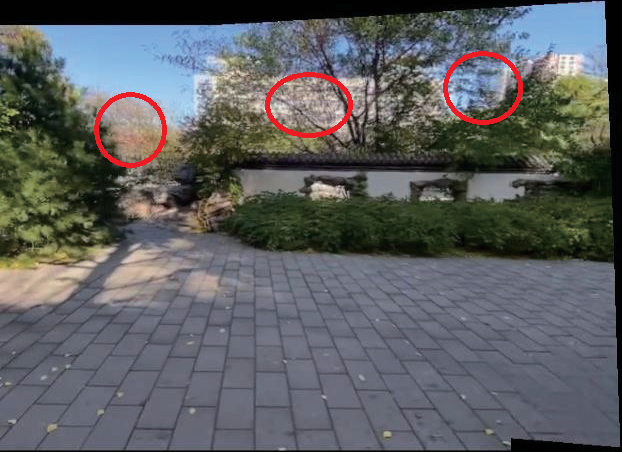}\\[-0.2em]
{\small ELA}\\[0.4em]

\includegraphics[width=\linewidth]{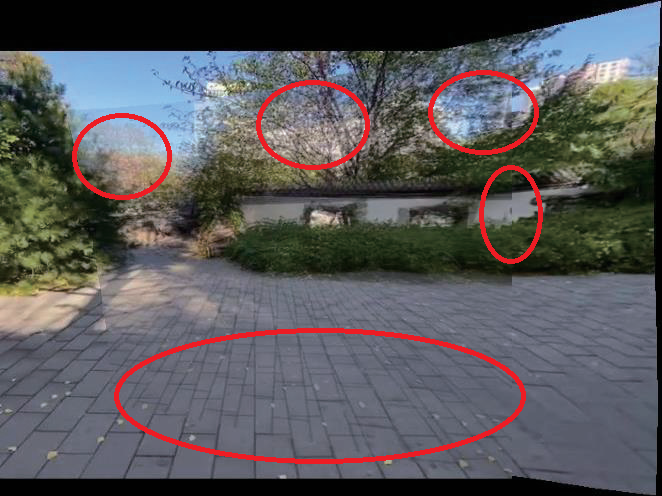}\\[-0.2em]
{\small LPC}\\[0.4em]
\centering

\includegraphics[width=\linewidth]{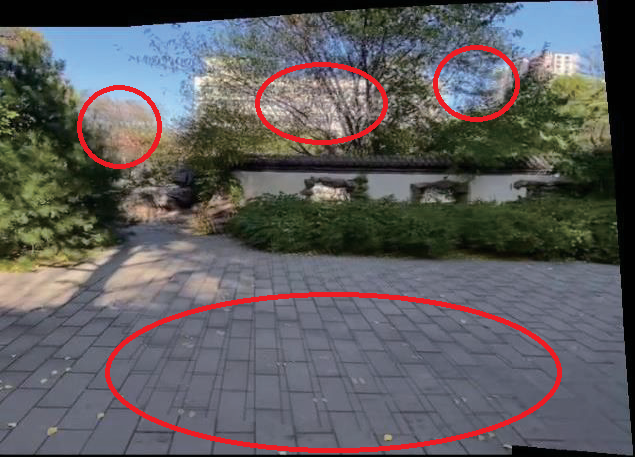}\\[-0.2em]
{\small SPW}\\[0.4em]

\includegraphics[width=\linewidth]{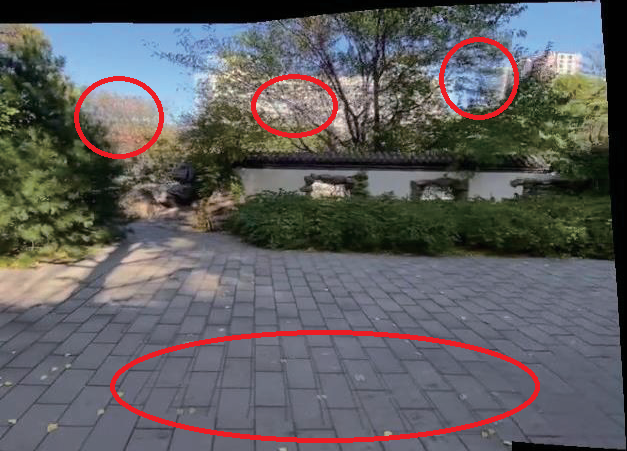}\\[-0.2em]
{\small \textbf{UDIS}}\\

\includegraphics[width=\linewidth]{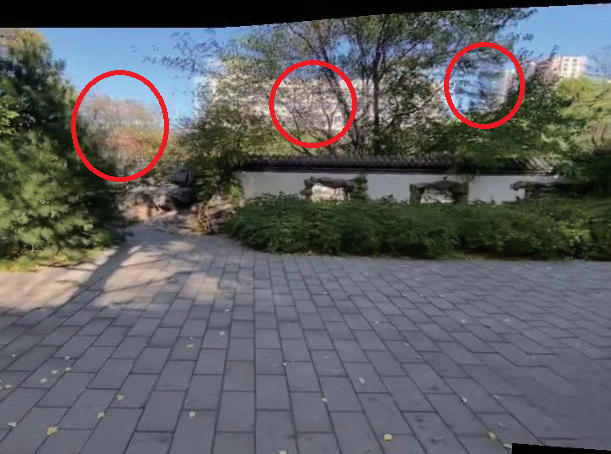}\\[-0.2em]
{\small \textbf{UDIS2}}\\

\includegraphics[width=\linewidth]{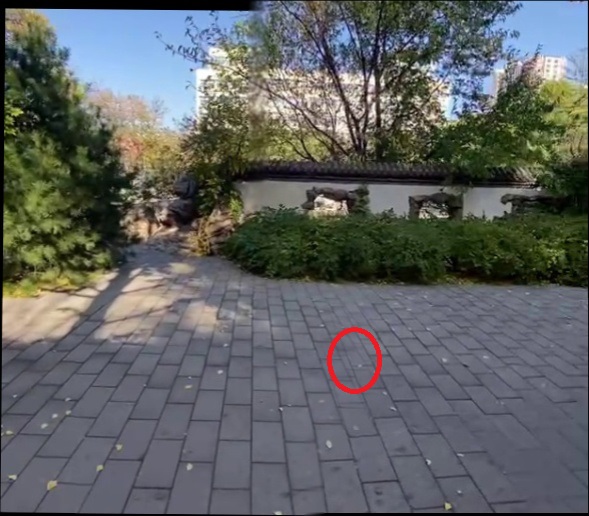}\\[-0.2em]
{\small \textbf{OURS}}\\

\end{multicols}

\caption{APAP, ELA, LPC, SPW, UDIS, UDIS++, and OURS.}
\label{fig:two_column_results}
\end{figure}

\begin{figure}[H]
  \centering

  \begin{minipage}[b]{0.40\linewidth}
    \centering
    \includegraphics[width=\linewidth]{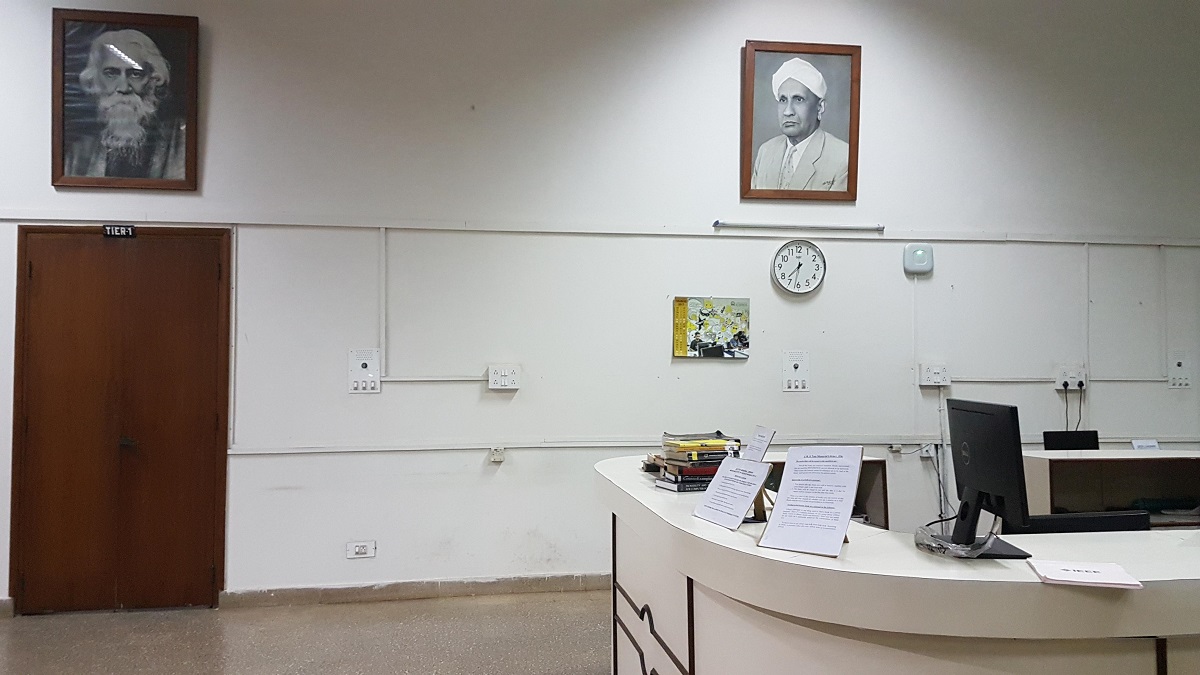}\\[2pt]
    \small source
  \end{minipage}\hfill
  \begin{minipage}[b]{0.40\linewidth}
    \centering
    \includegraphics[width=\linewidth]{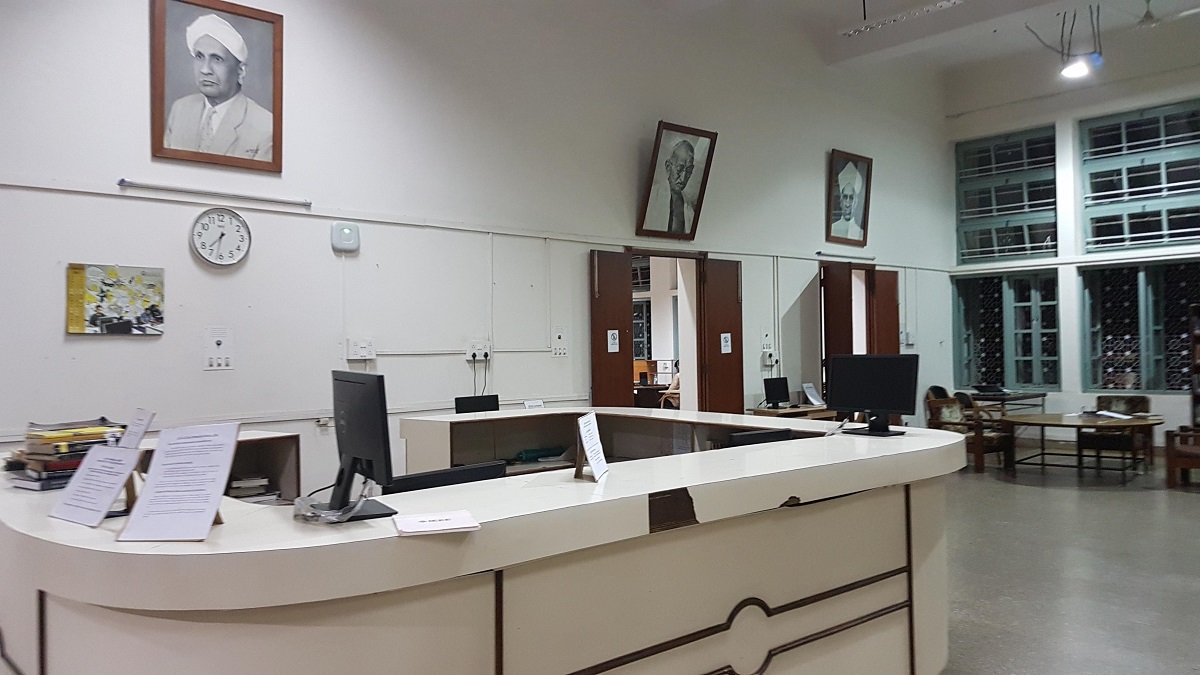}\\[2pt]
    \small target
  \end{minipage}

  \vspace{0.3em}

\begin{minipage}[b]{0.60\linewidth}
    \centering
    \includegraphics[width=\linewidth]{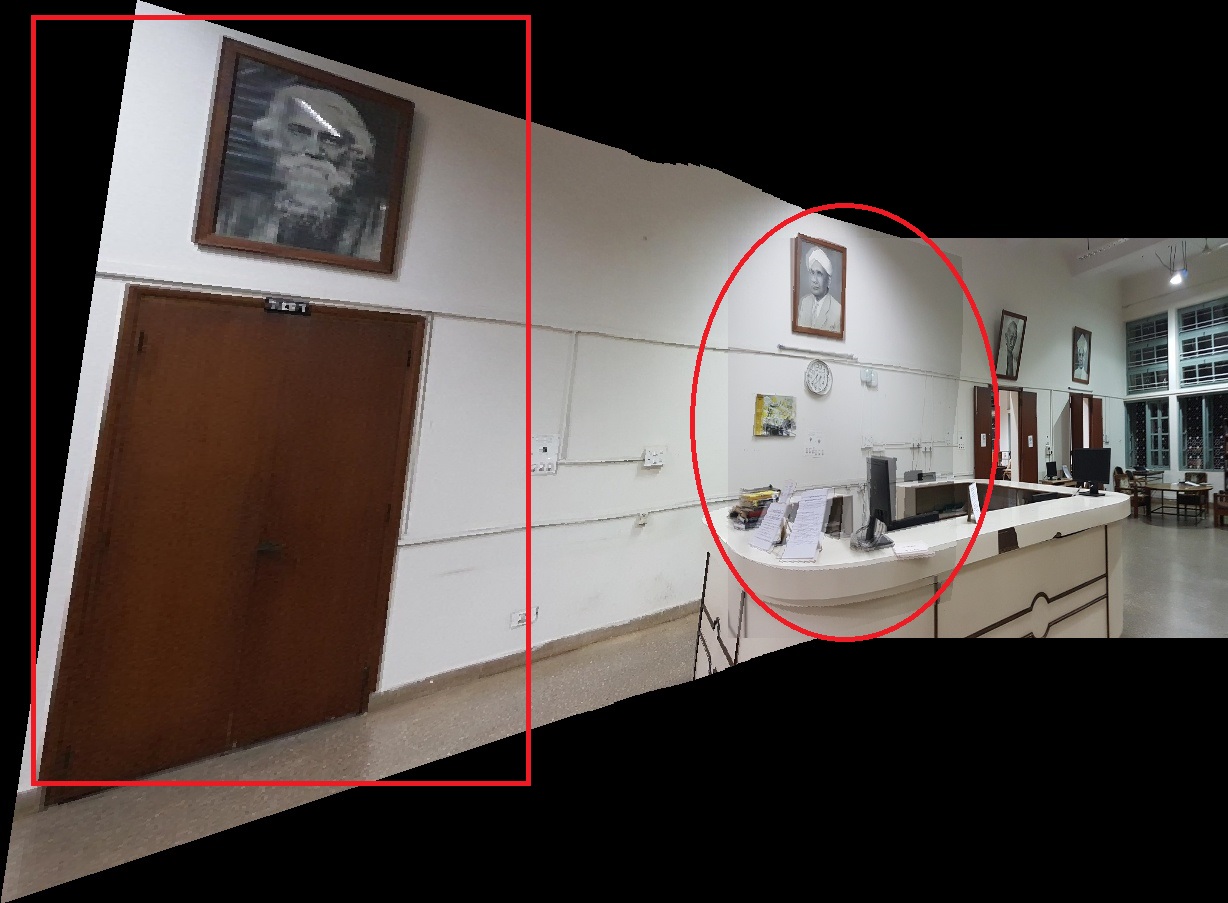}\\[2pt]
    \small APAP
  \end{minipage}

  \vspace{0.3em}

  \begin{minipage}[b]{0.60\linewidth}
    \centering
    \includegraphics[width=\linewidth]{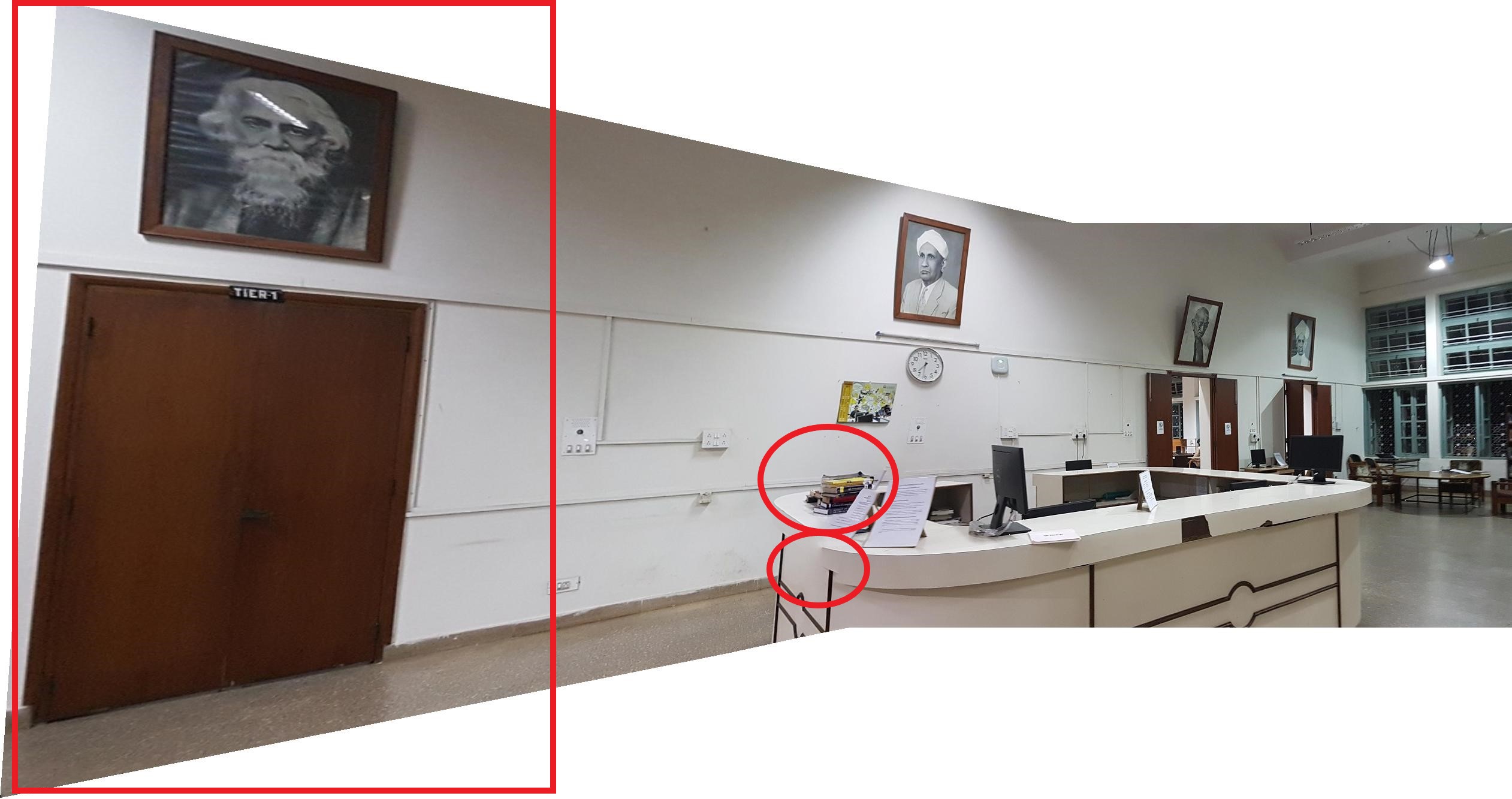}\\[2pt]
    \small LPC
  \end{minipage}

  \vspace{0.3em}

  \begin{minipage}[b]{0.60\linewidth}
    \centering
    \includegraphics[width=\linewidth]{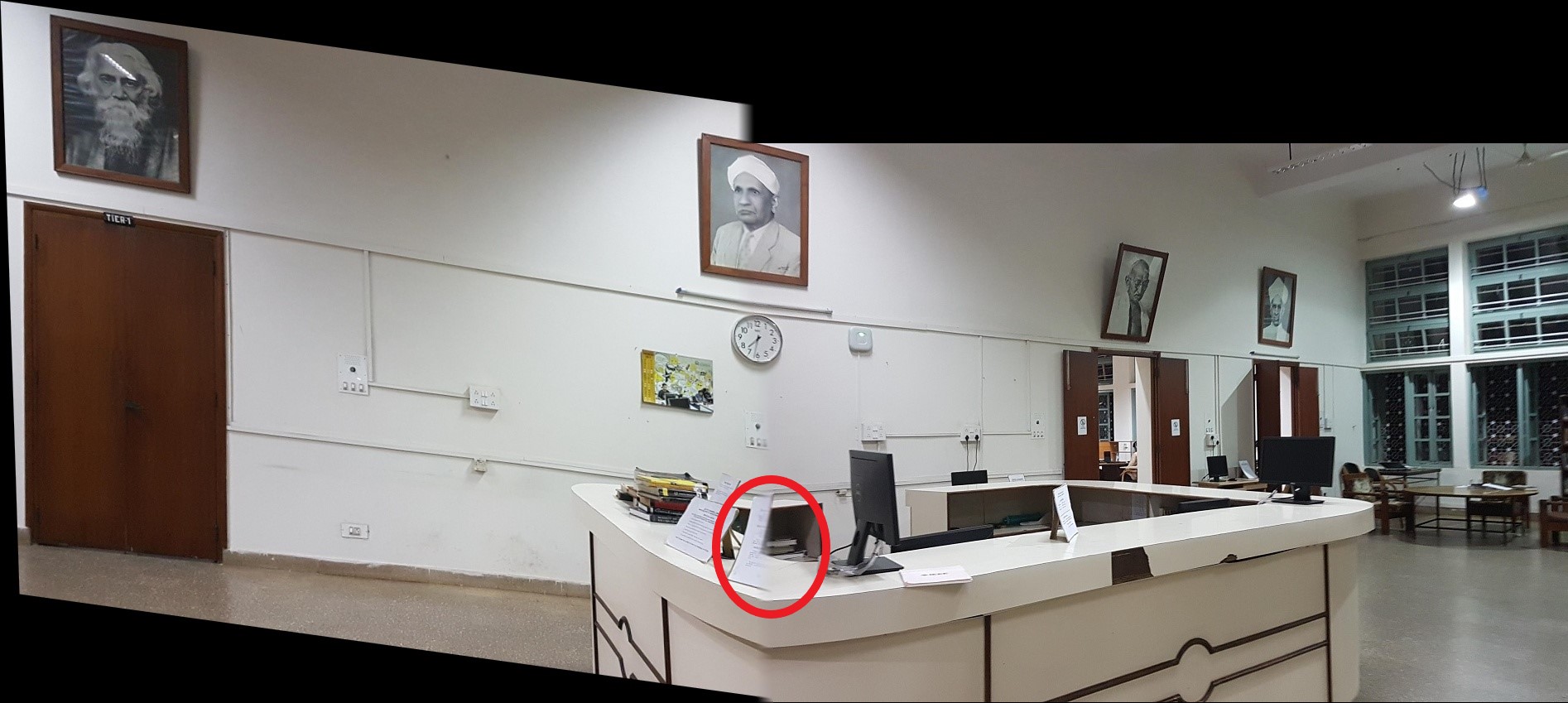}\\[2pt]
    \small OURS
  \end{minipage}

  \caption{Results of APAP \cite{zaragoza2013projective}, LPC \cite{jia2021leveraging},  and OURS.}
  \label{fig:comparison}
\end{figure}

\begin{figure}[H]
\centering

\newcommand{\ColW}{0.48\linewidth}
\newcommand{\SmallGap}{2pt}
\newcommand{\BigGap}{6pt}

\begin{minipage}[t]{\ColW}
  \centering


  \begin{minipage}{\linewidth}
    \centering
    \begin{minipage}[t]{0.45\linewidth}
      \centering
      \includegraphics[width=\linewidth]{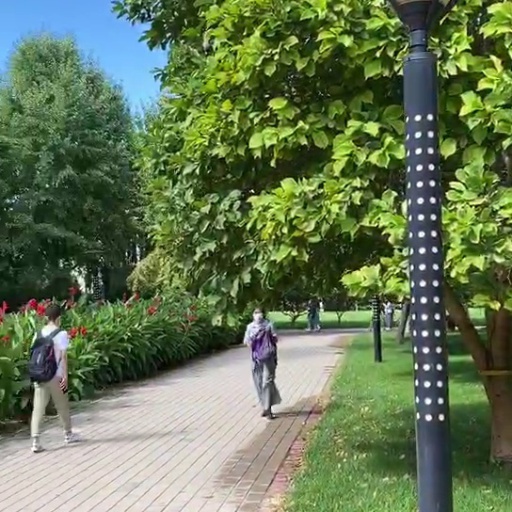}
    \end{minipage}\hfill
    \begin{minipage}[t]{0.45\linewidth}
      \centering
      \includegraphics[width=\linewidth]{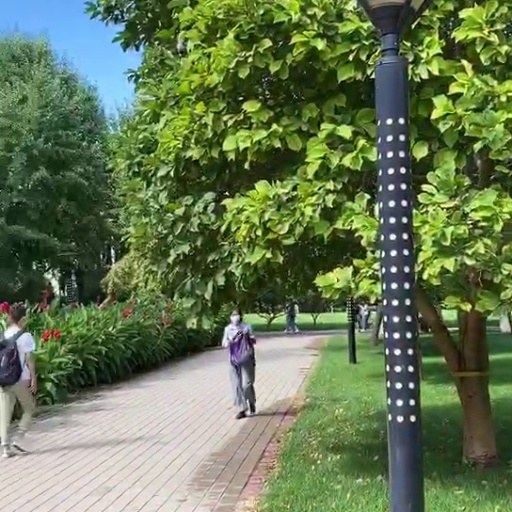}
    \end{minipage}\\[4pt]
    \small Input images
  \end{minipage}

  \vspace{\BigGap}

  \begin{minipage}{\linewidth}
    \centering
    \includegraphics[width=0.83\linewidth]{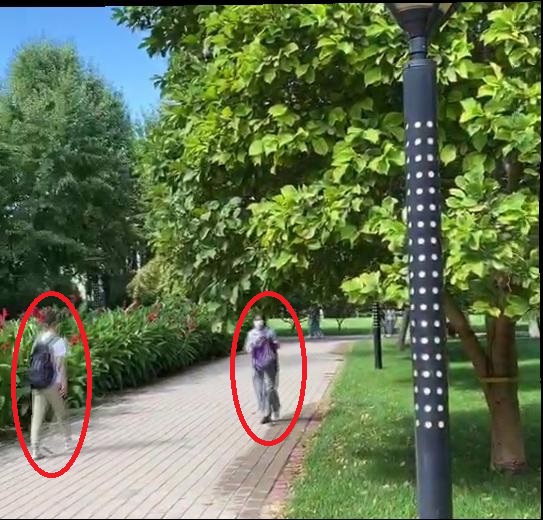}\\[2pt]
    \small APAP
  \end{minipage}\par\vspace{\BigGap}

  \begin{minipage}{\linewidth}
    \centering
    \includegraphics[width=0.83\linewidth]{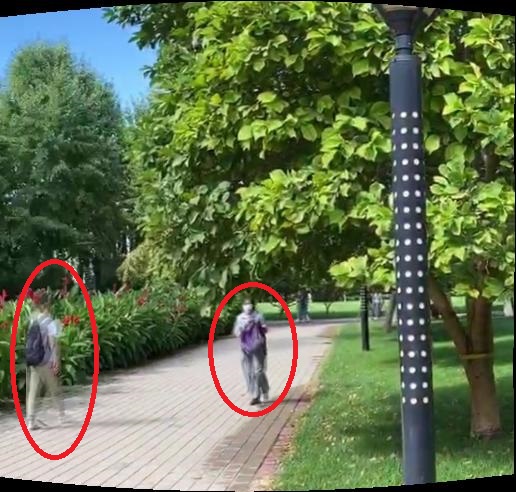}\\[2pt]
    \small ELA
  \end{minipage}\par\vspace{\BigGap}

  \begin{minipage}{\linewidth}
    \centering
    \includegraphics[width=0.83\linewidth]{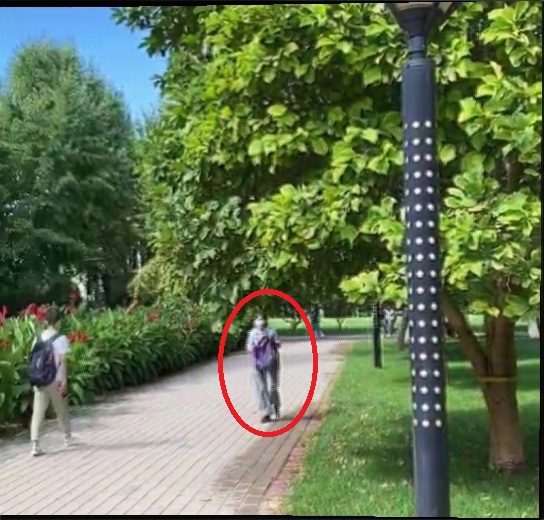}\\[2pt]
    \small UDIS
  \end{minipage}\par\vspace{\BigGap}

  \begin{minipage}{\linewidth}
    \centering
    \includegraphics[width=0.83\linewidth]{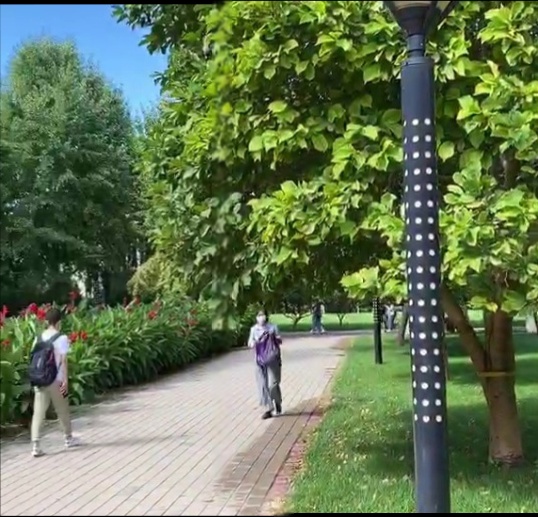}\\[2pt]
    \small OURS
  \end{minipage}

\end{minipage}%
\hfill%
\begin{minipage}[t]{\ColW}
  \centering


  \begin{minipage}{\linewidth}
    \centering
    \begin{minipage}[t]{0.45\linewidth}
      \centering
      \includegraphics[width=\linewidth]{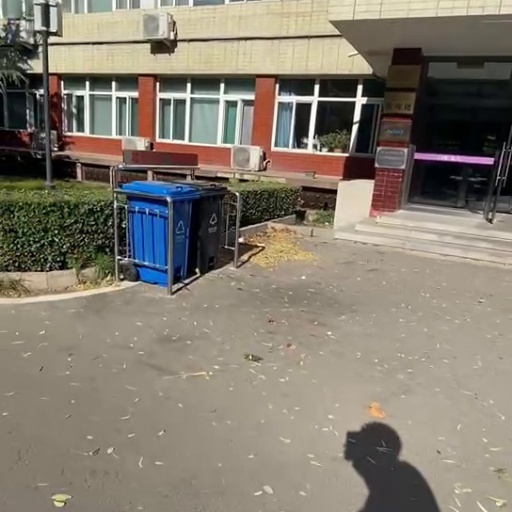}
    \end{minipage}\hfill
    \begin{minipage}[t]{0.45\linewidth}
      \centering
      \includegraphics[width=\linewidth]{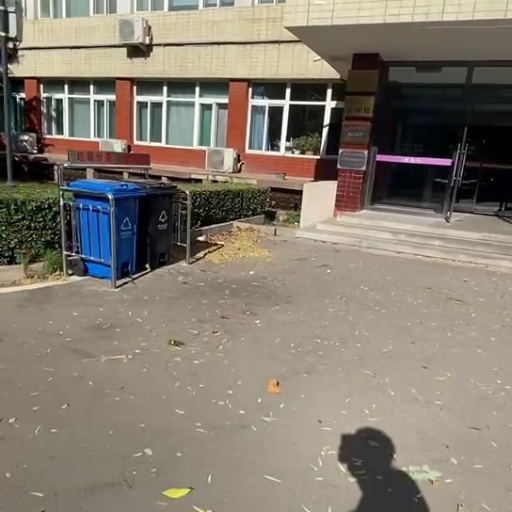}
    \end{minipage}\\[4pt]
    \small Input images
  \end{minipage}

  \vspace{\BigGap}

  \begin{minipage}{\linewidth}
    \centering
    \includegraphics[width=0.83\linewidth]{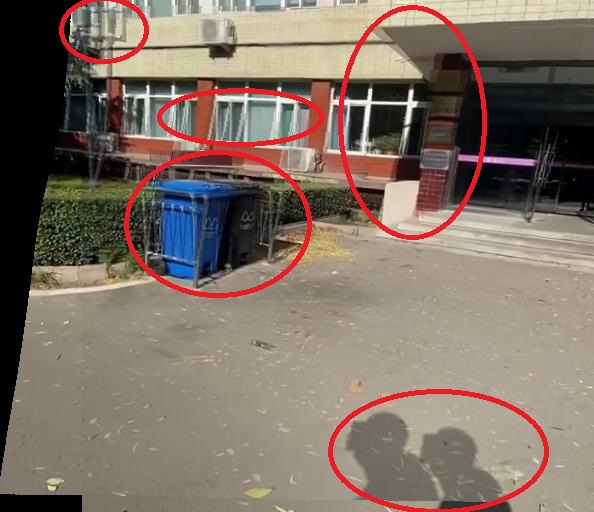}\\[2pt]
    \small APAP
  \end{minipage}\par\vspace{\BigGap}

  \begin{minipage}{\linewidth}
    \centering
    \includegraphics[width=0.83\linewidth]{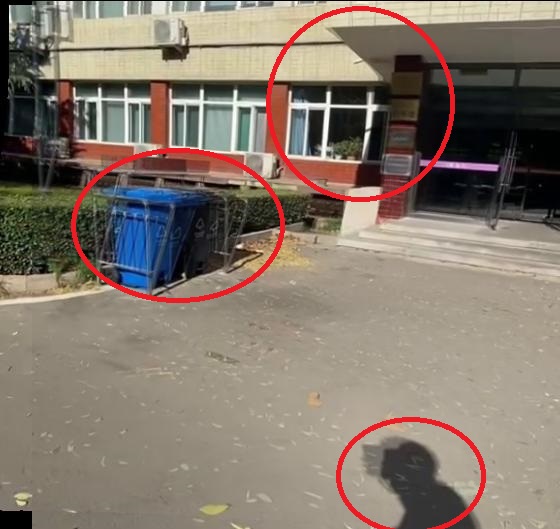}\\[2pt]
    \small ELA
  \end{minipage}\par\vspace{\BigGap}

  \begin{minipage}{\linewidth}
    \centering
    \includegraphics[width=0.83\linewidth]{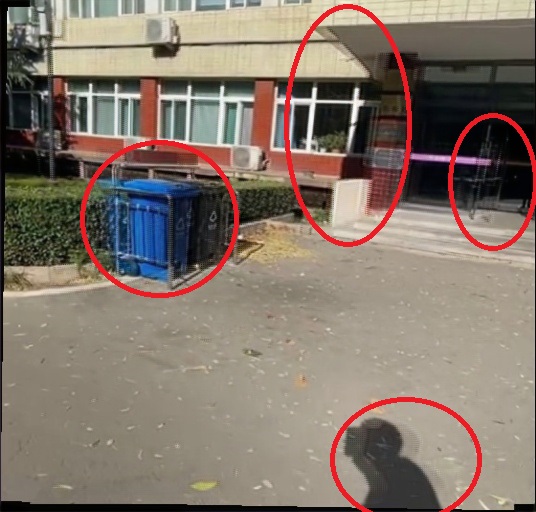}\\[2pt]
    \small UDIS
  \end{minipage}\par\vspace{\BigGap}

  \begin{minipage}{\linewidth}
    \centering
    \includegraphics[width=0.83\linewidth]{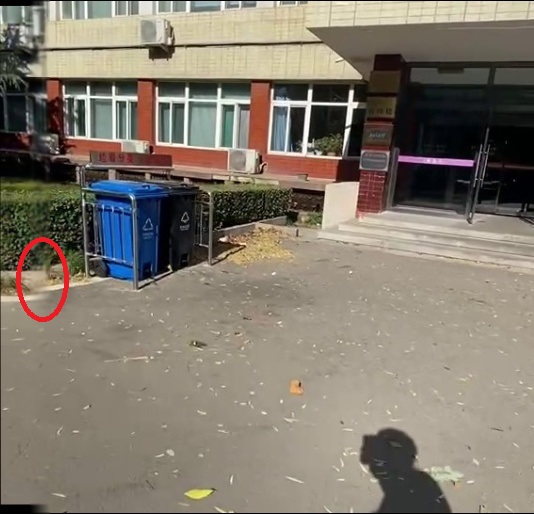}\\[2pt]
    \small OURS
  \end{minipage}

\end{minipage}

\caption{APAP, ELA, UDIS and OURS}
\label{fig:two_examples_4each_with_prefix}
\end{figure}

\bibliography{sn-bibliography}